\documentclass[12pt,english]{article}
\usepackage{lmodern}
\usepackage[T1]{fontenc}
\usepackage[latin9]{inputenc}
\usepackage{geometry}
\usepackage{amsmath}
\usepackage{amssymb}
\usepackage{esint}

\makeatletter
\usepackage{color}
\usepackage{hyperref}
\usepackage{cite}
\usepackage{babel}
\usepackage{mathdots}
\usepackage{amsfonts}
\usepackage{mathrsfs}
\usepackage{amsmath}
\usepackage{amssymb}
\usepackage{esint}
\usepackage{graphicx}
\usepackage{youngtab}
\usepackage{multicol}
\usepackage{multirow}
\usepackage{slashed}

\usepackage[titletoc]{appendix}

\allowdisplaybreaks

\numberwithin{equation}{section}

\date{}
\usepackage{datetime}

\usepackage[usenames,dvipsnames,svgnames,table]{xcolor}
\hypersetup{colorlinks=true, citecolor=red, linkcolor=blue, urlcolor=violet}

\usepackage{titlesec}
\titleformat{\section}{\large\bf}{\thesection}{1em}{}
\titleformat{\subsection}{\normalsize\bf}{\thesubsection}{1em}{}
\titleformat{\subsubsection}{\normalsize\it}{\thesubsubsection}{1em}{}

\makeatother

\usepackage{babel}

\begin{document}

\title{\textbf{\Large{Ho\v{r}ava-Lifshitz Gravity and Effective Theory\\of
the Fractional Quantum Hall Effect\\\bigskip{}\bigskip{}}}}

\author{\textbf{\normalsize{Chaolun Wu$^{\dagger}$ and Shao-Feng Wu$^{\star.\dagger}$}}}

\maketitle
\vspace{-16pt}

\begin{center}
\textit{$^{\dagger}$Kadanoff Center for Theoretical Physics and Enrico
Fermi Institute},
\par\end{center}

\begin{center}
\vspace{-35pt}

\par\end{center}

\begin{center}
\textit{University of Chicago, Chicago, Illinois 60637, USA}
\par\end{center}

\vspace{-6pt}

\begin{center}
\vspace{-35pt}

\par\end{center}

\begin{center}
\textit{$^{\star}$Department of Physics, Shanghai University, Shanghai
200444, China}
\par\end{center}

\vspace{-10pt}

\begin{center}
\textit{\small{Email: }}\texttt{\small{chaolunwu@uchicago.edu, sfwu@shu.edu.cn}}
\par\end{center}{\small \par}

\begin{center}
\thispagestyle{empty}
\par\end{center}

\begin{abstract}
We show that Ho\v{r}ava-Lifshitz gravity theory can be employed as
a covariant framework to build an effective field theory for the fractional
quantum Hall effect that respects all the spacetime symmetries such
as non-relativistic diffeomorphism invariance and anisotropic Weyl
invariance as well as the gauge symmetry. The key to this formalism
is a set of correspondence relations that maps all the field degrees
of freedom in the Ho\v{r}ava-Lifshitz gravity theory to external background
(source) fields among others in the effective action of the quantum
Hall effect, according to their symmetry transformation properties.
We originally derive the map as a holographic dictionary, but its
form is independent of the existence of holographic duality. This
paves the way for the application of Ho\v{r}ava-Lifshitz holography
on fractional quantum Hall effect. Using the simplest holographic
Chern-Simons model, we compute the low energy effective action at
leading orders and show that it captures universal electromagnetic
and geometric properties of quantum Hall states, including the Wen-Zee
shift, Hall viscosity, angular momentum density and their relations.
We identify the shift function in Ho\v{r}ava-Lifshitz gravity theory
as minus of guiding center velocity and conjugate to guiding center
momentum. This enables us to distinguish guiding center angular momentum
density from the internal one, which is the sum of Landau orbit spin
and intrinsic (topological) spin of the composite particles. Our effective
action shows that Hall viscosity is minus half of the internal angular
momentum density and proportional to Wen-Zee shift, and Hall bulk
viscosity is half of the guiding center angular momentum density. 
\end{abstract}

\newpage{}

{\hypersetup{linkcolor=black}

\tableofcontents{}

}

\bigskip{}
\bigskip{}
\bigskip{}

\section{Introduction}

\subsection{Aspects of the Fractional Quantum Hall Effect}

The fractional quantum Hall effect (FQHE) \cite{Tsui1982} has been
a fascinating subject that attracts physicists from many different
areas of both experimental and theoretical physics over the last three
decades. The crucial role played by interactions between electrons
poses a huge challenge to theorists and prevents the problem from
being solved exactly till these days. Despite this difficulty, many
different theoretical approaches to the problem has been developed
since Laughlin's ground-breaking work on the trial wave function \cite{Laughlin1983}.
These approaches employ many different concepts and theoretical tools
from various fields of modern physics and even mathematics, to reveal
diverse aspects of rich physics encoded in this simple phenomenon.

One class of these approaches is based on field theories \cite{ZHK1988,Lopez1991,Wen:1992uk,Zhang:1992eu,Halperin1993},
particularly the Chern-Simons gauge theory in 2+1 dimensions. However,
such approaches have shortages and limitations. One of the problems
is the difficulty to deal with the massless limit. In the massless
limit, the cyclotron frequency goes to infinity and the inter-Landau
level gap becomes much larger than the Coulomb energy gap, which only
depends on the magnetic field. This corresponds to the projection
to the lowest Landau level (LLL). In many field theoretical approaches,
such as the composite boson \cite{ZHK1988} and composite fermion
\cite{Lopez1991} theories, the inter-Landau level mixing causes the
LLL projection to be unnatural and many quantities which should depend
only on Coulomb energy scale are also sensitive to cyclotron frequency.
The smoothness of the massless limit is a requirement for a field
theory formalism to be an adequate description of the quantum Hall
effect. 

At the low energy (much lower compared to the Coulomb gap) effective
theory level, Chern-Simons field theory is believed to encode all
the universal properties of quantum Hall states \cite{Wen:1992uk}.
However, in addition to the gauge degrees of freedom related to electromagnetic
properties such as the quantized Hall conductivity \cite{Laughlin:1981,TKNN:1982,Niu:1985},
fractionally charged (anyonic) excitations \cite{Laughlin1983,Halperin1983,Arovas1984}
and chiral edge states \cite{Wen:1989mw,Wen:1990qp,Wen:1992vi} that
are well described by gauge Chern-Simons theory \cite{Frohlich:1990xz,Frohlich:1991wb},
the fractional quantum Hall states also possesses another sets of
degrees of freedom, which is called the ``geometric degrees of freedom''
by Haldane \cite{Haldane:2011ia}. This is another hallmark of the
fractional quantum Hall effect that makes it fundamentally different
from and more complicated than the integer quantum Hall effect \cite{vonKlitzing1980}.
The origin of these two sets of degree of freedom can be traced back
to the structure of Hilbert space of charged particles in magnetic
field \cite{Girvin1984,Dunne:1991cs} and the underlying symmetry
of $W_{\infty}$ algebra \cite{Cappelli:1992yv}. Early study related
to the geometric degrees of freedom dated back to the magneto-roton
theory of collective neutral excitations\cite{Girvin1986}. A new
quantum number, the shift $\mathcal{S}$, was introduced in \cite{Wen:1992ej}
as the coefficient of a new Chern-Simons-type term, the Wen-Zee term
\[
\frac{\mathcal{S}}{4\pi}\int\omega\wedge\mathrm{d}A
\]
where $A$ is the $U(1)$ gauge field and $\omega$ is the Abelian
spin connection associated with the 2-dimensional spatial manifold.
The shift is topological in nature and its role to characterize gravitational
response of fractional quantum Hall states is similar to that of the
quantized Hall conductivity to the electromagnetic response. The topological
phases of fractional quantum Hall states are determined by Hall conductivity
and the shift together, not the former alone. Although the shift is
originally introduced to describe quantum Hall states on curved manifolds
with non-trivial topology, it also manifests itself in states on flat
manifolds as a parity-odd dissipationless transport coefficient known
as Hall viscosity \cite{Avron:1995fg,Avron:1998,Tokatly0706,Tokatly0812,Read:2008rn,Read:2010epa,Hoyos:2011ez,Hidaka:2012rj,Wiegmann:1305,Biswas:2013eba,Fremling:2013kda,Hoyos:2014pba},%
\footnote{A closely related situation is Hall viscosity in chiral superfluids
\cite{Hoyos:2013eha,Moroz:2014ska,Shitade:2014iya,Hoyos:2014nua}
(in addition to \cite{Read:2008rn,Read:2010epa}). Hall viscosity
in relativistic systems has also been studied in \cite{Hughes:2011hv,Hughes:2012vg,Parrikar:2014usa},
\cite{Geracie:2014iva} and \cite{Golkar:2014wwa,Golkar:2014paa}.
The latter is related to quantum Hall effects in graphene. But since
the spacetime symmetry is different in these cases, they are not considered
here.%
} which in turn is related to the angular momentum density in gapped
systems \cite{Read:2010epa,Nicolis:2011ey,Bradlyn:2012ea,Hoyos:2014lla}.
Recently, Haldane proposed a geometric point of view \cite{Haldane:2011ia}
to renew the understanding of fractional quantum Hall dynamics, in
which the Hall viscosity \cite{Haldane:2009ke,Haldane:1112,Haldane1403}
and neutral collective excitations \cite{Haldane1201} are related
to the fluctuations of guiding center metric defined in the 2-dimensional
phase space of the non-commutative guiding center coordinates. A closely
related topic is the study of the nematic degrees of freedom, whose
dynamics is also governed by the Wen-Zee term. Inspired by early works
of \cite{Mulligan:2010wj,Mulligan:2011eg}, \cite{Maciejko:2013dia,You:2013soa}
propose field theoretical descriptions of the quantum Hall nematic
transition in which an isotropic Laughlin liquid undergoes phase transition
to a nematic state with the same filling factor but spontaneously
breaks rotational symmetry, and identify the order parameter as unimodular
spatial metric components.

\subsection{Symmetries and Quantum Hall Effective Field Theory}

In field theoretical approaches, symmetries play an important role
in building phenomenological models as well as solving and subtracting
physical information. Gauge invariance and Galilean symmetry, particularly
the former one, have been well known to be crucial to the understanding
of quantum Hall effects \cite{Frohlich:1993gs}. However, these are
not the only symmetries that are relevant and can be utilized in the
study. Non-relativistic systems possess a larger set of symmetries
than the aforementioned two. In \cite{Son:2005rv} the notion of \emph{non-relativistic
diffeomorphism invariance} is introduced, and further enlarged to
its maximal in \cite{Geracie:2014nka} for quantum Hall systems to
accommodate arbitrary spacetime diffeomorphism, a general gyromagnetic
factor and intrinsic spin. It is first used to construct a effective
action for quantum Hall effects in \cite{Hoyos:2011ez} to explain
the relation between Hall viscosity and the inhomogeneous magnetic
field correction to the Hall conductivity. \cite{Son:2013rqa} introduces
the Newton-Cartan geometry as a mathematical machinery that manifests
the non-relativistic diffeomorphism invariance in a covariant way,
and constructs an effective action that resolves the unnaturalness
problem of LLL projection and encodes most of the universal properties
of fractional quantum Hall states in a natural and unified way. The
Newton-Cartan formalism is further used to study spectral density
sum rules \cite{Golkar:2013gqa}, conservation laws and Ward identities
\cite{Geracie:2014nka}, hydrodynamics \cite{Geracie:2014TBA} of
quantum Hall fluids and extended to include torsion \cite{Geracie:2014nka,Gromov:2014vla}.
Recently, \cite{Jensen:2014aia} offers an extensive formal discussion
on algebraic aspects of Newton-Cartan geometry as a formalism of Galilean-invariant
field theories. Non-relativistic diffeomorphism invariant effective
actions can also be constructed without using Newton-Cartan formalism
\cite{Andreev:2013qsa,Abanov:2014ula,Gromov:2014gta,Andreev:2014gia}.
Some recent general discussions on non-relativistic spacetime symmetries
can be found in \cite{Banerjee:2014pya,Banerjee:2014nja,Brauner:2014jaa},
and similar ideas are applied to construct low energy effective action
for non-relativistic gapped systems in \cite{Bradlyn:2014wla}.

The idea behind the approach of \cite{Hoyos:2011ez,Son:2013rqa} is
that, a single gauge Chern-Simons term $\int A\wedge\mathrm{d}A$
alone as the leading order term of the low energy effective theory
does not respect the full spacetime symmetries of the underlying microscopic
field theory. The non-relativistic diffeomorphism invariance, along
with other constraints such as the naturalness of LLL projection,
requires a series of additional terms and fixes the relative coefficients
between them. This is how other universal properties of fractional
quantum Hall states mentioned earlier are embedded into the Chern-Simons
theory. Strictly speaking, the local non-relativistic diffeomorphism
invariance is just a spurionic symmetry of the system. It does not
induce a conservation law per Noether's theorem. It is not even uniquely
determined because to read it off the underlying microscopic field
theory has to be coupled to curved spacetime first and this coupling
is not unique. Only a subgroup of the full diffeomorphism invariance
that preserves a certain background (usually flat) is the true symmetry
(the Galilean or Schrödinger algebra). However, the full diffeomorphism
invariance is a powerful tool to constrain the form of the effective
action and to subtract physical information such as Ward identities
higher correlation functions and relations between transport coefficients
at the low energy level.

Another point to note is that, given a specific form of non-relativistic
diffeomorphism invariance (i.e. the coordinate transformation rules
for all fields and parameters in the theory that keep the action and
effective action invariant), there may exist more than one covariant
formalisms of the geometry one can use as the machinery to build up
the effective theory. Depending on the purpose, one may be more convenient
than another or vice versa. The Newton-Cartan formalism introduced
in \cite{Son:2013rqa}, with its non-relativistic nature \cite{Duval:2009vt}
and rich structure, becomes a very popular one to describe quantum
Hall effects. In this paper, however, we will employ a different covariant
formalism -- the Ho\v{r}ava-Lifshitz gravity theory, and show that
it is an equally powerful framework for quantum Hall effective theory.
It has one notable advantage over the Newton-Cartan formalism. It
is closely associated to relativistic gravity theories (for example,
through the khronon formalism \cite{Blas:2009yd,Germani:2009yt}),
thus provides a natural connection to the holographic approach, which
has been developed so far mostly within the frameworks of relativistic
gravity and string theories.

The idea of using Ho\v{r}ava-Lifshitz gravity theory to construct
effective actions with non-relativistic diffeomorphism invariance
(non-holographically) has already been introduced in \cite{Andreev:2013qsa}
(followed by \cite{Kluson:2014lya}), but their formalism is limited
in several aspects. Their non-relativistic diffeomorphism is not the
most general one for quantum Hall effect. Their construction relies
on taking the non-relativistic limit $c\rightarrow\infty$ of some
relativistic theory, which will usually not give the most general
results as to directly consider the diffeomorphism itself, thus is
over-constrained. Last but not least, in their metric ansatz they
identify the $U(1)$ field directly as the subleading order metric
components in $1/c$ expansion. We think the last part is the major
limitation and we will introduce the $U(1)$ field separately in our
formalism, independent of the graviton sector, and will identify the
shift function in the metric as something different in the context
of quantum Hall effects.

\subsection{Holography: Relativistic vs. Non-Relativistic Symmetries}

Holography, or the gauge/gravity duality \cite{Maldacena:1997re,Gubser:1998bc,Witten:1998qj},
has been a powerful tool to study strongly-correlated quantum systems
observed or theoretically proposed in many different areas of modern
physics. Its higher dimensional geometric approach gives dual descriptions
of quantum field theories at strong coupling regime, which is complementary
to the traditional perturbative methods. An early triumph of holography
in high energy physics is the prediction of the lower bound of shear
viscosity to entropy density ratio of strongly coupled quark-gluon
plasma \cite{Policastro:2002se,Kovtun:2003wp,Buchel:2003tz,Kovtun:2004de}.
Holography has also been applied to many condensed matter systems
and sheds new light on many age old problems, offering new perspectives
and efficient computational methods (for an incomplete list of reviews
of some most popular topics, we refer to \cite{Hartnoll:2009sz,Herzog:2009xv,McGreevy:2009xe,Horowitz:2010gk,Sachdev:2010ch,Iqbal:2011ae,Adams:2012th}).
A topic related to the current discussion is Lifshitz holography,
the holographic dual to field theories with Lifshitz scaling, starting
from \cite{Kachru:2008yh}. It typically employs relativistic gravity
coupled to various types of matter fields, among which the most popular
version is introduced in \cite{Taylor:2008tg}. Of particular interest
are the recent work of \cite{Christensen:2013lma,Christensen:2013rfa}
where the boundary geometry of $z=2$ Lifshitz holography under certain
conditions for the time-like vielbein is identified with Newton-Cartan
geometry with or without torsion, and of \cite{Chemissany:2014xpa,Chemissany:2014xsa}
where a systematic approach for constructing holographic dictionary
for Lifshitz holography is presented. Instead of using relativistic
gravity coupled to matters, we will use Ho\v{r}ava-Lifshitz gravity
to realize the Lifshitz holography. It has Lifshitz geometry as a
vacuum gravity solution without coupling to matters.

The fractional quantum Hall effect, a state of quantum fluids largely
due to the strong correlations of electrons through Coulomb interaction
in high magnetic field, is another natural playground for holography.
Early researches based on either bottom-up phenomenological approaches
or top-down string/brane settings turn out to be fruitful \cite{KeskiVakkuri:2008eb,Davis:2008nv,Fujita:2009kw,Hikida:2009tp,Alanen:2009cn,Bergman:2010gm,Bayntun:2010nx,Gubankova:2010rc,Jokela:2010nu,Fujita:2012fp,Melnikov:2012tb,Kristjansen:2012ny,Jokela:2013hta,Kristjansen:2013hma}.
However, one of the limitations of most of these studies is that as
they focus primarily on electromagnetic properties of the quantum
Hall phenomena, particularly the conductivity. Most of the geometric
properties that also characterize quantum Hall states, such as the
shift and Hall viscosity, are overlooked, thus the holographic approach
to quantum Hall effects so far is still incomplete. One purpose of
this paper is to fill in this gap. The origin of this problem can
be traced back to the spacetime symmetry of these holographic models.
Since they are all built in within the framework of string theory
or relativistic gravity theory, their spacetime symmetry is relativistic
by nature. This does not match the non-relativistic diffeomorphism
invariance of \cite{Son:2005rv,Son:2013rqa,Geracie:2014nka}, which
is more appropriate for the description of quantum Hall effects and
is crucial to the encoding of geometric properties into the effective
description. For example, as the (2+1)-dimensional gauge Chern-Simons
term $A\wedge\mathrm{d}A$ can easily find its dual description in
(3+1)-dimensional holographic model as the bulk Chern-Simons term
$\mathrm{d}A\wedge\mathrm{d}A$, the Wen-Zee term $\omega\wedge\mathrm{d}A$
can not because the spin connection in relativistic (3+1)-dimensional
bulk is naturally non-Abelian and a term like $\mathrm{d}\omega\wedge\mathrm{d}A$
is forbidden by symmetry. Our solution to this problem is to use a
non-relativistic gravity theory -- the Ho\v{r}ava-Lifshitz gravity
-- as the bulk description where a Wen-Zee term is not forbidden by
symmetry at least at the boundary and thus can be induced by the holographic
dictionary.

A guiding principle to construct a dual holographic model for a certain
quantum field theory is the matching of all symmetries on both sides.
This is especially important for the bottom-up approach where means
of constraining the form of the action are limited. In classic examples
of AdS/CFT correspondence \cite{Aharony:1999ti}, the full conformal
and flavor symmetries of the field theories are completely realized
by the background isometries of the dual gravity theories. This is
necessary for the top-down approach, but for the bottom-up approach
which we will follow here, especially for the applications to condensed
matter physics, this requirement that the full global symmetries are
realized by background isometries might be too strong and not necessary.
A weaker condition that the holographic on-shell action encodes all
the field theory symmetries is sufficient in practice. This means
that the bulk background isometry does not have to incarnate the full
symmetry algebra \cite{Balasubramanian:2010uw}. This weak requirement
is particularly useful for building up bottom-up models for systems
with non-relativistic symmetries to avoid certain complications such
as excessive extra dimensions. The difference between the strong and
weak requirements is that in the latter the holographic dictionary
will be more complicated and play a more vivid role to realize the
rest of the symmetry that is not realized by the isometry. In fact,
according to the standard holographic dictionary, 
\begin{equation}
\mathcal{I}_{\mathrm{grav}}[\bar{\phi}]=\mathcal{W}_{\mathrm{QFT}}\left[J\right]\:,\qquad\qquad\bar{\phi}=J\:,
\end{equation}
where $\mathcal{I}_{\mathrm{grav}}$ is the on-shell action of the
(weakly coupled or classical) gravity theory and $\mathcal{W}_{\mathrm{QFT}}$
the effective action%
\footnote{Throughout this paper, by ``effective action'' $\mathcal{W}$, we
simply mean the logarithm of the generating functional $\mathcal{Z}$
of the quantum field theory, as a functional of the source $J$: $\mathcal{W}\left[J\right]=-i\log\mathcal{Z}\left[J\right]$.%
} of the dual (strongly-coupled) quantum field theory. $\bar{\phi}$
denotes a collection of normalizable mode coefficients of the bulk
fields, which are usually called the boundary fields and identified
as the source field $J$ to some operator $\hat{\mathcal{O}}$ in
the quantum field theory. The form of $\mathcal{W}_{\mathrm{QFT}}\left[J\right]$
is highly constrained by the symmetries of the quantum field theory,
then so is that of $\mathcal{I}_{\mathrm{grav}}[\bar{\phi}]$ according
to the dictionary. Thus the necessary and sufficient condition for
the matching of symmetries is that the on-shell action $\mathcal{I}_{\mathrm{grav}}[\bar{\phi}]$
respects symmetries of the field theory. This is a combined result
of the bulk background isometry (particularly near the boundary) and
the holographic dictionary, not only the former one alone. In the
classic AdS/CFT correspondence \cite{Aharony:1999ti} where the holographic
dictionary takes a simple and trivial form (for example, the boundary
gauge field maps \emph{exactly} to the source of the conserved current
of the same gauge group in the field theory), the requirement of fulfilling
the symmetries falls upon completely on the background isometry. In
non-relativistic holography, this is not always true, or at least
certainly not as cheap as that for the relativistic. In early works
of \cite{Son:2008ye,Balasubramanian:2008dm},%
\footnote{For a review from algebraic point of view, we refer to \cite{Dobrev:2013kha}.%
} for example, the spacetime symmetry group for unitary fermions, the
Schrödinger group, is fully realized by the background isometry, but
at the price of introducing two extra dimensions rather than one in
relativistic cases. Furthermore, this type of models suffer problems
such as zero modes and infinite Kaluza-Klein towers in the mass spectrum
resulting from the discrete light-cone quantization (DLCQ). They can
hardly be the holographic duals of generic non-relativistic field
theories. A first step toward a more generic setup by getting rid
of the second extra dimension and giving up the strong requirement
on bulk isometry is made by \cite{Balasubramanian:2010uw}, where
an extra gauge field has to be introduced and the holographic dictionary
becomes less trivial by mixing the metric and the gauge field. \cite{Janiszewski:2012nb,Janiszewski:2012nf}
makes a further step by directly employing a non-relativistic bulk
gravity theory -- Ho\v{r}ava-Lifshitz gravity, and building the holographic
dictionary by matching to the non-relativistic diffeomorphism invariance
of \cite{Son:2005rv}. In this paper, we will follow closely the philosophy
of \cite{Janiszewski:2012nb,Janiszewski:2012nf} and extend their
approach to the parity-violating case by building a holographic dictionary
that matches the non-relativistic local spacetime and gauge symmetries
of \cite{Geracie:2014nka}. This offers a general platform for building
more detailed holographic models dual to strongly-coupled spin-polarized
charged particle systems in magnetic field, particularly the quantum
Hall effects.

\subsection{Ho\v{r}ava-Lifshitz Gravity Theory}

In the above when we are talking about matching the symmetries of
a field theory and its holographic dual, we mean the full local symmetries,
including the ``spurionic'' diffeomorphism invariance and local
gauge invariance, not just the global symmetries. A detailed explanation
of this point and its connection with building non-relativistic holographic
dictionary is presented in \cite{Janiszewski:2012nf}. In non-relativistic
systems, there is a preferred notion of time -- the global time, which
defines simultaneity and is a consequence of non-relativistic causality.
The general coordinate transformations that preserve the global time
foliation on the manifold involve an arbitrary time-dependent time
reparametrization and an arbitrary time- and space-dependent spatial
diffeomorphism:
\begin{equation}
\delta t=-\xi^{t}(t)\:,\qquad\delta x^{i}=-\xi^{i}(t,x^{j})\:,
\end{equation}
which is usually called foliation preserving diffeomorphism (FPD).
Here $i$ and $j$ run through all spatial directions. Non-relativistic
diffeomorphism invariance is a statement that physics is invariant
under these FPD. To build up the holographic dual, it is natural to
seek a gravity theory with the same causal structure, notion of global
time and diffeomorphism invariance. The candidate is Ho\v{r}ava-Lifshitz
gravity theory \cite{Horava:2008ih,Horava:2009uw}, a gravity theory
that is constructed to be invariant under the above FPD. The gravity
sector includes the lapse function $N$, shift function $N_{i}$ and
spatial metric $g_{ij}$, which can be viewed as the Arnowitt-Deser-Misner
(ADM) decomposition of the full spacetime metric \cite{Arnowitt:1962hi}:
\begin{equation}
ds^{2}=-N^{2}dt^{2}+g_{ij}\left(dx^{i}+N^{i}dt\right)\left(dx^{j}+N^{j}dt\right)\:.
\end{equation}
Their transformation under FPD will be listed later when we talk about
the bulk theory of the holography. Time derivative appears only in
the extrinsic curvature 
\begin{equation}
K_{ij}=\frac{1}{2N}\left(\partial_{t}g_{ij}-\nabla_{i}N_{j}-\nabla_{j}N_{i}\right)\:,
\end{equation}
where $\nabla_{i}$ is the covariant derivative associated with the
spatial metric $g_{ij}$. We define $K=g^{ij}K_{ij}$.

Ho\v{r}ava-Lifshitz gravity is originally proposed as a ultraviolet-complete
quantum gravity theory to describe the real world. For this purpose,
there are many phenomenological and cosmological issues under debates
where there exists an extensive literature (for some early reviews
and references, see \cite{Mukohyama:2010xz,Sotiriou:2010wn}). But
these are mostly irrelevant to us, since we are only interested in
using the theory as a framework to build up effective theories or
holographic duals for non-relativistic field theories. We do \emph{not}
need to assume \emph{projectability} or \emph{detailed balance} as
introduced in \cite{Horava:2009uw}. We are mostly interested in the
classical low energy limit (analog to the large $N_{c}$ limit of
the relativistic AdS/CFT correspondence) of the theory where the lowest
derivative terms dominate. The general form of the leading order graviton
action up to two-derivative terms is \cite{Blas:2009yd,Blas:2009qj,Blas:2010hb}:
\begin{equation}
S_{\mathrm{grav}}=\frac{1}{16\pi G_{N}}\int dtd^{d}\vec{x}\sqrt{g}N\left\{ K_{ij}K^{ij}-\left(1+\tilde{\lambda}\right)K^{2}+\beta\left(R-2\Lambda\right)+\alpha\frac{\left(\nabla_{i}N\right)\left(\nabla^{i}N\right)}{N^{2}}\right\} \:.\label{eq:HoraveAction_Graviton}
\end{equation}
where $R$ is the Ricci scalar associated with $g_{ij}$ and $\Lambda$
cosmological constant and spatial indices $i$ and $j$ are raised
and lowered by $g^{ij}$ and $g_{ij}$. $G_{N}$ is Newton's constant,
but as part of effective theory for condensed matter systems, it will
be mapped to some parameters in the problem under consideration. When
$\alpha=\tilde{\lambda}=0$ and $\beta=1$ the action goes back to
Einstein-Hilbert form. This low energy form of Ho\v{r}ava-Lifshitz
gravity is a limiting case \cite{Jacobson:2010mx,Jacobson:2013xta}
of the Einstein-aether theory \cite{Jacobson:2000xp,Eling:2003rd,Jacobson:2008aj}
with hypersurface orthogonal aether field. We will also need a $U(1)$
gauge field $V_{\mu}=\left(V_{t},V_{i}\right)$, whose low energy
action up to two derivatives has the general form \cite{Kiritsis:2009sh}
\begin{equation}
S_{\mathrm{gauge}}=\frac{1}{4g_{e}^{2}}\int dtd^{d}\vec{x}\sqrt{g}N\left\{ \frac{2}{N^{2}}g^{ij}\left(V_{ti}-N^{k}V_{ki}\right)\left(V_{tj}-N^{l}V_{lj}\right)+f\left[V_{1},V_{2}\right]\right\} \:,\label{eq:HoravaAction_Gauge}
\end{equation}
where $V_{\mu\nu}=\partial_{\mu}V_{\nu}-\partial_{\nu}V_{\mu}$, $g_{e}$
is the gauge coupling and $f$ is an arbitrary scalar functional of
its arguments. For $d=3$, $V_{1}=g^{ij}\mathfrak{B}_{i}\mathfrak{B}_{j}$,
$V_{2}=\nabla_{i}\mathfrak{B}_{j}\nabla^{i}\mathfrak{B}^{j}$ and
$\mathfrak{B}_{i}=\frac{1}{2}\varepsilon_{i}^{\phantom{i}jk}V_{jk}$.
For $d=2$, $V_{1}=\mathfrak{B}$, $V_{2}=\nabla_{i}\mathfrak{B}\nabla^{i}\mathfrak{B}$
and $\mathfrak{B}=\frac{1}{2}\varepsilon^{ij}V_{ij}$. Here $\varepsilon$
with indices is the Levi-Civita tensor associated with the metric
$g_{ij}$ in specific dimensions. The form of a scalar action is also
discussed in \cite{Kiritsis:2009sh}, but we will not need them explicitly
in this paper. The Chern-Simons terms can be added in a similar way
as in relativistic theories. We will discuss this in great details
in late sections. In 2+1 dimensions, a FPD invariant Wen-Zee term
can be added as 
\begin{equation}
S_{\mathrm{WZ}}=\frac{\mathcal{S}}{4\pi}\int\omega^{\prime}\wedge\mathrm{d}V\:.
\end{equation}
where $\omega^{\prime}=\omega+\ldots$ is a covariant Abelian spin
connection defined on the 2-dimensional spatial manifold. The $\omega^{\prime}$
is different from the Abelian spin connection $\omega$ in the classic
Wen-Zee term introduced in \cite{Wen:1992ej} by the ``$\ldots$''
part, because $\omega$ is not a FPD covariant 1-form. The precise
definition of $\omega^{\prime}$ can be found in (\ref{eq:CovSpinConn_wt})
and (\ref{eq:CovSpinConn_wi}).

In \cite{Horava:2010zj,daSilva:2010bm} the Ho\v{r}ava-Lifshitz gravity
described above are extended. A local $U(1)$ symmetry called $U(1)_{\Sigma}$
is introduced and the FPD invariance is upgraded to ``non-relativistic
general covariance''. Along with these is the introduction of some
additional fields into the theory, such as the ``Newton potential''
$A$ and ``Newton prepotential'' $\nu$. They are related to the
subleading terms in the metric in the non-relativistic limit where
the speed of light $c\rightarrow\infty$. \cite{Janiszewski:2012nb}
discusses their roles in holographic model building. Since our theories
have only FPD symmetry, not the non-relativistic general covariance,
we will not introduce these fields in our model. This agrees with
the spirit of \cite{Janiszewski:2012nb}.

Related to applications on holography, Ho\v{r}ava-Lifshitz gravity
theory admits Lifshitz solution \cite{Janiszewski:2012nb,Griffin:2012qx},
hyperscaling violating solutions \cite{Alishahiha:2012iy} and asymptotic
AdS or Lifshitz black hole solutions \cite{Janiszewski:2014iaa,Janiszewski:2014ewa,Li:2014fsa}.

We have a few more simplifications compared to \cite{Janiszewski:2012nb}.
They introduce a field $P_{i}$, the $O\left(c^{2}\right)$ order
of the shift function. This is related to the field $C_{i}$ introduced
in \cite{Geracie:2014nka} to source the energy flux when $\partial_{i}\xi^{t}\neq0$.
Since we are always staying in the global time coordinates, they can
be consistently set to zero. \cite{Janiszewski:2012nb} also introduces
a second $U(1)$ field called $b_{\mu}$, the $O\left(c^{2}\right)$
order of the gauge field in the parent relativistic theory. Its main
function is to provide a chemical potential related to the mass $m$
and it enters the holographic model mainly through the combination
$b_{t}/N$, which is a scalar. Thus we will not introduce this additional
$U(1)$ field, but instead just a scalar field $\phi$ functioning
as their $b_{t}/N$. The meaning of $\phi$ will be discussed one
its holographic map is built. Throughout this paper, we will not introduce
the speed of light $c$ in our construction, nor the non-relativistic
limit $c\rightarrow\infty$. In other words, we have already set $c=\infty$.
The FPD of fields and form of the effective action are not obtained
by taking non-relativistic limit of some relativistic parent theory,
such as in \cite{Andreev:2013qsa} and part of \cite{Janiszewski:2012nb}.
This allows maximal generality in the construction.

\subsection{Outline and Notations}

In this paper we show that Ho\v{r}ava-Lifshitz gravity can be used
as an effective theory framework to describe the fractional quantum
Hall effect, particularly at the low energy (hydrodynamic) regime.
(2+1)-dimensional Ho\v{r}ava-Lifshitz gravity can be directly used
to write down some FPD invariant effective actions, similar to what
is done in \cite{Hoyos:2011ez,Son:2013rqa,Andreev:2013qsa}, while
(3+1)-dimensional Ho\v{r}ava-Lifshitz gravity can serve as a holographic
dual description. Although both Ho\v{r}ava-Lifshitz gravity and the
non-relativistic quantum field theory for quantum Hall effects \cite{Geracie:2014nka}
(Section 2) are FPD invariant, their field contents and the FPD of
the fields are different. The foundation of the Ho\v{r}ava-Lifshitz
formalism for quantum Hall effects is the map between the field contents
and a matching of their FPD on both sides. This is one of the main
results of this paper, given in (\ref{eq:HolographicMap_N})-(\ref{eq:HolographicMap_phi}).
We originally derive this map as a holographic dictionary (Section
3), the key component in any holographic duality. But it is worthy
to emphasize that it can also exist independently from holography,
since it is entirely written in (2+1)-dimensional field language.
Thus is can be equally applies to (2+1)-dimensional Ho\v{r}ava-Lifshitz
gravity as an non-holographic effective theory (more in the traditional
sense) for quantum Hall effects.

We further derive a non-relativistic FPD invariant Chern-Simons low
energy effective action for the fractional quantum Hall effect using
Ho\v{r}ava-Lifshitz Holography with non-dynamical Chern-Simons terms
(Section 4). It produces physical results in agreement with previous
results obtained via other methods, for example, through Newton-Cartan
formalism \cite{Son:2013rqa}. It encodes many universal geometric
properties such as the shift and Hall viscosity. Using our results,
we further clarify a mystery about the relation between Hall viscosity
$\eta_{H}$ and angular momentum density $\ell$: 
\begin{equation}
\eta_{H}=-\frac{1}{2}\ell\:.
\end{equation}
This relation has been well established in the context of quantum
Hall effects using other methods \cite{Read:2008rn,Read:2010epa,Nicolis:2011ey,Hoyos:2014lla},
but not confirmed from the Chern-Simons effective theory as far as
we know, including in holography \cite{Saremi:2011ab,Chen:2011fs,Chen:2012ti,Liu:2012zm,Zou:2013fua,Son:2013xra,Liu:2013cha,Wu:2013vya,Liu:2014gto}.%
\footnote{So far as we know, most holographic studies on Hall viscosity \cite{Saremi:2011ab,Chen:2011fs,Chen:2012ti,Zou:2013fua,Son:2013xra}
assume no background magnetic field but finite temperature. Thus the
results are for thermal Hall viscosity, not in the context of quantum
Hall effects. An exception is \cite{Leigh:2012jv}, where vorticity
is present. %
} We distinguish two types of angular momentum densities from the Chern-Simons
effective theory and show that how the above relation arises (Section
5). This inspires us to interpret the velocity field $v^{i}$ commonly
appearing in quantum Hall effective field theory ($v^{i}$ is related
to torsion in the Newton-Cartan formalism \cite{Son:2013rqa,Christensen:2013lma,Christensen:2013rfa}
and to the shift function $N^{i}$ in our Ho\v{r}ava-Lifshitz formalism)
as guiding center velocity field whose conjugate quantity is the guiding
center momentum density, in conjecture with Haldane's geometric formalism
\cite{Haldane:2009ke,Haldane:2011ia,Haldane:1112,Haldane1403} of
the fractional quantum Hall effect.

\emph{Notations}: The field theory lives in a (2+1)-dimensional spacetime,
with coordinates labeled by time $t$ and two spatial directions $\vec{x}=(x,y)$.
The bulk radial coordinate is labeled by $r$, with the boundary located
at $r=0$. Volume elements are $d^{3}x=dtd^{2}\vec{x}$ and $d^{4}x=d^{3}xdr$.
Indices $i$, $j$, $k$ and $l$ run through $x$ and $y$ (boundary
spatial coordinates); $I$, $J$ and $K$ run through $x$, $y$ and
$r$ (bulk spatial coordinates); $\mu$ and $\nu$ run through $t$,
$x$ and $y$ (boundary spacetime coordinates); $M$, $N$ and $P$
run thought all four bulk coordinates. We use $\varepsilon$ with
upper or lower indices to denote Levi-Civita tensors (with the corresponding
determinant of the metric built in) and $\epsilon$ with indices Levi-Civita
symbol (whose components are $\pm1$ or $0$); while $\epsilon$ appears
with no index is the infinitesimal book-keeping parameter for the
derivative expansion in late sections. We use $a$ and $b$ as the
vielbein frame indices on the 2-dimensional manifold spanned by $x^{i}$,
i.e. labeling the directions in the tangent space of $x$ and $y$,
while $A$ and $B$ are the 3-dimensional counterpart in the tangent
space of $x$, $y$ and $r$. A ``$\bar{\phantom{\sigma}}$'' on
a bulk field indicates it is the near-boundary ($r\rightarrow0$)
independent leading order term of the corresponding bulk field. We
will use the symbol ``$\Rightarrow$'' to denote the near boundary
behavior, i.e. the $r\rightarrow0$ limit, where we keep the finite
and possibly singular terms and drop the sub-leading terms that are
vanishing in this limit. 

\bigskip{}

\section{A Non-Relativistic Field Theory in 2+1 Dimensions}

\subsection{Microscopic Action and Local Symmetries}

The microscopic action that defines the quantum field theory for the
quantum Hall effect in flat spacetime that we will study in this paper
is given by 
\begin{equation}
S_{\mathrm{NR}}=\int d^{3}x\left\{ \frac{i}{2}\left(\psi^{\dagger}(D_{t}\psi)-\psi(D_{t}\psi)^{\dagger}\right)-\frac{\delta^{ij}}{2m}\left(D_{i}\psi\right)^{\dagger}\left(D_{j}\psi\right)+\frac{\mathrm{g}B}{4m}\psi^{\dagger}\psi+\textrm{interactions}\right\} \:,
\end{equation}
with $D_{\mu}=\partial_{\mu}-iA_{\mu}$ and the magnetic field $B=\epsilon^{ij}\partial_{i}A_{j}$.
$\mathrm{g}$ is the gyromagnetic factor. To analyze local spacetime
symmetries, we need to couple it to curved spacetime.

First, let us define the vielbein $e_{i}^{a}$ associated with the
2-dimensional spatial metric $g_{ij}$:
\begin{equation}
g_{ij}=\delta_{ab}e_{i}^{a}e_{j}^{b}\:,\qquad g^{ij}=\delta_{ab}e^{ai}e^{bj}\:,
\end{equation}
where $a,b=1,2$ are vielbein indices and $\delta_{ab}=(+,+)$ is
the flat 2-dimensional Euclidean metric. $\epsilon_{ab}$ is the totally
anti-symmetric Levi-Civita symbol and $\varepsilon^{ij}=\epsilon^{ij}/\sqrt{g}$
is the Levi-Civita tensor associated with the metric $g_{ij}$, and
$g=\det\left(g_{ij}\right)$. The $U(1)$ spin connection vector $\omega_{\mu}=(\omega_{t},\omega_{i})$
is defined as 
\begin{eqnarray}
\omega_{t} & = & \frac{1}{2}\epsilon_{ab}e^{ak}\partial_{t}e_{k}^{b}\:,\\
\omega_{i} & = & \frac{1}{2}\left(\epsilon_{ab}e^{ak}\partial_{i}e_{k}^{b}-\varepsilon^{jk}\partial_{j}g_{ki}\right)\:.
\end{eqnarray}
It transforms under diffeomorphism and Weyl transformation (to be
introduced below) as 
\begin{eqnarray}
\delta\omega_{t} & = & \xi^{\mu}\partial_{\mu}\omega_{t}+\omega_{\mu}\partial_{t}\xi^{\mu}+\frac{1}{2}\varepsilon^{jk}\partial_{j}\left(g_{kl}\partial_{t}\xi^{l}\right)\:,\\
\delta\omega_{i} & = & \xi^{\mu}\partial_{\mu}\omega_{i}+\omega_{k}\partial_{i}\xi^{k}-\varepsilon^{jk}g_{ki}\partial_{j}\sigma\:,
\end{eqnarray}
The scalar curvature of the 2-dimensional Euclidean space is $\mathcal{R}^{(2)}=\varepsilon^{ij}\partial_{i}\omega_{j}$
and transforms as a scalar under diffeomorphism. The magnetic field
is now defined as $B=\varepsilon^{ij}\partial_{i}A_{j}$ in curved
spacetime.

We choose microscopic action of the non-relativistic field theory
in curved spacetime to be $S_{\textrm{NR}}=S_{\textrm{0}}+S_{\textrm{int}}$
with 
\begin{equation}
S_{\textrm{0}}=\int d^{3}x\sqrt{g}e^{-\Phi}\left\{ \frac{i}{2}e^{\Phi}\left(\psi^{\dagger}(D_{t}\psi)-\psi(D_{t}\psi)^{\dagger}\right)-\frac{1}{2m}\left(g^{ij}+i\varepsilon^{ij}\right)\left(D_{i}\psi\right)^{\dagger}\left(D_{j}\psi\right)\right\} \:,\label{eq:FTAction_S0}
\end{equation}
where the covariant derivative is 
\begin{eqnarray}
D_{t} & = & \partial_{t}-iA_{t}+is\omega_{t}-i\frac{(\mathrm{g}-2)e^{-\Phi}}{4m}\left[B+(1-s)\mathcal{R}^{(2)}\right]\:,\\
D_{i} & = & \partial_{i}-iA_{i}+is\omega_{i}\:,
\end{eqnarray}
where $s$ is intrinsic spin of $\psi$ (for spin-polarized electrons,
i.e. components of Dirac spinor in vacuum, $s=1/2$). It is easy to
see that $S_{0}$ goes back to the flat-spacetime action introduced
at the beginning of this section. Possible choices for the interaction
action $S_{\textrm{int}}$ include 4-fermion interaction and Coulomb
interaction. Our choice of $D_{t}$ such that $(\mathrm{g}-2)B/4m$
appears together with $A_{t}$ and $g^{ij}+i\varepsilon^{ij}$ appears
in a combination in the standard Pauli form has been employed, for
example, in \cite{Alicki1994}. This is different from \cite{Geracie:2014nka},
where $g^{ij}+i\mathrm{\frac{g}{2}}\varepsilon^{ij}$ appears together.

Under diffeomorphism $\delta t=-\xi^{t}(t)$, $\delta x^{i}=-\xi^{i}(t,\vec{x})$,
the action (\ref{eq:FTAction_S0}) is invariant up to boundary terms
if the fields transform as following: 
\begin{eqnarray}
\delta_{\xi}\psi & = & \xi^{\mu}\partial_{\mu}\psi\:,\label{eq:FTDiffeo_psi}\\
\delta_{\xi}A_{t} & = & \xi^{\mu}\partial_{\mu}A_{t}+A_{\mu}\partial_{t}\xi^{\mu}-\frac{\mathrm{g}-2s}{4}\varepsilon^{jk}\partial_{j}\left(g_{kl}\partial_{t}\xi^{l}\right)\label{eq:FTDiffeo_At}\\
 &  & -\frac{\mathrm{g}-2}{4}\varepsilon^{ij}\left[\partial_{i}\log\left(me^{\Phi}\right)\right]g_{jk}\partial_{t}\xi^{k}\:,\nonumber \\
\delta_{\xi}A_{i} & = & \xi^{\mu}\partial_{\mu}A_{i}+A_{k}\partial_{i}\xi^{k}+me^{\Phi}g_{ik}\partial_{t}\xi^{k}\:,\label{eq:FTDiffeo_Ai}\\
\delta_{\xi}\Phi & = & \xi^{\mu}\partial_{\mu}\Phi-\partial_{t}\xi^{t}\:,\label{eq:FTDiffeo_Phi}\\
\delta_{\xi}e_{i}^{a} & = & \xi^{\mu}\partial_{\mu}e_{i}^{a}+e_{k}^{a}\partial_{i}\xi^{k}\:,\\
\delta_{\xi}g_{ij} & = & \xi^{\mu}\partial_{\mu}g_{ij}+g_{ik}\partial_{j}\xi^{k}+g_{jk}\partial_{i}\xi^{k}\:,\label{eq:FTDiffeo_hij}\\
\delta_{\xi}m & = & \xi^{\mu}\partial_{\mu}m\:,\\
\delta_{\xi}g & = & \xi^{\mu}\partial_{\mu}g\:,\\
\delta_{\xi}s & = & \xi^{\mu}\partial_{\mu}s\:,
\end{eqnarray}
Here we have assumed that the parameter $m$, $\mathrm{g}$ and $s$
can also be external scalar fields whose values can vary at different
spacetime points. This is usually true in the materials due to the
presence of the medium. For example, elections' interaction with the
lattice and medium around can not only shift the values of these parameter
from their vacuum values, but also make them vary by positions. In
this sense, the action shall be viewed as a renormalized effective
action with effective mass $m$ etc \cite{Kittel_Book}.

So far we have not specified the precise meaning of $\psi$, or whether
it is bosonic or fermionic. We want to keep it as general as possible.
It just represents an underlying microscopic field degree of freedom
which is to be path-integrated out when computing the effective action.
I can be the electrons as well as other composite particles. Recently,
\cite{Cho:2014vfl} uses a similar action with the same coupling to
gravity via spin connection for both composite fermion \cite{Lopez1991}
and boson \cite{ZHK1988} theories to modify the conventional flux
attachment procedure and derive the Wen-Zee shift and Hall viscosity.
Our $\psi$ can be viewed in the same way, to denote the composite
bosons or fermions, with corresponding interaction term $S_{\mathrm{int}}$.
In this case $s$ will be the topological spin of the composite particles,
rather than the intrinsic or renormalized spin of the electrons. We
will compare our results with those of \cite{Cho:2014vfl} in Section
5.4.

Under local anisotropic Weyl transformation $\sigma=\sigma(t,\vec{x})$,
the action (\ref{eq:FTAction_S0}) is invariant if the fields transform
as 
\begin{eqnarray}
\delta_{\sigma}\psi & = & -\sigma\psi\:,\label{eq:FTWeyl_psi}\\
\delta_{\sigma}\Phi & = & -z\sigma\:,\label{eq:FTWeyl_Phi}\\
\delta_{\sigma}e_{i}^{a} & = & \sigma e_{i}^{a}\:,\\
\delta_{\sigma}g_{ij} & = & 2\sigma g_{ij}\:,\label{eq:FTWeyl_hij}\\
\delta_{\sigma}A_{t} & = & 0\:,\label{eq:FTWeyl_At}\\
\delta_{\sigma}A_{i} & = & \left(1-s\right)\varepsilon^{jk}g_{ki}\partial_{j}\sigma\:,\label{eq:FTWeyl_Ai}\\
\delta_{\sigma}m & = & (z-2)\sigma m\:,\label{eq:FTWeyl_m}
\end{eqnarray}
The form of $S_{\textrm{int}}$ will be chosen to respect the above
symmetries as well. \cite{Bradlyn:2014wla} contains a brief discussion
on $S_{\textrm{int}}$ in curved space. The form of $S_{\textrm{int}}$
for quantum Hall effects is typically bi-local:
\[
S_{\textrm{int}}=-\frac{1}{2}\iint d^{3}x_{1}d^{3}x_{2}\sqrt{g(x_{1})g(x_{2})}e^{-\Phi(x_{1})-\Phi(x_{2})}\psi^{\dagger}(x_{1})\psi(x_{1})V(x_{1},x_{2})\psi^{\dagger}(x_{2})\psi(x_{2})\:.
\]
The form of $V(x_{1},x_{2})$ shall be chosen to respect the above
symmetries. We will not discuss how to do this. One point we want
to highlight is that, choosing different $V(x_{1},x_{2})$ may or
may not introduce additional scales into the problem through its coupling
coefficients. For example, if $V(x_{1},x_{2})\sim\lambda_{\mathrm{4F}}\delta^{(3)}(x_{1}-x_{2})$
up to an inverse measure, the interaction becomes the contact 4-Fermi
interaction \cite{Son:2008ye,Geracie:2014nka}: 
\[
S_{\textrm{int}}=-\frac{\lambda_{\mathrm{4F}}}{2}\int d^{3}x\sqrt{g}e^{-\Phi}\left(\psi^{\dagger}\psi\right)^{2}\:.
\]
Under Weyl transformation, 
\begin{equation}
\delta\lambda_{\mathrm{4F}}=-(z-2)\sigma\lambda_{\mathrm{4F}}\:.\label{eq:FTWeyl_4Fcoupling}
\end{equation}
For $z=2$ it preserves the scale invariance, same as $m$ does. Another
way to look at it is, $\lambda_{\mathrm{4F}}$ does not have a length
scale associated with it when $z=2$ (i.e. dimensionless and marginal).%
\footnote{$\psi\sim[\textrm{length}]^{-1}$, $t\sim[\textrm{length}]^{z}$.
Then $\int d^{3}x\sqrt{g}e^{-\Phi}\left(\psi^{\dagger}\psi\right)^{2}\sim[\textrm{length}]^{z-2}$.
Thus $\lambda_{\mathrm{4F}}\sim[\textrm{length}]^{2-z}$.%
} In this case, $\lambda_{\mathrm{4F}}$ can still run logarithmically
through quantum renormalization \cite{Nikolic:2007zz}, but at the
LLL projection ($m\rightarrow0$ limit) which is the most interesting
case for us, this running will disappear. The interactions can also
be introduced through auxiliary fields, such as that illustrated in
\cite{Son:2005rv} for Yukawa interaction. Other interactions, such
as Coulomb and Yukawa interactions, usually have some dimensionful
couplings. Once being included, these couplings will introduce additional
length scales into the theory. Their Weyl transformations can be dealt
in a similar fashion as for $m$ and $\lambda_{\mathrm{4F}}$, although
their forms will be very different.

Without the interaction $S_{\textrm{int}}$, $m$ and $\Phi$ only
appear together as a combination $me^{\Phi}$ in $S_{\textrm{0}}$,
which scales under Weyl transformation by $-2\sigma$. In this case,
talking the Weyl scaling of $m$ and $\Phi$ separately does not make
much sense since one can always shift part of the scaling from one
to the other, or equivalently, define $z=2$ and $\delta_{\sigma}m=0$.
In other words, for free non-relativistic fields, $z$ always equals
to $2$, and there is no Lifshitz scaling. This changes when the interactions
such as that in $S_{\textrm{int}}$ is turned on, because now $\Phi$
appears separately in the interactions (for example, as part of the
measure $e^{-\Phi}$), thus the Weyl scaling of $m$ and $\Phi$ can
no longer be shifted between each other. In this case we can have
$z\neq2$ Lifshitz scaling physically. The fact that $me^{\Phi}$
appears as a combination in $S_{0}$ is an accident of the free theory.
Since we are considering interacting theories in general, we will
always think that $m$ and $\Phi$ appear separately 

We notice that $\mathrm{g}=2$ and $s=1$ are special values when
the symmetry transformations are particularly simple and all the parity-violating
terms are gone. This is also noticed in \cite{Geracie:2014nka} and
it corresponds to the parity-preserving case considered in early works
of \cite{Son:2005rv,Son:2008ye,Balasubramanian:2008dm,Janiszewski:2012nb}.
Hence the holographic dictionary we are going to build will inherit
this feature and goes back to that of \cite{Janiszewski:2012nb} in
this special case.

The action is also invariant under arbitrary shifts of parameters
$\mathrm{g}$ and $s$ if the gauge field is shifted in the following
way as well: 
\begin{eqnarray}
\Delta A_{t} & = & \omega_{t}\Delta s-\frac{e^{-\Phi}}{4m}\left\{ \left[B+(1-s)\mathcal{R}^{(2)}\right]\Delta\mathrm{g}-\mathcal{R}^{(2)}(\mathrm{g}-2+\Delta\mathrm{g})\Delta s\right\} \:,\\
\Delta A_{i} & = & \omega_{i}\Delta s\:.
\end{eqnarray}
This is essentially a field redefinition of $A_{\mu}$. It reflects
the fact that in the microscopic action (\ref{eq:FTAction_S0}), $\mathrm{g}$
and $s$ only appear in the combination 
\begin{eqnarray}
\tilde{A}_{t} & = & A_{t}-s\omega_{t}+\frac{(\mathrm{g}-2)e^{-\Phi}}{4m}\left[B+(1-s)\mathcal{R}^{(2)}\right]\:,\\
\tilde{A}_{i} & = & A_{i}-s\omega_{i}\:.
\end{eqnarray}
Thus after path-integrating out the microscopic field $\psi$, in
the effective action, $\mathrm{g}$ and $s$ will still only appear
in the above combination of $\tilde{A}_{\mu}$, not separately. This
imposes another useful constraint on the possible forms of the effective
action. The combination $\tilde{A}_{\mu}$ transforms under diffeomorphism
and Weyl transformation as 
\begin{eqnarray}
\delta\tilde{A}_{t} & = & \xi^{\mu}\partial_{\mu}\tilde{A}_{t}+\tilde{A}_{\mu}\partial_{t}\xi^{\mu}-\frac{1}{2}\varepsilon^{ij}\partial_{i}\left(g_{jk}\partial_{t}\xi^{k}\right)\:,\label{eq:FTTransf_At_tilde}\\
\delta\tilde{A}_{i} & = & \xi^{\mu}\partial_{\mu}\tilde{A}_{i}+\tilde{A}_{k}\partial_{i}\xi^{k}+me^{\Phi}g_{ik}\partial_{t}\xi^{k}+\varepsilon^{jk}g_{ki}\partial_{j}\sigma\:.\label{eq:FTTransf_Ai_tilde}
\end{eqnarray}
The covariant derivative can be written as $D_{\mu}=\partial_{\mu}-i\tilde{A}_{\mu}$.

There are two local $U(1)$ symmetry transformations under which the
action is invariant. One is the local gauge transformation for the
gauge field $A_{\mu}$, $\Lambda=\Lambda(t,\vec{x})$: 
\begin{equation}
\delta A_{\mu}=\delta\tilde{A}_{\mu}=-\partial_{\mu}\Lambda\:,
\end{equation}
with a shift of the phase factor $\psi\rightarrow e^{-i\Lambda}\psi$.
The other is a local rotation of the vielbein frame, $\delta e_{i}^{a}=\Lambda_{\phantom{a}b}^{a}e_{i}^{b}$,
$\Lambda_{\phantom{a}b}^{a}=\theta(t,\vec{x})\epsilon_{\phantom{a}b}^{a}$:
\begin{equation}
\delta\omega_{\mu}=-\partial_{\mu}\theta\:,\qquad\delta\tilde{A}_{\mu}=s\partial_{\mu}\theta\:.
\end{equation}
This rotation corresponds to a spin rotation of the field $\psi\rightarrow e^{is\theta}\psi$.

The effective action (strictly speaking the logarithm of the generating
functional) $\mathcal{W}$ is a functional of $\Phi$, $g_{ij}$,
$A_{\mu}$, $m$, $\mathrm{g}$, $s$ among other external fields
and parameters, defined after the path integral over $\psi$: 
\begin{equation}
\mathcal{W}\left[\Phi,g_{ij},A_{\mu},\ldots;m,\mathrm{g},s,\ldots\right]=-i\log\int\mathcal{D}\psi\mathcal{D}\psi^{\dagger}e^{i\left(S_{\textrm{0}}+S_{\textrm{int}}\right)}\:.\label{eq:FTEffectiveAction}
\end{equation}
The expectation values of conserved current $J^{\mu}$, stress tensor
$T^{ij}$ and energy density $\mathcal{E}^{0}$ can then be obtained
as the following: 
\begin{equation}
\delta\mathcal{W}=\int dtd^{2}\vec{x}\sqrt{g}e^{-\Phi}\left\{ \langle J^{\mu}\rangle\delta A_{\mu}+\frac{1}{2}\langle T^{ij}\rangle\delta g_{ij}+\langle\mathcal{E}^{0}\rangle\delta\Phi\right\} \:.\label{eq:ConservedQuantities}
\end{equation}
The main goal of this paper is to find the holographic dual description
of the effective action $\mathcal{W}\left[\Phi,g_{ij},A_{\mu},\ldots;m,\mathrm{g},s,\ldots\right]$.

\subsection{Comments on Coupling to Curved Spacetime}

It is necessary to note here that the way to couple a flat-spacetime
action to curved spacetime is not unique. The curved-spacetime action
(\ref{eq:FTAction_S0}) we choose here is slightly different from
that in \cite{Geracie:2014nka}, where a minimal coupling is chosen.
The differences are the form of $D_{t}$ and the position of $\mathrm{g}$.
There are two reasons that motivate us to choose the present form.
We choose $\mathrm{g}B$ term to appear side-by-side with $A_{t}$
in $D_{t}$ to reflect the same fact in flat-spacetime action, where
this is not manifest in \cite{Geracie:2014nka} and complicated field
redefinitions are needed if one wants to shift $\mathrm{g}$. Consequently,
the diffeomorphism of $A_{t}$, (\ref{eq:FTDiffeo_At}) is a little
different than \cite{Geracie:2014nka}, and the relationship between
momentum density $p^{i}$ and conserved current are modified as well.
Following the same procedure as in \cite{Geracie:2014nka}, we can
derive the relation for our current case: 
\begin{equation}
\langle p^{i}\rangle=m\langle J^{i}\rangle-\varepsilon^{ij}\partial_{j}\left(\frac{\mathrm{g}-2s}{4}e^{-\Phi}\langle J^{t}\rangle\right)+\frac{\mathrm{g}-2}{4}e^{-\Phi}\langle J^{t}\rangle\varepsilon^{ij}\partial_{j}\log\left(me^{\Phi}\right)\:.\label{eq:Momentum-Current}
\end{equation}
We also add $\mathcal{R}^{(2)}$ to $D_{t}$ to simplify the Weyl
transformation of $A_{t}$, (\ref{eq:FTWeyl_At}) so that it is identically
zero. This simplifies the conformal Ward identity: 
\begin{equation}
\langle\mathcal{E}^{0}\rangle=\frac{1}{2}g_{ij}\langle T^{ij}\rangle-\frac{1-s}{2}e^{\Phi}\varepsilon^{ij}\nabla_{i}\left(e^{-\Phi}\langle J_{j}\rangle\right)\label{eq:ConformalWardID}
\end{equation}
when $z=2$. Here $\nabla_{i}$ is the covariant derivative compatible
with metric $g_{ij}$: $\nabla_{k}g_{ij}=0$. Thus our version of
the curved spacetime action is more toward the \emph{conformal} coupling
rather than the minimal coupling in \cite{Geracie:2014nka}. Following
the procedures in \cite{Geracie:2014nka}, one can derive all Ward
identities for energy density, stress tensor and conserved current
using the local symmetries discussed in the previous subsection.

\subsection{Global Time and Relationship to Newton-Cartan Geometry}

Here we briefly address the issue related to global time, which is
a hall-mark of non-relativistic theories. A more comprehensive discussion
has already been presented in \cite{Geracie:2014nka}. We outline
some of the key points here for self-containedness of later discussions.
In the above discussion, energy flux $\mathcal{E}^{i}$ and its source
(let us call it $C_{i}$) are missing. \cite{Geracie:2014nka} shows
that $C_{i}$ can be introduced to the formalism by modifying spatial
quantities like $\partial_{i}$ and $D_{i}$ to $\tilde{\partial}_{i}=\partial_{i}+C_{i}\partial_{t}$
and $\tilde{D}_{i}=D_{i}+C_{i}D_{t}$. The diffeomorphism of $C_{i}$
is 
\begin{equation}
\delta C_{i}=\xi^{\mu}\partial_{\mu}C_{i}+C_{k}\tilde{\partial}_{i}\xi^{k}-\tilde{\partial}_{i}\xi^{t}\:,\label{eq:FTDiffeo_Ci}
\end{equation}
and $n_{\mu}=\left(e^{-\Phi},-e^{-\Phi}C_{i}\right)$ transforms as
a Lorentz vector. $C_{i}$ is invariant under local Weyl transformation.
When 
\begin{equation}
\varepsilon^{ij}(\partial_{i}+C_{i}\partial_{t})C_{j}=0\:,\qquad\textrm{i.e.}\qquad n\wedge\mathrm{d}n=0\:,
\end{equation}
there exist sets of \emph{global time coordinates} (GTCs) in which
$C_{i}=0$. This is generally required by the causality of non-relativistic
theories. Diffeomorphisms that satisfy $\partial_{i}\xi^{t}=0$ (which
keeps $C_{i}=0$) are the transformations between different GTCs.
This also implies that in GTCs energy is globally conserved as time
evolves and there is no notion of energy flux needed. In our discussion
of the field theory, we have set $C_{i}=0$ and $\partial_{i}\xi^{t}=0$
throughout. This means that we have been assuming the existence of
GTCs and work only in them. Since we are interested in non-relativistic
field theories, this is reasonable and economical, and we will work
in this way throughout the rest of the paper. We will leave the study
of holographic dual to the case when $C_{i}\neq0$ or even $n\wedge\mathrm{d}n\neq0$
to the future.

\cite{Christensen:2013lma,Christensen:2013rfa} and \cite{Geracie:2014nka}
show that for the case $n\wedge\mathrm{d}n=0$ (of which $C_{i}=0$
is a special case), the geometry can be covariantly described by Newton-Cartan
geometry, with or without torsion. However, we want to emphasize that
it is not always necessary to introduce such a formalism. One could
just work with the spacetime symmetries (\ref{eq:FTDiffeo_psi})-(\ref{eq:FTDiffeo_hij})
and (\ref{eq:FTWeyl_psi})-(\ref{eq:FTWeyl_m}) directly with all
those anomalous terms in the transformation rules. This is tabbed
as ``non-covariant'' approach in \cite{Geracie:2014nka}. The maths
is usually cumbersome and laborious, but direct, and one gets all
the physical results at the end. The Newton-Cartan geometry is a convenient
way to covariantize the formalism, while other equivalent ways exist
as well, with different notions of covariance and different structures
in the formalisms. Since we assume the existence of GTCs, another
convenient covariant formalism is Ho\v{r}ava-Lifshitz gravity. It
has already been employed in \cite{Andreev:2013qsa}. The notion of
covariance and metric and connection structures of Ho\v{r}ava-Lifshitz
gravity are closer to those of relativistic gravity theories than
to Newton-Cartan geometry, with additional assumption of the existence
of global time that breaks the Lorentz symmetry between time and space.
Since relativistic holography has been very well studied, given its
similarity to relativistic gravity theory, Ho\v{r}ava-Lifshitz gravity
is a natural choice as the covariant formalism we will use to covariantize
the geometry of the non-relativistic field theory discussed in this
section, because the generalization to the holographic bulk is ready
\cite{Janiszewski:2012nb,Janiszewski:2012nf}. The holographic dual
we are studying here is an extension of \cite{Janiszewski:2012nb}'s
to parity violating cases, and its map to the boundary, which is the
main results of our paper, can be viewed as another covariant description
of the geometry of the non-relativistic field theory using Ho\v{r}ava-Lifshitz
gravity, parallel to the popular Newton-Cartan formalism.

\bigskip{}

\section{Dual Gravity Theory in 3+1 Dimensional Bulk}

\subsection{Bulk Fields and Their Diffeomorphism}

The bulk theory, Ho\v{r}ava-Lifshitz gravity, assumes the existence
of global time coordinates. The foliation of global time is preserved
under diffeomorphism $x^{M}=-\hat{\xi}^{M}$ with the constraint 
\begin{equation}
\partial_{I}\hat{\xi}^{t}=0\:,\quad\textrm{i.e.}\quad\hat{\xi}^{M}=\left(\hat{\xi}^{t}(t),\hat{\xi}^{I}(t,x^{I})\right)\:.\label{eq:BulkDiffeoNRConstraint}
\end{equation}
This is the bulk counterpart of the condition $\partial_{i}\xi^{t}=0$
in the non-relativistic field theory discussed in the previous section,
now extended to include the bulk radial direction as well.

The graviton sector of the bulk theory consists of a lapse function
$N$, a shift function $N_{I}=(N_{i},N_{r})$ and a spatial metric
$G_{IJ}$. We define $G^{IJ}$ is the matrix inverse of $G_{IJ}$
and the index of the shift function $N_{I}$ is raised and lowered
using $G^{IJ}$ and $G_{IJ}$. The bulk vielbein associated with $G_{IJ}$
is $E_{I}^{A}$, with $G_{IJ}=\delta_{AB}E_{I}^{A}E_{J}^{B}$. They
transform under diffeomorphism as 
\begin{eqnarray}
\delta N & = & \hat{\xi}^{M}\partial_{M}N+N\partial_{t}\hat{\xi}^{t}\:,\\
\delta N_{I} & = & \hat{\xi}^{M}\partial_{M}N_{I}+N_{I}\partial_{t}\hat{\xi}^{t}+N_{K}\partial_{I}\hat{\xi}^{K}+G_{IK}\partial_{t}\hat{\xi}^{K}\:,\\
\delta G_{IJ} & = & \hat{\xi}^{M}\partial_{M}G_{IJ}+G_{IK}\partial_{J}\hat{\xi}^{K}+G_{JK}\partial_{I}\hat{\xi}^{K}\:,\\
\delta E_{I}^{A} & = & \hat{\xi}^{M}\partial_{M}E_{I}^{A}+E_{K}^{A}\partial_{I}\hat{\xi}^{K}\:.
\end{eqnarray}
There is a bulk $U(1)$ gauge field $V_{M}$ which transforms as 
\begin{eqnarray}
\delta V_{t} & = & \hat{\xi}^{M}\partial_{M}V_{t}+V_{M}\partial_{t}\hat{\xi}^{M}\:,\\
\delta V_{I} & = & \hat{\xi}^{M}\partial_{M}V_{I}+V_{K}\partial_{I}\hat{\xi}^{K}\:.
\end{eqnarray}
and a scalar $\phi$ which transforms as 
\begin{equation}
\delta\phi=\hat{\xi}^{M}\partial_{M}\phi\:.
\end{equation}
The rotations of the vielbein frame is 
\begin{equation}
\delta E_{I}^{A}=\hat{\Lambda}_{\phantom{A}B}^{A}E_{I}^{B}\:,\qquad\hat{\Lambda}_{\phantom{A}B}^{A}=-\hat{\Lambda}_{\phantom{B}A}^{B}\:,
\end{equation}
where the anti-symmetric $\hat{\Lambda}_{\phantom{A}B}^{A}$ are the
generators of rotations. The above transformations are all infinitesimal
transformations. The gauge transformation for $V_{M}$ is 
\begin{equation}
\delta V_{M}=-\partial_{M}\hat{\Lambda}\:.
\end{equation}
We put a ``$\hat{\phantom{a}}$'' on every bulk gauge parameter
to distinguish it from the corresponding one in the field theory.
The bulk action shall be invariant under all the above transformations.

\subsection{Gauge Conditions and Residual Gauge Transformations}

Next, we choose gauge conditions for fields in the bulk. The time-like
boundary is located at $r=0$ and breaks the translational invariance
along the spatial direction (denoted by the Poincaré radial coordinate
$r$ here) perpendicular to the boundary. In the bulk, the isometry
of the AdS-type geometry also naturally splits the 3-dimensional spatial
manifold with metric $G_{IJ}$ into the radial direction labeled by
$r$ and the transverse directions (labeled by $x$ and $y$) with
translational and $SO(2)$ rotational symmetries. These introduce
a co-dimension one foliation in the bulk labeled by $r=\textrm{constant}$.
We choose gauge conditions for the bulk fields that manifest this
foliation:
\begin{eqnarray}
E_{r}^{3} & = & \Upsilon(r)\:,\label{eq:BulkGaugeCondition_1}\\
E_{x}^{3}=E_{y}^{3} & = & 0\:,\\
E_{r}^{1}=E_{r}^{2} & = & 0\:,\label{eq:BulkGaugeCondition_3}
\end{eqnarray}
Here we also fix the orientation of the vielbein frame in the tangent
space such that $3$-direction is along $r$-direction and $1$- and
$2$-directions are perpendicular to it. These give rise to the usual
gauge conditions for the bulk metric: $G_{ir}=0$ and $G_{rr}=\Upsilon(r)^{2}$.
Given an arbitrary choice of coordinates and frame to start with,
these five gauge equations can be achieved by a combination of diffeomorphism
and vielbein frame rotations within $1$-3 and $2$-$3$ planes, which
fix $\xi^{I}$ and two of the three Euler angles of the rotations,
related to $\Lambda_{\phantom{1}3}^{1}$ and $\Lambda_{\phantom{2}3}^{2}$,
up to some residual gauge transformations to be discussed later. Similar
gauge conditions have been used in \cite{Griffin:2011xs}. In relativistic
holography, another gauge condition $N_{r}=0$ is usually imposed,
but here it can not be achieved, because this requires a diffeomorphism
of $r$-dependent $\xi^{t}$, which is forbidden by (\ref{eq:BulkDiffeoNRConstraint}).
Thus we will not impose a gauge condition for $N_{r}$ in the bulk.
For the bulk $U(1)$ field, we choose the usual radial gauge condition
\begin{equation}
V_{r}=0\:.
\end{equation}

The above gauge conditions do not completely fix the gauge degrees
of freedom. There are residual gauge transformations that leave these
gauge conditions unchanged. The residual diffeomorphism that respect
the gauge condition for metric $G_{ir}=0$ and $G_{rr}=\Upsilon(r)^{2}$
is 
\begin{equation}
\left\{ \begin{aligned} & \hat{\xi}^{t}=\bar{\xi}^{t}(t)\\
 & \hat{\xi}^{i}=\bar{\xi}^{i}(x)+L\int dr\Upsilon(r)G^{ij}(x,r)\partial_{j}\bar{\sigma}(x)\\
 & \hat{\xi}^{r}=-\frac{L}{\Upsilon(r)}\bar{\sigma}(x)
\end{aligned}
\:,\right.\label{eq:BulkResidualDiffeo}
\end{equation}
where $x=(t,\vec{x})$ and $G^{ij}(x,r)$ is matrix inverse of $G_{ij}(x,r)$.
$L$ is a length scale which will be set to the AdS radius later.
However, the above diffeomorphism does not respect the gauge conditions
for vielbein. It leaves $E_{r}^{3}$ invariant, but changes the others:
$\delta E_{i}^{3}=-L\partial_{i}\bar{\sigma}$ and $\delta E_{r}^{a}=L\Upsilon(r)E^{ai}\partial_{i}\bar{\sigma}$.
To compensate these shifts, we need to perform a residual rotation
of the vielbein frame at the same time, with the following rotation
parameters: 
\begin{equation}
\hat{\Lambda}_{\phantom{a}3}^{a}=-LE^{ai}(x,r)\partial_{i}\bar{\sigma}(x)\:.\label{eq:BulkResidualRotation}
\end{equation}
Meanwhile, the rotation within $1$-$2$ plane is still arbitrary:
$\hat{\Lambda}_{\phantom{1}2}^{1}=\hat{\Lambda}_{\phantom{1}2}^{1}(x,r)$.
The residual diffeomorphism also changes $V_{r}$ by $\delta V_{r}=L\Upsilon(r)G^{ij}V_{i}\partial_{j}\bar{\sigma}$.
To compensate this shift, the residual gauge transformation for the
$U(1)$ field is 
\begin{equation}
\hat{\Lambda}=\bar{\Lambda}(x)+L\int dr\Upsilon(r)G^{ij}(x,r)V_{i}(x,r)\partial_{j}\bar{\sigma}(x)\:.
\end{equation}

We shall note here that one is entitled to choose different gauge
conditions in the bulk other than the ones we specify above. But we
need that at least near the boundary the gauge conditions agree with
the form specified here, because the near-boundary residual symmetry
transformations and hence the holographic dictionary are built on
the choice of the above gauge conditions near the boundary, and these
gauge conditions do reflect the fact that the existence of the time-like
boundary at $r=0$ breaks the isometry along $r$-direction. The consequences
of this set of gauge conditions near the boundary is what we will
explore next.

\subsection{Near-Boundary Behaviors of Fields and Transformations}

Near the boundary $r=0$, we choose the background metric to have
the asymptotic Lifshitz form \cite{Horava:2009vy,Griffin:2011xs,Janiszewski:2012nb,Griffin:2012qx}
in Poincaré coordinates:%
\footnote{This metric form shall be understood as the following asymptotic conditions
for the background functions: 
\[
N\Rightarrow\left(\frac{L}{r}\right)^{z}\:,\qquad N_{I}\Rightarrow0\:,\qquad G_{IJ}=\left(\frac{L}{r}\right)^{2}\delta_{IJ}\:.
\]
\cite{Janiszewski:2012nb} has a more general setup that $N\Rightarrow\left(L/r\right)^{\gamma}$
with $\gamma\neq z$. For simplicity, we will assume $\gamma=z$ to
keep our holographic map neat. $\gamma\neq z$ can always be achieved
by tuning the asymptotic indices of scalar fields/components in the
holographic map, at the price of making it appear more complicated,
as shown in \cite{Janiszewski:2012nb}.%
} 
\begin{equation}
ds^{2}\Rightarrow-\left(\frac{L}{r}\right)^{2z}dt^{2}+\left(\frac{L}{r}\right)^{2}\left(d\vec{x}^{2}+dr^{2}\right)\:,\label{eq:AdS-Lifshitz_metric}
\end{equation}
where $L$ is AdS radius. This implies $\Upsilon(r)=L/r$ in the gauge
conditions. The near-boundary behaviors of the bulk graviton fields
take the following form: 
\begin{eqnarray}
N & \Rightarrow & \left(\frac{L}{r}\right)^{z}\left[\bar{N}(x)+O\left(r\right)\right]\:,\\
N_{I} & \Rightarrow & \left(\frac{L}{r}\right)^{2}\left[\bar{N}_{I}(x)+O\left(r\right)\right]\:,\\
G_{ij} & \Rightarrow & \left(\frac{L}{r}\right)^{2}\left[\bar{G}_{ij}(x)+O\left(r\right)\right]\:,\\
E_{i}^{a} & \Rightarrow & \left(\frac{L}{r}\right)\left[\bar{E}_{i}^{a}(x)+O\left(r\right)\right]\:.\label{eq:NBAsymp_Vielbein}
\end{eqnarray}
The asymptotic indices $z$ and $2$ of $N$ and $G_{ij}$ are determined
by the above background Lifshitz metric. The index of $N_{I}$ is
in general not fully determined by the background metric, but also
by other free parameters of Ho\v{r}ava-Lifshitz gravity, and thus
can be tuned for a desired value. We choose the value to be $2$,
for reasons to be explained later. The near-boundary behavior of the
$U(1)$ field is 
\begin{equation}
V_{\mu}\Rightarrow\bar{V}_{\mu}(x)+O\left(r\right)\:,
\end{equation}
i.e. the asymptotic $r$ index is $0$ for $V_{\mu}$. The reason
for that is this index corresponds to the coefficient of Weyl parameter
$\sigma$ in the global Weyl transformation in the dual field theory.
For the $U(1)$ field in the field theory, it does not transform under
global Weyl transformation, i.e. the coefficient is $0$, thus we
choose the $r$ index to be $0$ in the bulk dual. This can always
be achieved by tuning the parameters of the bulk theory. We choose
the scalar to have the following asymptotic behavior: 
\begin{equation}
\phi\Rightarrow\left(\frac{L}{r}\right)^{\Delta_{\phi}}\left[\bar{\phi}(x)+O(r)\right]\:,
\end{equation}
where $\Delta_{\phi}$ is related to the conformal dimension of the
dual operator and can be tuned by the mass of the scalar in the bulk
theory.

Then near the boundary, the bulk residual diffeomorphism becomes 
\begin{equation}
\left\{ \begin{aligned} & \hat{\xi}^{t}\Rightarrow\bar{\xi}^{t}(t)\\
 & \hat{\xi}^{i}\Rightarrow\bar{\xi}^{i}(x)+\frac{1}{2}r^{2}\bar{G}^{ij}(x)\partial_{j}\bar{\sigma}(x)+O\left(r^{3}\right)\\
 & \hat{\xi}^{r}\Rightarrow-r\bar{\sigma}(x)
\end{aligned}
\:,\right.\label{eq:NBResidualDiffeo}
\end{equation}
and the residual vielbein rotation and $U(1)$ gauge transformation
become 
\begin{eqnarray}
\hat{\Lambda}_{\phantom{a}3}^{a} & \Rightarrow & -r\bar{E}^{ai}(x)\partial_{i}\bar{\sigma}(x)\:,\label{eq:NBResidualRotation}\\
\hat{\Lambda} & \Rightarrow & \bar{\Lambda}(x)+\frac{1}{2}r^{2}\bar{G}^{ij}(x)\bar{V}_{i}(x)\partial_{j}\bar{\sigma}(x)+O\left(r^{3}\right)\:,\label{eq:NBResidualGaugeTransf}
\end{eqnarray}
where $\bar{G}^{ij}$ is the matrix inverse of $\bar{G}_{ij}$. To
maintain the near-boundary asymptotic behavior of the bulk vielbein
in (\ref{eq:NBAsymp_Vielbein}), we need 
\begin{equation}
\hat{\Lambda}_{\phantom{a}b}^{a}\Rightarrow\bar{\Lambda}_{\phantom{a}b}^{a}(x)+O\left(r\right)\:.
\end{equation}
We will call fields with ``$\bar{\phantom{v}}$'' boundary fields.
$i$, $j$, $k$ and $l$ indices of the boundary fields are raised
and lowered by $\bar{G}^{ij}$ and $\bar{G}_{ij}$, e.g. $\bar{E}^{ai}=\bar{G}^{ij}\bar{E}_{j}^{a}$. 

From the near-boundary asymptotic behaviors of the bulk fields and
the bulk diffeomorphism, we can work out the residual symmetry transformations
of the boundary fields at the boundary $r=0$: 
\begin{eqnarray}
\delta\bar{N} & = & \bar{\xi}^{\mu}\partial_{\mu}\bar{N}+\bar{N}\partial_{t}\xi^{t}+z\bar{\sigma}\bar{N}\:,\label{eq:NBDiffeo_N}\\
\delta\bar{N}_{i} & = & \bar{\xi}^{\mu}\partial_{\mu}\bar{N}_{i}+\bar{N}_{i}\partial_{t}\bar{\xi}^{t}+\bar{N}_{k}\partial_{i}\bar{\xi}^{k}+\bar{G}_{ik}\partial_{t}\bar{\xi}^{k}+2\bar{\sigma}\bar{N}_{i}\:,\label{eq:NBDiffeo_Ni}\\
\delta\bar{N}_{r} & = & \bar{\xi}^{\mu}\partial_{\mu}\bar{N}_{r}+\bar{N}_{r}\partial_{t}\bar{\xi}^{t}+\bar{\sigma}\bar{N}_{r}\:,\label{eq:NBDiffeo_Nr}\\
\delta\bar{G}_{ij} & = & \bar{\xi}^{\mu}\partial_{\mu}\bar{G}_{ij}+\bar{G}_{ik}\partial_{j}\bar{\xi}^{k}+\bar{G}_{jk}\partial_{i}\bar{\xi}^{k}+2\bar{\sigma}\bar{G}_{ij}\:.\label{eq:NBDiffeo_Gij}\\
\delta\bar{E}_{i}^{a} & = & \bar{\xi}^{\mu}\partial_{\mu}\bar{E}_{i}^{a}+\bar{E}_{k}^{a}\partial_{i}\bar{\xi}^{k}+\bar{\sigma}\bar{E}_{i}^{a}+\bar{\Lambda}_{\phantom{a}b}^{a}\bar{E}_{i}^{b}\:,\label{eq:NBDiffeo_Vielbein}\\
\delta\bar{V}_{\mu} & = & \bar{\xi}^{\nu}\partial_{\nu}\bar{V}_{\mu}+\bar{V}_{\nu}\partial_{\mu}\bar{\xi}^{\nu}-\partial_{\mu}\bar{\Lambda}\:,\label{eq:NBDiffeo_Vector}\\
\delta\bar{\phi} & = & \bar{\xi}^{\mu}\partial_{\mu}\bar{\phi}+\Delta_{\phi}\bar{\sigma}\bar{\phi}\:.\label{eq:NBDiffeo_Scalar}
\end{eqnarray}
We define $\bar{\varepsilon}^{ij}$ is the Levi-Civita tensor associated
with the boundary metric $\bar{G}_{ij}$. There are additional vectorial
structures we can make up using the above boundary fields, whose near-boundary
diffeomorphism transformations are collected in Appendix (B).

\subsection{Holographic Dictionary}

Now we are in a position to map the boundary fields of the bulk gravity
theory to the source fields of the field theory. The guiding principle
is the matching of symmetry transformations of the fields on both
sides. First, we map the symmetry transformation parameters, following
the idea of \cite{Janiszewski:2012nf}. The near-boundary residual
diffeomorphism parameters $\bar{\xi}^{\mu}$ and $\bar{\sigma}$ of
the bulk theory are mapped to the diffeomorphism and Weyl parameters
$\xi^{\mu}$ and $\sigma$ in the field theory, while the residual
gauge parameter $\bar{\Lambda}$ is mapped to the gauge parameter
$\Lambda$.

It immediately follows that $\bar{G}_{ij}$ and $\bar{E}_{i}^{a}$
are mapped to $g_{ij}$ and $e_{i}^{a}$ in the field theory, since
their diffeomorphism and Weyl transformations match perfectly. The
vielbein frame rotation $\bar{\Lambda}_{\phantom{a}b}^{a}(x)$ is
mapped to that in the field theory which rotates the frame of $e_{i}^{a}$.
The near-boundary diffeomorphism of $\bar{N}$ matches to $e^{-\Phi}$
perfectly as well. There is no obvious field theory dual for $\bar{N}_{r}$.
In fact, in relativistic holography, its counterpart the bulk field
$G_{tr}$ is usually set identically to zero by a gauge condition
thus its boundary field $\bar{G}_{tr}$ is automatically vanishing.
Here we do not gauge away the bulk field $N_{r}$ because of the non-relativistic
nature of the bulk gravity theory: $\partial_{I}\hat{\xi}^{t}=0$
which forbids this action. However, we are still allowed to set its
boundary value $\bar{N}_{r}$ to zero as a boundary condition and
this is indeed consistent (invariant) with respect to the residual
diffeomorphism (\ref{eq:NBDiffeo_Nr}).

Now the only remaining part is to map $\bar{N}_{i}$ and $\bar{V}_{\mu}$
to $\tilde{A}_{\mu}$ in the field theory. None of their diffeomorphism
(\ref{eq:NBDiffeo_Ni}) and (\ref{eq:NBDiffeo_Vector}) matches directly
with the transformations (\ref{eq:FTTransf_At_tilde}) and (\ref{eq:FTTransf_Ai_tilde})
of $\tilde{A}_{\mu}$. Since both $\bar{V}_{\mu}$ and $\tilde{A}_{\mu}$
are $U(1)$ gauge fields, and the residual gauge transformation $\bar{\Lambda}(x)$
in (\ref{eq:NBResidualGaugeTransf}) is already mapped to the gauge
transformation of $\tilde{A}_{\mu}$, it is natural to primarily map
$\bar{V}_{\mu}$ to $\tilde{A}_{\mu}$, i.e. $\bar{V}_{\mu}=\tilde{A}_{\mu}+\ldots$,
with some additional structures ``$\ldots$'' in the map so that
their diffeomorphism and Weyl transformation match perfectly. We have
collected these additional structures and their diffeomorphism transformations
in Appendix (B). For example, one of the structures we add to $\tilde{A}_{i}$
is the field theory dual of the near-boundary term $\bar{\phi}\bar{N}_{i}/\bar{N}$,
whose diffeomorphism is given in (\ref{eq:NBVectorStruct_2}). By
requiring $\bar{G}_{ik}\partial_{t}\bar{\xi}^{k}$ term to cancel
the $me^{\Phi}g_{ik}\partial_{t}\xi^{k}$ term in (\ref{eq:FTTransf_Ai_tilde})
and the Weyl index to be zero, we can identify the map for the scalar:
$\bar{\phi}=m$ with conformal dimension $\Delta_{\phi}=z-2$, which
also matches that of $m$'s in (\ref{eq:FTWeyl_m}). The physical
meaning of the bulk scalar $\phi$ is clear now: it is the bulk dual
of the mass $m$. When $z=2$, it is consistent to set it to a constant
in the bulk theory, i.e. just a parameter with no dynamics. When $z\neq2$,
we know in the field theory that $m$ is not scale-invariant (from
its Weyl transformation) and it will have a non-trivial renormalization
group (RG) flow from its UV value (or function) in (\ref{eq:FTAction_S0})
to some other value (function) in the IR. In this case, the dual bulk
field $\phi$ can be dynamical and develop some non-trivial profile
in the bulk, which is a geometrization of the RG flow of $m$ in the
field theory. The other structures involving $\varepsilon^{ij}$ in
Appendix (B) are used to cancel the other anomalous terms $-\frac{1}{2}\varepsilon^{ij}\partial_{i}\left(g_{jk}\partial_{t}\xi^{k}\right)$
in (\ref{eq:FTTransf_At_tilde}) and $\varepsilon^{jk}g_{ki}\partial_{j}\sigma$
in (\ref{eq:FTTransf_Ai_tilde}).

The last question remaining is what $\bar{N}_{i}$ maps to in the
field theory? So far every source field in the field theory action
(\ref{eq:FTAction_S0}) has been mapped from the near-boundary fields
of the dual gravity theory, and it seems there is no target image
left for $\bar{N}_{i}$. Even though $\bar{N}_{i}$ appears as additional
vectorial structures in the map between $\bar{V}_{\mu}$ and $\tilde{A}_{\mu}$,
we can view this as primarily a one-to-one and onto map between $\bar{V}_{\mu}$
and $\tilde{A}_{\mu}$, so $\bar{N}_{i}$ itself is still unmapped.
Alternatively, we can think the spatial components of this map maps
a combination of $\bar{V}_{i}$ and $\bar{N}_{i}$ to $\tilde{A}_{i}$,
which is a two-to-one map, thus we still need to map the other independent
combination of $\bar{V}_{i}$ and $\bar{N}_{i}$ to something in the
field theory. At this time, let us just assume that $\bar{N}_{i}$
maps to some field, called $-v_{i}$, in the field theory, whose transformation
is exactly the same as $\bar{N}_{i}$'s given in (\ref{eq:NBDiffeo_Ni}).
We will discuss the consequence of this and the meaning of $v_{i}$
in detail in the next subsection and late sections. Here for the construction
of holographic map, the presence of $\bar{N}_{i}$ is crucial, because
without it we would not be able to add additional structures, as those
shown in Appendix (B), to the map between $\bar{V}_{\mu}$ and $A_{\mu}$
to match their symmetry transformations on both sides.

Before presenting the holographic map, we notice that the following
modified definition for spin connection 
\begin{eqnarray}
\omega_{t}^{\prime} & = & \omega_{t}+\frac{1}{2}\varepsilon^{ij}\partial_{i}v_{j}-\frac{1}{2}\varepsilon^{ij}v_{i}\partial_{j}\log\left(me^{\Phi}\right)\label{eq:CovSpinConn_wt}\\
\omega_{i}^{\prime} & = & \omega_{i}+\frac{1}{2}\varepsilon^{jk}g_{ij}\partial_{k}\log\left(me^{\Phi}\right)\label{eq:CovSpinConn_wi}
\end{eqnarray}
transforms as an exact Lorentz vector under diffeomorphism and invariantly
under local Weyl transformation, i.e. same as $\bar{V}_{\mu}$ in
(\ref{eq:NBDiffeo_Vector}). Thus we can always add $\omega_{\mu}^{\prime}$
to the map of $\bar{V}_{\mu}$ with an arbitrary coefficient (we call
$s^{\prime}$) while still preserve the transformation properties
of the map.

We now present the holographic dictionary. The maps for the fields
are 
\begin{eqnarray}
\bar{N} & = & e^{-\Phi}\:,\label{eq:HolographicMap_N}\\
\bar{G}_{ij} & = & g_{ij}\:,\\
\bar{E}_{i}^{a} & = & e_{i}^{a}\:,\\
\bar{N}_{i} & = & -v_{i}\:,\label{eq:HolographicMap_Ni}\\
\bar{N}_{r} & = & 0\:,\\
\bar{V}_{t} & = & \tilde{A}_{t}-s^{\prime}\omega_{t}^{\prime}-\frac{1}{2}me^{\Phi}v^{2}-\frac{1}{2}\varepsilon^{ij}\partial_{i}v_{j}+\frac{1}{2}\varepsilon^{ij}v_{i}\partial_{j}\log\left(me^{\Phi}\right)\:,\label{eq:HolographicMap_Vt}\\
\bar{V}_{i} & = & \tilde{A}_{i}-s^{\prime}\omega_{i}^{\prime}+me^{\Phi}v_{i}-\frac{1}{2}\varepsilon^{jk}g_{ij}\partial_{k}\log\left(me^{\Phi}\right)\:,\label{eq:HolographicMap_Vi}\\
\bar{\phi} & = & m\:,\label{eq:HolographicMap_phi}
\end{eqnarray}
and the near-boundary index of $\phi$ to be 
\begin{equation}
\Delta_{\phi}=z-2\:.
\end{equation}
Here $v^{2}=g^{ij}v_{i}v_{j}$. Since in the field theory we allow
$\mathrm{g}$ and $s$ to be scalar functions which are Weyl invariant,
they, and the new $s^{\prime}$, can also be dual to some bulk scalar
fields with $r^{0}$ asymptotic behavior near the boundary. So far
$s^{\prime}$ is just a parameter appearing in the map which is allowed
by the symmetries. Its physical meaning in the context of fractional
quantum Hall effect will be explained in Section 5.4 when we compare
our results with the composite fermion and boson theories.

There can be additional parameters in the field theory, particularly
the couplings of interactions in $S_{\mathrm{int}}$. They can be
dual to additional bulk scalars, whose near-boundary behavior (conformal
dimension or bulk mass) shall match the corresponding Weyl transformation
in the field theory. For example, the 4-Fermi interaction coupling
$\lambda_{\mathrm{4F}}$ in (\ref{eq:FTWeyl_4Fcoupling}) shall dual
to a bulk scalar with asymptotic $r^{z-2}$ leading order behavior
near the boundary. When $z=2$, both $m$ and $\lambda_{\mathrm{4F}}$
are marginal by dimensional counting. If they are truly constant in
a conformal field theory (i.e. no running through renormalization),
then their bulk dual scalars can also be set to the same constants
without dynamics. However, if they are renormalized \cite{Nikolic:2007zz},
their bulk dual shall still be some dynamic fields that can develop
some non-trivial profiles in the bulk to reflect this running of couplings.

\subsection{Additional Constraints}

Now let us discuss what $v_{i}$ (i.e. $-\bar{N}_{i}$) means and
how we can deal with it. These are actually two separate questions.
In relativistic holography, from the perspective of bottom-up approaches,
because the full relativistic Poincaré symmetry is very powerful to
limit the allowed bulk action terms that one can start with, the bulk
dynamics is very restrictive. Typically, $\bar{N}_{i}$ is an independent
boundary field and appears in the on-shell boundary action. On contrary,
for a non-relativistic gravity theory, such as Ho\v{r}ava-Lifshitz
gravity, the symmetry is less restrictive. Thus one has more freedom
to add additional terms to the action with arbitrary coefficients.
These coefficients act as external tunable parameters that one can
use to fine-tune the bulk dynamics to the desired form that is appropriate
for the problem in consideration. In this large landscape of bulk
dynamics and possible forms of on-shell boundary action, $\bar{N}_{i}$
can behave quite differently, thus gives rises to different interpretations
and treatments of its field theory dual $v_{i}$. There exist many
different scenarios. There are certain scenarios in which the question
of how to deal with $\bar{N}_{i}$ becomes trivial. Two examples are
as following.
\begin{enumerate}
\item For some tactically constructed model bulk actions (possibly in the
help of additional symmetries such as the $U(1)_{\Sigma}$ symmetry
of \cite{Horava:2010zj,daSilva:2010bm}) with certain choices of parameters,
$\bar{N}_{i}$ (i.e. $-v_{i}$) may completely disappear after the
holographic dictionary is applied. This possibility is considered
first in \cite{Janiszewski:2012nb}.
\item For certain choices of the parameters of the bulk action the bulk
equations of motion near the boundary may constrain $\bar{N}_{i}$
to be either vanishing or (local or non-local) functional of the other
boundary fields, thus it is not independent. Similar phenomena are
not unfamiliar in relativistic Lifshitz holography \cite{Ross:2009ar,Ross:2011gu},
especially for $z>2$. Consequently, $v_{i}$ is a functional of other
source fields such as $g_{ij}$ and $\Phi$. In these cases, we only
need to find out what this functional is by solving the equations
of motion. $\bar{N}_{i}$ is essentially mapped to other source fields
by this functional. There is no additional \emph{independent} source
field $v_{i}$ appearing in the field theory through the holographic
dictionary. In these cases, there is no puzzle of how to deal with
$v_{i}$, although the question of how to interpret it still exists.
\end{enumerate}
In the rest of this paper, we will focus on another possible scenario
similar to the relativistic case, that is, when $\bar{N}_{i}$ is
independent of the other boundary fields in the bulk theory, and after
the holographic dictionary is applied, it still appears (as $-v_{i}$)
in the effective action. We then need to find a way to interpret the
meaning of $v_{i}$ and the operator $\hat{\mathcal{O}}^{i}$ it sources
(if it does), and to find a way to determine it. We think this is
the case most relevant to quantum Hall effects and later we will show
that it does yield results in nice agreement with the quantum Hall
literature.

\subsubsection{How not to interpret $v_{i}$}

First let us try to find out what could \emph{not} be the possible
interpretation of $v_{i}$. A reasonable speculation is that since
it comes from $\bar{N}_{i}$, which is essentially the $ti$-component
of the full bulk spacetime metric near the boundary, by analogy to
the relativistic holographic dictionary, it sources energy flux $\mathcal{E}^{i}$
and/or momentum density $p^{i}$ of the field theory. 

For energy flux $\mathcal{E}^{i}$, we have discussed in the previous
section that it is sourced by the field $C_{i}$ in the field theory.
But the diffeomorphism of $C_{i}$, (\ref{eq:FTDiffeo_Ci}) is different
from $\bar{N}_{i}$'s in (\ref{eq:NBDiffeo_Ni}). Thus $v_{i}$ can
not be identified with the source field for energy flux. Furthermore,
we have worked in GTCs throughout, in which $C_{i}=0$. In GTCs energy
is globally conserved and there is no need for the notion of energy
flux in the field theory we are considering. Even we want to match
$v_{i}$ to $C_{i}$, it must be zero in GTCs. But $v_{i}=0$ is not
invariant under residual diffeomorphism (\ref{eq:NBDiffeo_Ni}), thus
can not be a valid map or constraint. On the other hand, from the
holographic point of view, \cite{Janiszewski:2012nb} has conjectured
that the energy flux source $C_{i}$ (what is called $B_{i}$ there)
is dual to what they call the field $P_{i}$ in the bulk, the $O\left(c^{2}\right)$
order of $ti$-component of the full spacetime Lorentz metric if the
dependence on speed of light $c$ is restored: $G_{tI}=c^{2}P_{I}+N_{I}$.
Diffeomorphism of $P_{I}$ involves explicitly $\partial_{I}\hat{\xi}^{t}$.
The non-relativistic theory can be viewed as a limit of $c\rightarrow\infty$.
For us, since we have restricted ourselves to GTCs for both field
theory and its bulk dual, in which $\partial_{I}\hat{\xi}^{t}=0$
and $P_{I}$ vanishes identically, $v_{i}$ can not be related to
energy flux.

How about momentum density $p^{i}$? The answer is again negative.
As we discussed in the previous section, in the non-relativistic field
theory, the momentum density is not an independent quantity, but rather
expressed in terms of the conserved current as given in (\ref{eq:Momentum-Current}).
Thus there is no need for a separate field in the field theory to
source the momentum density $\langle p_{i}\rangle$. The only thing
one need to do is to compute $\langle J^{\mu}\rangle$, then use (\ref{eq:Momentum-Current})
to calculate $\langle p^{i}\rangle$. Thus it is not necessary to
have a separate source field $v_{i}$ to source momentum density.

\subsubsection{How not to deal with $v_{i}$}

In fact, an object similar to $v_{i}$ has appeared in the literature.
In both \cite{Son:2013rqa} and \cite{Geracie:2014nka} it is the
so called ``velocity vector'' (also denoted by $v^{i}$ there) in
the Newton-Cartan geometry which is employed to aid to construct the
effective action of the field theory based on symmetries. This velocity
vector essentially acts like a dual shift function for Newton-Cartan
geometry, similar to the role of $N^{I}$ in our bulk theory. In fact
if we raise $v_{i}$'s index by $g^{ij}$, the resulting transformation
for $v^{i}$ is 
\begin{equation}
\delta v^{i}=\xi^{\mu}\partial_{\mu}v^{i}+v^{i}\partial_{t}\xi^{t}-v^{k}\partial_{k}\xi^{i}-\partial_{t}\xi^{i}\:.
\end{equation}
This is the same as the transformation rule for $v^{i}$ in \cite{Son:2013rqa,Geracie:2014nka}.
The second term $v^{i}\partial_{t}\xi^{t}$ is absent in \cite{Son:2013rqa},
because $\Phi=0$ and $\partial_{t}\xi^{t}=0$ there. Thus, we can
say that our $v^{i}$ is essentially the same as that appears in \cite{Son:2013rqa,Geracie:2014nka}.
(The same object appears in Lifshitz holography as well \cite{Christensen:2013lma,Christensen:2013rfa}.)
In neither \cite{Son:2013rqa} nor \cite{Geracie:2014nka} does $v^{i}$
act like a fundamental external source field. In \cite{Geracie:2014nka}
the introduction of $v^{i}$ to the field theory is to modify the
seemingly non-covariant vector potential $A_{\mu}$ to be a Newton-Cartan
covariant vector and thus to reformulate the effective action and
Ward identities in a covariant manner. This is similar to what we
are doing here: we can view the map (\ref{eq:HolographicMap_Vt})
and (\ref{eq:HolographicMap_Vi}) is a modification from the non-covariant
$A_{\mu}$ (in aid of $v_{i}$) to the covariant vector $\bar{V}_{\mu}$,
and the on-shell action calculated from the bulk dynamics automatically
offers a covariant way to organize the effective action as a functional
of the covariant sources $\bar{V}_{\mu}$, $\bar{N}$, $\bar{N}_{i}$,
$\bar{G}_{ij}$ among others. The difference between our case and
\cite{Geracie:2014nka}'s is the notion of covariance -- Lorentzian
for us (strictly speaking Ho\v{r}ava, which is Lorentzian with $\partial_{i}\xi^{t}=0$)
while Newton-Cartan for \cite{Geracie:2014nka}, but other than these
technical details, the philosophy is the same. Thus we can identify
our $v^{i}$ as the same $v^{i}$ appearing in \cite{Son:2013rqa,Geracie:2014nka}.
Based on the fact that $v^{i}$ is only introduced formally and shall
not affect the physics, \cite{Geracie:2014nka} assumes that the effective
action shall be independent of $v^{i}$ once it is written out explicitly
in terms of the physical sources $\Phi$, $g_{ij}$, $A_{\mu}$ etc.
However, effective action obtained from holography can rarely satisfy
such a stringent constraint, so we assume our effective action will
depend on $v^{i}$ in general, unless we impose an additional constraint
besides the holographic map (\ref{eq:HolographicMap_N})-(\ref{eq:HolographicMap_phi}).
This additional constraint is precisely the subject under discussion
here.

On the other hand, \cite{Son:2013rqa} views $v^{i}$ as a dynamical
field with respect to which the effective action is extremized. From
the field theoretical point of view, $\mathcal{W}$ in (\ref{eq:FTEffectiveAction})
is the extremum of another quasi-effective action $\mathcal{I}\left[v^{i},A_{\mu},g_{ij},\Phi,\ldots\right]$
with respect to $v^{i}$. In other words, $v^{i}$ has been path-integrated
out in $\mathcal{I}$ which can be classically done by imposing Euler-Lagrange
equation for $v^{i}$. What is the origin of $v^{i}$ in field theory?
It is possibly an auxiliary field introduced to facilitate path-integral
for the microscopic field $\psi$. For example, in \cite{Son:2013rqa}
it is shown to arise from the Hubbard-Stratonovich transformation
for $D_{i}\varphi$, where $\varphi$ is the phase of $\psi$. After
such a transformation, one can integrate out $\psi$ and $\psi^{\dagger}$
(i.e. $|\psi|$ and $\varphi$), leaving an quasi-effective action
$\mathcal{I}$ which depends explicitly on $v^{i}$. The path-integral
of $v^{i}$ remains to be carried out to yield the final answer to
the effective action $\mathcal{W}$. This can be formally written
as 
\[
e^{i\mathcal{W}\left[A_{\mu},g_{ij},\Phi,\ldots\right]}=\int\mathcal{D}v^{i}e^{i\mathcal{I}\left[v^{i},A_{\mu},g_{ij},\Phi,\ldots\right]}\:,
\]
where 
\[
e^{i\mathcal{I}\left[v^{i},A_{\mu},g_{ij},\Phi,\ldots\right]}=\int\mathcal{D}\psi\mathcal{D}\psi^{\dagger}e^{iS'\left[\psi,v^{i},A_{\mu},g_{ij},\Phi,\ldots\right]}
\]
and $S'\left[\psi,v^{i},\ldots\right]$ is related to the original
microscopic action $S\left[\psi,\ldots\right]$ by a Hubbard-Stratonovich
transformation. The path-integral over $v^{i}$ can be done classically
by imposing the Euler-Lagrange equation for $v^{i}$ 
\begin{equation}
\frac{\delta}{\delta v^{i}}\mathcal{I}\left[v^{i},A_{\mu},g_{ij},\Phi,\ldots\right]=0\label{eq:Extremum_vi}
\end{equation}
and finding the extremum of $\mathcal{I}$. 

In holographic approach, after calculating the bulk dynamics, we will
end up with an on-shell action as a functional of the boundary fields:
$\bar{\mathcal{I}}\left[\bar{N}_{i},\bar{V}_{\mu},\cdots\right]$.
Here every thing is written in terms of the bulk theory language,
and the near-boundary field $\bar{N}_{i}$, $\bar{V}_{\mu}$ and others
are viewed as independent variables of the functional $\bar{\mathcal{I}}$.
We denote the other boundary fields such as $\bar{N}$, $\bar{G}_{ij}$
etc as ``$\cdots$''. Next we use our holographic dictionary (\ref{eq:HolographicMap_N})-(\ref{eq:HolographicMap_phi})
to write the action explicitly in terms of the physical sources $\Phi$,
$g_{ij}$, $A_{\mu}$ and the extra field $v^{i}$, and identify this
functional as $\mathcal{I}\left[v^{i},A_{\mu},g_{ij},\Phi,\ldots\right]$
introduced before, whose independent variables are now $v^{i}$ and
$A_{\mu}$, $g_{ij}$ and $\Phi$: 
\[
\bar{\mathcal{I}}\left[\bar{N}_{i}\left[v_{i}\right],\bar{V}_{\mu}\left[v_{i},A_{\mu}\ldots\right],\cdots\right]=\mathcal{I}\left[v^{i},A_{\mu},g_{ij},\Phi,\ldots\right]\:.
\]
Now if we naively apply the philosophy of \cite{Son:2013rqa}, we
can impose (\ref{eq:Extremum_vi}) as a constraint, find its solution
$v_{\star}^{i}$, then the effective action 
\[
\mathcal{W}\left[A_{\mu},g_{ij},\Phi,\ldots\right]=\mathcal{I}\left[v_{\star}^{i},A_{\mu},g_{ij},\Phi,\ldots\right]\:.
\]

From the holographic point of view, this approach seems fine, and
we do get a result for $\mathcal{W}$, which is an effective action
of a certain quantum field theory with the same non-relativistic spacetime
symmetries discussed in the previous section. But from the field theoretical
point of view, is it the one for the field theory we discussed in
the previous section which describes the quantum Hall effect? Not
quite! A subtlety is that in \cite{Son:2013rqa} the path integral
is not only carried out for $v^{i}$, but also for other auxiliary
fields $\rho$, $\varphi$ and $a_{\mu}$. Thus even we are doing
the same integral for $v^{i}$, we are not completely in parallel
with \cite{Son:2013rqa}. A consequence is that our results obtained
from this way will not have a well-defined projection to the lowest
Landau level (LLL). This can be shown by a straightforward calculation
which we will not present here. The projection to LLL is defined as
the limit $\mathrm{g}=2$ and $m\rightarrow0$, and the effective
action $\mathcal{W}$ shall be regular in this limit. \cite{Geracie:2014nka}
has shown that field theories defined by microscopic actions such
as (\ref{eq:FTAction_S0}) do have a smooth LLL projection. However,
if we impose (\ref{eq:Extremum_vi}) as the additional constraint
in holography, the resulting effective action is either singular or
over-constrained in LLL projection. Thus it can not be a valid constraint.
This choice is ruled out not by holographic considerations, but by
a field theoretical consideration -- the LLL projection, which is
very special to the quantum Hall problem we are considering here.
Had one considered a holographic dual for another quantum system,
this constraint might be a viable one.

\subsubsection{Choice (I): A Path Integral Constraint}

We have just argued that the path integral over $v^{i}$ does not
work, but the idea of integrating out the extra fields still work.
The key point is that we need to find out the right fields to integrate
out. In addition to the requirement that it reproduces the right features
of the quantum Hall problem, such as the LLL projection, a more fundamental
criterion is that the path integral over these fields shall also be
gauge and diffeomorphism invariant. More specifically, path integral
over some fields is classically equivalent to imposing the Euler-Lagrange
equations for those fields, similar to the extremum condition (\ref{eq:Extremum_vi})
for $v^{i}$, and these equations shall be invariant under gauge transformation
and diffeomorphism. Instead of looking for such constraints for the
field theory action $\mathcal{I}\left[v^{i},A_{\mu},g_{ij},\Phi,\ldots\right]$,
it is much easier to find ones for the holographic action $\bar{\mathcal{I}}\left[\bar{N}_{i},\bar{V}_{\mu},\cdots\right]$.
Since the problem we are facing is the two-to-one map from $\bar{N}_{i}$
and $\bar{V}_{i}$ to $A_{i}$, and the map (\ref{eq:HolographicMap_Vi})
contains one linear combination of $\bar{N}_{i}$ and $\bar{V}_{i}$,
we will seek to find another independent linear combination of $\bar{N}_{i}$
and $\bar{V}_{i}$ -- let us call it $\bar{X}^{i}$ -- to integrate
out. We want the equation of motion for $\bar{X}^{i}$ -- let us call
it $\bar{Y}_{i}=0$, where $\bar{Y}_{i}\sim\delta\bar{\mathcal{I}}/\delta\bar{X}^{i}$
up to a volume measure -- to be diffeomorphism invariant. If $\bar{Y}_{i}$
transforms exactly like spatial components of a Lorentz 1-form, similar
to that of $\bar{V}_{i}$: 
\[
\delta\bar{Y}_{i}=\bar{\xi}^{\mu}\partial_{\mu}\bar{Y}_{i}+\bar{Y}_{k}\partial_{i}\bar{\xi}^{k}\:,
\]
then $\bar{Y}_{i}=0$ is diffeomorphism invariant. Since $\bar{Y}_{i}$
is a functional derivative with respect to $\bar{X}^{i}$, this requires
$\bar{X}^{i}$ to transform exactly as spatial components of Lorentz
vector $\bar{X}^{\mu}$: 
\[
\delta\bar{X}^{\mu}=\bar{\xi}^{\nu}\partial_{\nu}\bar{X}^{\mu}-\bar{X}^{\nu}\partial_{\nu}\bar{\xi}^{\mu}\:.
\]
From holography we have already had a good Lorentz 1-form $\bar{V}_{\mu}$,
it is not hard to construct a Lorentz vector from it -- just by raising
the index $\mu$ with the full spacetime metric in ADM form. The result
is 
\begin{eqnarray}
\bar{X}^{\mu} & = & \left(-\frac{1}{\bar{N}^{2}}\left[\bar{V}_{t}-\bar{N}^{k}\bar{V}_{k}\right]\,,\,\bar{G}^{ij}\bar{V}_{j}+\frac{1}{\bar{N}^{2}}\bar{N}^{i}\left[\bar{V}_{t}-\bar{N}^{j}\bar{V}_{j}\right]\right)\:,\label{eq:PathIntegral_Vector}\\
\bar{V}_{\mu} & = & \left(\left[-\bar{N}^{2}+\bar{N}^{k}\bar{N}_{k}\right]\bar{X}^{t}+\bar{N}_{k}\bar{X}^{k}\,,\,\bar{N_{i}}\bar{X}^{t}+\bar{G}_{ij}\bar{X}^{j}\right)\:,
\end{eqnarray}
where the second line is the inverse of the first line. It is straightforward
to check $\bar{X}^{\mu}$ defined in this way satisfies the above
diffeomorphism transformation using (\ref{eq:NBDiffeo_N})-(\ref{eq:NBDiffeo_Vector}).
Then using the second line of the above equations, the condition $\delta\bar{\mathcal{I}}/\delta\bar{X}^{i}=0$
can be written explicitly as 
\begin{equation}
\bar{N}_{i}\frac{\delta\bar{\mathcal{I}}}{\delta\bar{V}_{t}}+\bar{G}_{ij}\frac{\delta\bar{\mathcal{I}}}{\delta\bar{V}_{j}}=0\:,\label{eq:PathIntegral_Condition}
\end{equation}
where the holographic on-shell action $\bar{\mathcal{I}}=\bar{\mathcal{I}}\left[\bar{V}_{\mu},\bar{N},\bar{N}_{i},\bar{G}_{ij},\bar{\phi}\right]$.
This constraint equation means that we are integrating out $\bar{X}^{i}$
in $\bar{\mathcal{I}}$ while holding $\bar{X}^{t}$, $\bar{N}_{i}$,
$\bar{N}$, $\bar{G}_{ij}$ and $\bar{\phi}$ fixed. It is diffeomorphism
invariant as advertised, Weyl invariant (this is trivial, since its
right hand side is zero) as well as gauge invariant (by analog to
the definition of conserved current). In practice, it can be viewed
as an equation for $\bar{N}_{i}$, which allows one to solve $\bar{N}_{i}$
(i.e. $v_{i}$) in terms of other fields. Thus it shall be viewed
as part of the holographic dictionary. Together with (\ref{eq:HolographicMap_N})-(\ref{eq:HolographicMap_phi}),
they completely determine the holographic map from the boundary fields
of the bulk theory to the source fields of the non-relativistic field
theory, allowing one to obtain a unique field theory effective action
$\mathcal{W}$ from the holographic on-shell action $\bar{\mathcal{I}}$.
In the next section, we will show that the effective action obtained
this way is a good description of the quantum Hall effect, in agreement
with previous results obtained from non-holographic approaches, and
has a good LLL projection. From that we will further uncover the true
physical interpretation of $v^{i}$ in the context of fractional quantum
Hall effect.

\subsubsection{Choice (II): A Field Constraint}

(\ref{eq:PathIntegral_Condition}) is not the only choice for constraint
equation that is diffeomorphism and gauge invariant. Other than this
type of constraint written in terms of functional derivatives of the
action, which acquires a natural interpretation in term of path integral,
we can write down another type of constraint using only the fields,
without the action. It it straightforward to check that the following
equation 
\begin{equation}
\bar{V}_{ti}+\bar{V}_{ij}\bar{G}^{jk}\bar{N}_{k}=0\label{eq:Field_Constraint}
\end{equation}
is also diffeomorphism, Weyl and gauge invariant. Here $\bar{V}_{\mu\nu}=\partial_{\mu}\bar{V}_{\nu}-\partial_{\nu}\bar{V}_{\mu}$.
Even though the possible interpretation of this equation is different
from (\ref{eq:PathIntegral_Condition})'s, it functions in the same
way as the latter does: it is an equation for $\bar{N}_{i}$ and together
with the rest of the holographic dictionary (\ref{eq:HolographicMap_N})-(\ref{eq:HolographicMap_phi})
completely and uniquely determines the field theory effective action
$\mathcal{W}$ in terms of the sources $\Phi$, $g_{ij}$ and $A_{\mu}$.
From holographic point of view, it can not be ruled out in favor of
(\ref{eq:PathIntegral_Condition}). Nor can it be from field theoretical
considerations for quantum Hall problems, at least at the primitive
level regarding only the universal transport properties, which we
will discuss in detail in the next section. For example, it also has
a good LLL projection. Actually, despite the seemingly different appearances
of (\ref{eq:PathIntegral_Condition}) and (\ref{eq:Field_Constraint}),
for the quantum Hall effect at leading orders in derivative expansion
where the action $\bar{\mathcal{I}}$ is dominated by gauge Chern-Simons
term, the major part of the leading order of (\ref{eq:PathIntegral_Condition})
and (\ref{eq:Field_Constraint}) are the same and the correlation
functions they produce are also the same at leading orders. Thus unless
one goes to a more detailed calculation involving higher order corrections
or comes up with other physical considerations for the quantum Hall
effect which favor one over the other, at the current level of discussion,
we will view both constraints (\ref{eq:PathIntegral_Condition}) and
(\ref{eq:Field_Constraint}) are viable.

It is interesting to note that the left hand side of the above constraint
(\ref{eq:Field_Constraint}) takes the same form as the combination
appearing in the first term of the low energy gauge action (\ref{eq:HoravaAction_Gauge})
in Ho\v{r}ava-Lifshitz gravity theory. This part is roughly the energy
density stored in the uniform electric field, and it depends on the
shift function $\bar{N}_{i}$. The rest of the action is roughly minus
of the energy density stored in the magnetic field, and is independent
of $\bar{N}_{i}$. The former, the electric energy density, is a square
of the ``electric field'', i.e. the left hand side of (\ref{eq:Field_Constraint}),
thus is always non-negative. It only minimizes the action when it
is zero, which implies the above constraint.

\bigskip{}

\section{Low Energy Chern-Simons Effective Action of FQHE}

In this section we study a simple holographic model with non-dynamical
Chern-Simons terms to show how the holographic dictionary we have
just built can be used to derive an effective action for quantum Hall
fluids, from which many universal properties inaccessible before can
be extracted now. A similar holographic model with non-dynamical gauge
Chern-Simons term is used in \cite{KeskiVakkuri:2008eb}. The infrared
physics of quantum Hall effect at scale much larger than the magnetic
length $1/\sqrt{B}$ is dominated by the Chern-Simons term from the
point of view of effective theory \cite{Wen:1992uk}. This give rise
to many universal properties which are insensitive to local dynamics
(at least at this scale). We will focus on extracting these universal
properties at leading and, in some occasions, sub-leading orders,
and they are always dominated by the Chern-Simons term. Terms in the
effective action generated from the local dynamics -- the contributions
from non-Chern-Simons terms in effective theory or in holography,
like those from (\ref{eq:HoraveAction_Graviton}) and (\ref{eq:HoravaAction_Gauge})
-- are in general non-local. But for quantum Hall effects where $B$
field is large and system is gapped, these terms can be written as
an series of derivative expansions and every term is local now.%
\footnote{In the following we will refer to this type of terms as ``local terms'',
sometimes with a subscript ``$_{\mathrm{loc}}$'', on contrary to
terms obtained from non-dynamical Chern-Simons terms, which are topological
and non-local by nature.%
} Almost every such term is sub-leading in the derivative expansion,
except for one term which we will denote by a general functional $f\left[B\right]$.
The non-dynamical Chern-Simons terms in 3+1 dimensional holography
are essentially all boundary terms. This gives us several advantages.
Since they are boundary terms, they are insensitive to the bulk dynamics
which can not be uniquely determined from bottom-up approaches, but
preserve the universality that we are mostly interested in. This allows
us not to specify the local bulk dynamics, or specify it in a way
as general as possible (for example, we assume only a general functional
$f\left[B\right]$ later on to denote some possible leading order
contributions from local terms without knowing or solving its explicit
form). This grants maximum flexibility for later model-building when
the universal properties we find here can still be valid. Last but
not least, it simplifies the calculation and makes simple analytic
results possible. More complicated models such as those involving
dynamical Chern-Simons terms coupled to axion fields are beyond the
scope of this paper and we will leave them to future research.

\subsection{Holographic Gauge Chern-Simons Term}

The bulk gauge Chern-Simons term \cite{Wilczek:1987,Carroll:1989vb}
is 
\begin{equation}
\hat{\mathcal{I}}_{\mathrm{CS}}=\frac{\nu}{16\pi}\int d^{4}x\epsilon^{MNPQ}V_{MN}V_{PQ}\:,
\end{equation}
where $V_{MN}=\partial_{M}V_{N}-\partial_{N}V_{M}$. For simplicity,
we choose the coupling to be a constant $\nu$, rather than a dynamical
axion. Then this term is essentially a boundary term: 
\begin{equation}
\bar{\mathcal{I}}_{\mathrm{CS}}=\frac{\nu}{4\pi}\int_{r=0}d^{3}x\epsilon^{ij}\left\{ \bar{V}_{t}\partial_{i}\bar{V}_{j}+\bar{V}_{i}\left(\partial_{j}\bar{V}_{t}-\partial_{t}\bar{V}_{j}\right)\right\} \:.\label{eq:BoundaryAction_CS}
\end{equation}
We denote the total on-shell action of the bulk theory to be 
\begin{equation}
\bar{\mathcal{I}}\left[\bar{V}_{\mu},\bar{N},\bar{N}_{i},\bar{G}_{ij},\bar{\phi}\right]=\bar{\mathcal{I}}_{\mathrm{CS}}\left[\bar{V}_{\mu}\right]+\bar{\mathcal{I}}_{\mathrm{loc}}\left[\bar{V}_{\mu},\bar{N},\bar{N}_{i},\bar{G}_{ij},\bar{\phi}\right]\:,\label{eq:BoundaryAction_total}
\end{equation}
where $\bar{\mathcal{I}}_{\mathrm{loc}}$ is the collection of all
the other terms in the action, e.g. those involving local dynamics
of the graviton and gauge field. 

We will solve the model perturbatively by a derivative expansion.
First, we adopt the power counting scheme commonly used in the context
of quantum Hall effective theories (e.g. in \cite{Hoyos:2011ez}).
Let $\epsilon$ be an infinitesimal book-keeping parameter that labels
the orders of the derivative expansion (not to be confused with Levi-Civita
symbol). We assume 
\[
\partial_{i}\sim O\left(\epsilon\right)\:,\quad\partial_{t}\sim O\left(\epsilon^{2}\right)\:,\quad A_{i}\sim O\left(\epsilon^{-1}\right)\:,\quad A_{t},g_{ij},\Phi\sim O\left(1\right)\:.
\]
Then $B\sim O\left(1\right)$, $E_{i}\equiv\partial_{i}A_{t}-\partial_{t}A_{i}\sim O\left(\epsilon\right)$,
$\omega_{\mu}\sim O\left(\epsilon\right)$, $\mathcal{R}^{(2)}\sim O\left(\epsilon^{2}\right)$.
Thus according to the holographic dictionary, $\bar{V}_{i}\sim O\left(\epsilon^{-1}\right)$,
$\bar{V}_{t}\sim O\left(1\right)$, $\bar{V}_{ij}\sim O\left(1\right)$,
$\bar{V}_{ti}\sim O\left(\epsilon\right)$. We further assume 
\[
v_{i}\sim O\left(\epsilon\right)\:,\qquad\textrm{i.e.}\qquad\bar{N}_{i}\sim O\left(\epsilon\right)\:.
\]
We will later need to check that this assumption is self-consistent.
Even though at this time we do not know the specific form of $\bar{\mathcal{I}}_{\mathrm{loc}}$,
we know at least their basic structures allowed by symmetries. The
leading orders in derivative expansion will have forms given in (\ref{eq:HoraveAction_Graviton})
and (\ref{eq:HoravaAction_Gauge}) \cite{Kiritsis:2009sh}. This implies
the following order estimation: 
\[
\frac{\delta\bar{\mathcal{I}}_{\mathrm{loc}}}{\delta\bar{V}_{i}}\sim O\left(\epsilon\right)\:,\quad\frac{\delta\bar{\mathcal{I}}_{\mathrm{loc}}}{\delta\bar{V}_{t}}\sim O\left(\epsilon^{2}\right)\:,\quad\frac{\delta\bar{\mathcal{I}}_{\mathrm{loc}}}{\delta\bar{N}_{i}}\sim O\left(\epsilon\right)\:.
\]
We further single out one particular term in $\bar{\mathcal{I}}_{\mathrm{loc}}$,
denoted by $f\left[\frac{1}{2}\bar{\varepsilon}^{ij}\bar{V}_{ij}\right]$:
\begin{equation}
\bar{\mathcal{I}}_{\mathrm{loc}}\supset\int d^{3}x\bar{N}\sqrt{\bar{G}}f\left[\frac{1}{2}\bar{\varepsilon}^{ij}\bar{V}_{ij}\right]\label{eq:LocalAction_f[B]}
\end{equation}
where $f$ is an arbitrary scalar functional of $\frac{1}{2}\bar{\varepsilon}^{ij}\bar{V}_{ij}$.
The function $-f\left[B\right]$ is related to the energy density
$\langle\mathcal{E}^{0}\rangle$ as a function of the magnetic field.%
\footnote{Our notation $f\left[B\right]$ emphasize that it has no dependence
on mass $m$ and other source fields other than the metric-dependence
through $B$. It can, however, depend on other couplings introduced
through the interactions, such as the dimensionless $\lambda_{\mathrm{4F}}$
in (\ref{eq:FTWeyl_4Fcoupling}) or other dimensionful couplings like
that of the Coulomb interaction. For the 4-Fermi interaction which
is scale invariant at $z=2$, just by dimensional analysis $f\left[B\right]$
must take the form 
\begin{equation}
f\left[B\right]\sim\lambda_{\mathrm{4F}}B^{2}\:.\label{eq:f[B]_4Fermi}
\end{equation}
} This is the only term that contribute to $O\left(\epsilon\right)$
order from $\delta\bar{\mathcal{I}}_{\mathrm{loc}}/\delta\bar{V}_{i}$:
\[
\frac{\delta\bar{\mathcal{I}}_{\mathrm{loc}}}{\delta\bar{V}_{i}}=\bar{N}f^{\prime\prime}\left[\frac{1}{2}\bar{\varepsilon}^{k'l'}\bar{V}_{k'l'}\right]\epsilon^{ij}\partial_{j}\left(\frac{1}{2}\bar{\varepsilon}^{kl}\bar{V}_{kl}\right)+f^{\prime}\left[\frac{1}{2}\bar{\varepsilon}^{k'l'}\bar{V}_{k'l'}\right]\epsilon^{ij}\partial_{j}\bar{N}+O\left(\epsilon^{2}\right)\:.
\]
Here one ``$^{\prime}$'' is one derivative of $f$ with respect
to its argument.

Using the holographic dictionary (\ref{eq:HolographicMap_N})-(\ref{eq:HolographicMap_phi}),
the Chern-Simons term can be expanded as 
\begin{align}
\bar{\mathcal{I}}_{\mathrm{CS}} & =\int d^{3}x\Big\{\frac{\nu}{4\pi}\epsilon^{\rho\mu\nu}A_{\rho}\partial_{\mu}A_{\nu}-\frac{\nu}{2\pi}(s+s^{\prime})\epsilon^{\rho\mu\nu}\omega_{\rho}\partial_{\mu}A_{\nu}+\frac{\nu}{4\pi}(s+s^{\prime})^{2}\epsilon^{\rho\mu\nu}\omega_{\rho}\partial_{\mu}\omega_{\nu}\Big\}\nonumber \\
 & +\frac{\nu}{4\pi}\int d^{3}x\sqrt{g}e^{-\Phi}\Big\{\left[me^{2\Phi}\left(-2\varepsilon^{ij}E_{i}v_{j}-Bv^{2}\right)+\left(s^{\prime}+1\right)e^{\Phi}E^{i}\partial_{i}\log\left(me^{\Phi}\right)\right]\nonumber \\
 & +\frac{(\mathrm{g}-2)}{2m}B\left[B+\left(1-2s-s^{\prime}\right)\mathcal{R}^{(2)}+\varepsilon^{ij}\omega_{i}\partial_{j}\left(s+s^{\prime}\right)-\varepsilon^{ij}\partial_{i}\left(\frac{s^{\prime}+1}{2}\varepsilon^{kl}g_{jk}\partial_{l}\log\left(me^{\Phi}\right)\right)\right]\nonumber \\
 & -\left(2-\frac{\mathrm{g}}{2}+s^{\prime}\right)e^{\Phi}B\varepsilon^{ij}\left[\partial_{i}v_{j}+v_{j}\partial_{i}\log\left(me^{\Phi}\right)\right]\Big\}+O\left(\epsilon^{3}\right)\:.\label{eq:W_CS_with_vi}
\end{align}
By our power counting scheme, the third terms in the first line is
of order $O\left(\epsilon^{4}\right)$, but since it takes the form
of gravitational Chern-Simons form and its structure is unique compare
to other possible terms at the same order, we write it down explicitly.
(\ref{eq:LocalAction_f[B]}) can be expanded as 
\begin{eqnarray}
\bar{\mathcal{I}}_{\mathrm{loc}} & \supset & \int d^{3}xe^{-\Phi}\sqrt{g}\Big\{ f\left[B\right]+f^{\prime}\left[B\right]\Big[-\varepsilon^{ij}\partial_{i}\left(\left(s+s^{\prime}\right)\omega_{j}\right)+\varepsilon^{ij}\partial_{i}\left(me^{\Phi}v_{j}\right)\nonumber \\
 &  & \qquad+\nabla^{2}\left(\frac{s^{\prime}+1}{2}\log\left(me^{\Phi}\right)\right)\Big]+O\left(\epsilon^{3}\right)\Big\}\:.\label{eq:W_loc_with_vi}
\end{eqnarray}

\subsection{Under the Path Integral Constraint}

In this subsection, we use the path integral constraint (\ref{eq:PathIntegral_Condition})
to calculate the effective action. Using (\ref{eq:BoundaryAction_total})
it can be written as 
\begin{equation}
-\frac{\nu}{2\pi}\epsilon^{ij}\bar{V}_{tj}+\frac{\delta\bar{\mathcal{I}}_{\mathrm{loc}}}{\delta\bar{V}_{i}}+\bar{N}^{i}\left(\frac{\nu}{4\pi}\epsilon^{kl}\bar{V}_{kl}+\frac{\delta\bar{\mathcal{I}}_{\mathrm{loc}}}{\delta\bar{V}_{t}}\right)=0\:.
\end{equation}
At $O\left(\epsilon\right)$ order, this gives 
\begin{eqnarray*}
\bar{N}^{i} & = & \frac{\bar{\varepsilon}^{ij}\bar{V}_{tj}}{\frac{1}{2}\bar{\varepsilon}^{kl}\bar{V}_{kl}}-\frac{2\pi}{\nu}\bar{N}f^{\prime\prime}\left[\frac{1}{2}\bar{\varepsilon}^{k'l'}\bar{V}_{k'l'}\right]\bar{\varepsilon}^{ij}\partial_{j}\log\left(\frac{1}{2}\bar{\varepsilon}^{kl}\bar{V}_{kl}\right)\\
 &  & -\frac{2\pi}{\nu}\frac{1}{\frac{1}{2}\bar{\varepsilon}^{kl}\bar{V}_{kl}}f^{\prime}\left[\frac{1}{2}\bar{\varepsilon}^{k'l'}\bar{V}_{k'l'}\right]\bar{\varepsilon}^{ij}\partial_{j}\bar{N}+O\left(\epsilon^{2}\right)\:.
\end{eqnarray*}
Applying the holographic dictionary, we can translate it into field
theory language: 
\begin{eqnarray}
v^{i} & = & \frac{\varepsilon^{ij}E_{j}}{B}+\frac{(\mathrm{g}-2)e^{-\Phi}}{4m}\varepsilon^{ij}\partial_{j}\log\left(\frac{B}{m}e^{-\Phi}\right)+\frac{e^{-\Phi}}{4m}\varepsilon^{ij}\partial_{j}\mathrm{g}\nonumber \\
 &  & +\frac{2\pi}{\nu}e^{-\Phi}\left(f^{\prime\prime}\left[B\right]\varepsilon^{ij}\partial_{j}\log B-\frac{1}{B}f^{\prime}\left[B\right]\bar{\varepsilon}^{ij}\partial_{j}\Phi\right)+O\left(\epsilon^{2}\right)\:.\label{eq:DriftVelocity_PIC}
\end{eqnarray}
Plugging this into (\ref{eq:W_CS_with_vi}) and (\ref{eq:W_loc_with_vi})
we will get the final results for the effective action. The full expressions
are very lengthy so we will not record it here. But when $\Phi=0$
and $m$, $\mathrm{g}$, $s$ and $s^{\prime}$ are all constants,
the expressions simplify to 
\begin{align}
\mathcal{W}_{\mathrm{CS}} & =\int d^{3}x\Big\{\frac{\nu}{4\pi}\epsilon^{\rho\mu\nu}A_{\rho}\partial_{\mu}A_{\nu}-\frac{\nu}{2\pi}(s+s^{\prime})\epsilon^{\rho\mu\nu}\omega_{\rho}\partial_{\mu}A_{\nu}+\frac{\nu}{4\pi}(s+s^{\prime})^{2}\epsilon^{\rho\mu\nu}\omega_{\rho}\partial_{\mu}\omega_{\nu}\Big\}\nonumber \\
 & +\frac{\nu}{4\pi}\int dtd^{2}\vec{x}\sqrt{g}\Big\{ m\frac{E^{2}}{B}-\left(2-\frac{\mathrm{g}}{2}+s^{\prime}\right)E^{i}\partial_{i}\log B+\frac{(\mathrm{g}-2)}{2m}B\left[B+\left(1-2s-s^{\prime}\right)\mathcal{R}^{(2)}\right]\nonumber \\
 & -\left[\frac{\mathrm{g}-2}{4}\left(\frac{3}{2}-\frac{\mathrm{g}}{4}+s^{\prime}\right)\frac{B}{m}+\left(1+s^{\prime}\right)\frac{2\pi}{\nu}Bf^{\prime\prime}\left[B\right]+\frac{m}{B}\left(\frac{2\pi}{\nu}Bf^{\prime\prime}\left[B\right]\right)^{2}\right]\left(\partial_{i}\log B\right)^{2}\Big\}\nonumber \\
 & \qquad+O\left(\epsilon^{3}\right)\:.
\end{align}
and 
\begin{eqnarray}
\mathcal{W}_{\mathrm{loc}} & \supset & \int dtd^{2}\vec{x}\sqrt{g}\Big\{ f\left[B\right]-\left(s+s^{\prime}\right)f^{\prime}\left[B\right]\mathcal{R}^{(2)}+mf^{\prime\prime}\left[B\right]E^{i}\partial_{i}\log B\nonumber \\
 &  & \qquad+\left(\frac{\mathrm{g}-2}{4}Bf^{\prime\prime}\left[B\right]+\frac{2\pi}{\nu}mBf^{\prime\prime}\left[B\right]^{2}\right)\left(\partial_{i}\log B\right)^{2}+O\left(\epsilon^{3}\right)\Big\}\:.
\end{eqnarray}

For simplicity, for now we assume the above terms are the only contributions
to the effective action, i.e. $\mathcal{W}_{\mathrm{loc}}$ contains
only (\ref{eq:LocalAction_f[B]}). It is not hard to compute additional
terms if there are some, so we temporarily ignore them here. Then
the total action takes the form 
\begin{align}
\mathcal{W} & =\int d^{3}x\Big\{\frac{\nu}{4\pi}\epsilon^{\rho\mu\nu}A_{\rho}\partial_{\mu}A_{\nu}-\frac{\nu}{2\pi}(s+s^{\prime})\epsilon^{\rho\mu\nu}A_{\rho}\partial_{\mu}\omega_{\nu}+\frac{\nu}{4\pi}(s+s^{\prime})^{2}\epsilon^{\rho\mu\nu}\omega_{\rho}\partial_{\mu}\omega_{\nu}\Big\}\nonumber \\
 & +\int d^{3}x\sqrt{g}\left\{ \alpha\left[B,E^{2}\right]+f_{0}\left[B,\mathcal{R}^{(2)}\right]+f_{1}\left[B\right]E^{i}\partial_{i}\log B+f_{2}\left[B\right]\left(\partial_{i}\log B\right)^{2}+O\left(\epsilon^{3}\right)\right\} \:,\label{eq:EffectiveAction_template}
\end{align}
where 
\begin{eqnarray}
\alpha & = & \frac{\nu mE^{2}}{4\pi B}\:,\\
f_{0} & = & f\left[B\right]+\frac{\nu(\mathrm{g}-2)}{8\pi m}B\left[B+\left(1-2s-s^{\prime}\right)\mathcal{R}^{(2)}\right]-\left(s+s^{\prime}\right)f^{\prime}\left[B\right]\mathcal{R}^{(2)}\:,\\
f_{1} & = & -\frac{\nu}{2\pi}\left(1-\frac{\mathrm{g}}{4}+\frac{s^{\prime}}{2}\right)+mf^{\prime\prime}\left[B\right]\:,\\
f_{2} & = & -\frac{\nu(\mathrm{g}-2)}{16\pi}\left(\frac{3}{2}-\frac{\mathrm{g}}{4}+s^{\prime}\right)\frac{B}{m}-\left(1-\frac{\mathrm{g}}{4}+\frac{s^{\prime}}{2}\right)Bf^{\prime\prime}\left[B\right]+\frac{\pi m}{\nu}Bf^{\prime\prime}\left[B\right]^{2}\:.
\end{eqnarray}
The three terms in the first line can be combined into a single ``complete
square'' term of $A-(s+s^{\prime})\omega$, the same structure appearing
in \cite{Wen:1992ej,Abanov:2014ula}.

\subsection{Correlation Functions of Electromagnetic Response}

Using formulae given in Appendix (A), it is straightforward to compute
the correlation functions. The 1-point functions of the conserved
current is 
\begin{eqnarray}
\langle J^{t}\rangle & = & \frac{\nu}{2\pi}\left[B-\left(s+s^{\prime}\right)\mathcal{R}^{(2)}-m\nabla_{i}\left(\frac{E^{i}}{B}\right)\right]-\nabla_{i}\left(f_{1}\nabla^{i}\log B\right)+O\left(\epsilon^{3}\right)\:,\label{eq:1PointFunction_Jt}\\
\langle J^{i}\rangle & = & \frac{\nu}{2\pi}\varepsilon^{ij}\left[E_{j}-\left(s+s^{\prime}\right)\left(\partial_{j}\omega_{t}-\partial_{t}\omega_{j}\right)\right]+\frac{1}{\sqrt{g}}\partial_{t}\left[\sqrt{g}\left(\frac{\nu mE^{i}}{2\pi B}+f_{1}\nabla^{i}\log B\right)\right]\nonumber \\
 &  & +\varepsilon^{ij}\nabla_{j}\left[-\frac{\nu mE^{2}}{4\pi B^{2}}+f_{0}^{\prime}-\frac{f_{1}}{B}\nabla_{k}E^{k}-f_{2}^{\prime}\left(\nabla_{k}\log B\right)^{2}-2\frac{f_{2}}{B}\nabla^{2}\log B\right]+O\left(\epsilon^{4}\right)\:.\nonumber \\
\label{eq:1PointFunction_Ji}
\end{eqnarray}
For flat background with constant $B$ and $E_{i}$ fields, the 2-point
functions have the following structures governed by Ward Identities
\begin{eqnarray}
\langle J^{t}(x)J^{t}(0)\rangle & = & i\Pi_{0}\vec{\partial}^{2}\delta(x)\:,\label{eq:2PointFunction_template_1}\\
\langle J^{t}(x)J^{i}(0)\rangle & = & -i\Pi_{0}\partial^{i}\partial_{t}\delta(x)+i\Pi_{1}\epsilon^{ij}\partial_{j}\delta(x)-i\Pi_{3}\epsilon^{ij}E^{k}\partial_{k}\partial_{j}\delta(x)\:,\\
\langle J^{i}(x)J^{t}(0)\rangle & = & -i\Pi_{0}\partial^{i}\partial_{t}\delta(x)-i\Pi_{1}\epsilon^{ij}\partial_{j}\delta(x)-i\Pi_{3}\epsilon^{ij}E^{k}\partial_{k}\partial_{j}\delta(x)\:,\\
\langle J^{i}(x)J^{j}(0)\rangle & = & i\Pi_{0}\delta^{ij}\partial_{t}^{2}\delta(x)+i\Pi_{1}\epsilon^{ij}\partial_{t}\delta(x)+i\Pi_{2}\left(\delta^{ij}\vec{\partial}^{2}-\partial^{i}\partial^{j}\right)\delta(x)\nonumber \\
 &  & +i\Pi_{3}\left(\epsilon^{jk}E^{i}+\epsilon^{ik}E^{j}\right)\partial_{k}\partial_{t}\delta(x)\:,\label{eq:2PointFunction_template_4}
\end{eqnarray}
with 
\begin{eqnarray}
\Pi_{0} & = & \frac{\nu m}{2\pi B}+O\left(\epsilon\right)\:,\\
\Pi_{1} & = & \frac{\nu}{2\pi}+\frac{\nu}{2\pi B}\left[\left(1-\frac{\mathrm{g}}{4}+\frac{s^{\prime}}{2}\right)-\frac{2\pi m}{\nu}f^{\prime\prime}\left[B\right]\right]\vec{\partial}^{2}+O\left(\epsilon^{3}\right)\:,\\
\Pi_{2} & = & f^{\prime\prime}\left[B\right]+\frac{\nu(\mathrm{g}-2)}{4\pi m}\nonumber \\
 &  & +\left\{ \frac{\nu(\mathrm{g}-2)}{4\pi mB}\left(\frac{3}{4}-\frac{\mathrm{g}}{8}+\frac{s^{\prime}}{2}\right)+2\frac{f^{\prime\prime}\left[B\right]}{B}\left[\left(1-\frac{\mathrm{g}}{4}+\frac{s^{\prime}}{2}\right)-\frac{\pi m}{\nu}f^{\prime\prime}\left[B\right]\right]\right\} \vec{\partial}^{2}\nonumber \\
 &  & +\frac{\nu mE^{2}}{2\pi B^{3}}+O\left(\epsilon^{3}\right)\:,\\
\Pi_{3} & = & -\frac{\nu m}{2\pi B^{2}}+O\left(\epsilon\right)\:.
\end{eqnarray}
Here $\vec{\partial}^{2}\equiv\delta^{ij}\partial_{i}\partial_{j}$.
These are the same as those derived in \cite{Son:2013rqa}.%
\footnote{To compare with \cite{Son:2013rqa}, we need $s^{\prime}\Rightarrow s$,
$E_{i}\Rightarrow0$, $f\left[B\right]\Rightarrow-\epsilon\left[\rho\right]$
with $\frac{\nu}{2\pi}B\Rightarrow\rho$, where on the right hand
side of each ``$\Rightarrow$'' are the notations of \cite{Son:2013rqa}.%
} Notice that $\Pi_{1}$ is actually the Hall conductivity $\sigma_{H}$
defined in the usual way. Here we largely ignore the response to curvature
by setting the background to be flat, but they can also be calculated
explicitly and compared with the literature, for example \cite{Can:2014ota,Gromov:2014gta}.

\subsection{Under the Field Constraint}

In parallel to the previous subsection, in this subsection we switch
to the field constraint (\ref{eq:Field_Constraint}). It can be solved
as 
\[
\bar{N}^{i}=\frac{\bar{\varepsilon}^{ij}\bar{V}_{tj}}{\frac{1}{2}\bar{\varepsilon}^{kl}\bar{V}_{kl}}\:.
\]
Applying the holographic dictionary, we can translate it into field
theory language as 
\begin{equation}
v^{i}=\frac{\varepsilon^{ij}E_{j}}{B}+\frac{(\mathrm{g}-2)e^{-\Phi}}{4m}\varepsilon^{ij}\partial_{j}\log\left(\frac{B}{m}e^{-\Phi}\right)+\frac{e^{-\Phi}}{4m}\varepsilon^{ij}\partial_{j}\mathrm{g}+O\left(\epsilon^{2}\right)\:.\label{eq:DriftVelocity_FC}
\end{equation}
Again plugging this into (\ref{eq:W_CS_with_vi}) and (\ref{eq:W_loc_with_vi})
we will get lengthy expressions for the effective action. When $\Phi=0$
and $m$, $\mathrm{g}$, $s$ and $s^{\prime}$ are all constants,
the results are 
\begin{align}
\mathcal{W}_{\mathrm{CS}} & =\int d^{3}x\Big\{\frac{\nu}{4\pi}\epsilon^{\rho\mu\nu}A_{\rho}\partial_{\mu}A_{\nu}-\frac{\nu}{2\pi}(s+s^{\prime})\epsilon^{\rho\mu\nu}\omega_{\rho}\partial_{\mu}A_{\nu}+\frac{\nu}{4\pi}(s+s^{\prime})^{2}\epsilon^{\rho\mu\nu}\omega_{\rho}\partial_{\mu}\omega_{\nu}\Big\}\nonumber \\
 & +\frac{\nu}{4\pi}\int d^{3}x\sqrt{g}\Big\{ m\frac{E^{2}}{B}-\left(2-\frac{\mathrm{g}}{2}+s^{\prime}\right)E^{i}\partial_{i}\log B+\frac{(\mathrm{g}-2)}{2m}B\left[B+\left(1-2s-s^{\prime}\right)\mathcal{R}^{(2)}\right]\nonumber \\
 & \qquad-\left[\frac{\mathrm{g}-2}{4}\left(\frac{3}{2}-\frac{\mathrm{g}}{4}+s^{\prime}\right)\frac{B}{m}\right]\left(\partial_{i}\log B\right)^{2}\Big\}+O\left(\epsilon^{3}\right)\:.
\end{align}
and 
\begin{eqnarray}
\mathcal{W}_{\mathrm{loc}} & \supset & \int d^{3}x\sqrt{g}\Big\{ f\left[B\right]-\left(s+s^{\prime}\right)f^{\prime}\left[B\right]\mathcal{R}^{(2)}+mf^{\prime\prime}\left[B\right]E^{i}\partial_{i}\log B\nonumber \\
 &  & \qquad+\frac{\mathrm{g}-2}{4}Bf^{\prime\prime}\left[B\right]\left(\partial_{i}\log B\right)^{2}+O\left(\epsilon^{3}\right)\Big\}\:.
\end{eqnarray}
Comparing with the corresponding results in the previous subsection
using path integral constraint (\ref{eq:PathIntegral_Condition}),
we see the only difference is that contributions from the local terms
of the action (those involving $f\left[B\right]$ and its derivatives)
are absent in the constraint equation, and consequently in the expressions
of $v^{i}$ and Chern-Simons action. The field constraint (\ref{eq:Field_Constraint})
is equivalent to the path integral constraint (\ref{eq:PathIntegral_Condition})
when only the Chern-Simons action is considered in the latter. This
is the reason that for the quantum Hall effect, these two constraints
are equivalent at the leading order when magnetic field is big and
Chern-Simons term dominates. If one considers other problems where
Chern-Simons term is not important, these two constraints will give
quite different results. Then one has to weigh which constraint is
more appropriate under other physical considerations.

Now same as in the previous subsection, let us assume these are the
only contributions to the effective action. The effective action still
takes the same form as in (\ref{eq:EffectiveAction_template}), so
do the functions $\alpha\left[B,E^{2}\right]$, $f_{0}\left[B,\mathcal{R}^{(2)}\right]$
and $f_{1}\left[B\right]$. The only difference is the form of $f_{2}\left[B\right]$,
which is now 
\begin{equation}
f_{2}=-\frac{\nu(\mathrm{g}-2)}{16\pi}\left(\frac{3}{2}-\frac{\mathrm{g}}{4}+s^{\prime}\right)\frac{B}{m}+\frac{\mathrm{g}-2}{4}Bf^{\prime\prime}\left[B\right]\:.
\end{equation}
For 1-point functions, $\langle J^{t}\rangle$ remains the same as
before, so do most of the 2-point functions in (\ref{eq:2PointFunction_template_1})-(\ref{eq:2PointFunction_template_4}).
The only differences are $O\left(\epsilon^{3}\right)$ order of $\langle J^{i}\rangle$
and $O\left(\epsilon^{2}\right)$ order of $\Pi_{2}$ in $\langle J^{i}(x)J^{j}(0)\rangle$,
which is now 
\begin{eqnarray}
\Pi_{2} & = & f^{\prime\prime}\left[B\right]+\frac{\nu(\mathrm{g}-2)}{4\pi m}+\left[\frac{\nu(\mathrm{g}-2)}{4\pi mB}\left(\frac{3}{4}-\frac{\mathrm{g}}{8}+\frac{s^{\prime}}{2}\right)-\left(\frac{\mathrm{g}-2}{2}\right)\frac{f^{\prime\prime}\left[B\right]}{B}\right]\vec{\partial}^{2}\nonumber \\
 &  & +\frac{\nu mE^{2}}{2\pi B^{3}}+O\left(\epsilon^{3}\right)\:.
\end{eqnarray}
The functional expression of $\langle J^{i}\rangle$ in terms of $f_{2}$
and others are the same as (\ref{eq:1PointFunction_Ji}).

To the orders of perturbative expansion we have computed, the two
different constraints, the path integral one (\ref{eq:PathIntegral_Condition})
and the field one (\ref{eq:Field_Constraint}) yield different results
only at $O\left(m^{0}\right)$ and higher order of $m$ for $f_{2}$,
thus for $\langle J^{i}\rangle$ and $\Pi_{2}$ at the corresponding
orders. However, these differences are non-universal parts of the
correlation functions as discussed in \cite{Son:2013rqa}. For the
universal parts, for example, the $O\left(m^{-1}\right)$ order terms,
the two constraints always produce the same results. Thus at the level
of our current discussion, we do not favor one constraint over the
other. We view both of them are viable. One of them may be a better
choice when a specific problem is studied using a more detailed model,
but we will leave this to the future.

\subsection{Holographic Gravitational Chern-Simons Term for $z>1$}

We can also add the gravitational Chern-Simons term \cite{Jackiw:2003pm}
to our holographic model: 
\begin{equation}
\hat{\mathcal{I}}_{\textrm{PY}}=\frac{c_{g}}{384\pi}\int d^{4}x\epsilon^{PQRS}R_{\phantom{MN}RS}^{MN}R_{NMPQ}\:,\label{eq:ChernSimons:BulkAction_grav}
\end{equation}
where $R_{MNPQ}$ is the Riemann tensor constructed from the full
4-dimensional spacetime metric and indices $M$, $N$, $P$, $Q$,
$R$ and $S$ run through $t$, $x$, $y$ and $r$. The subscript
``$_{\mathrm{PY}}$'' stands for ``Pontryagin''. Here for simplicity,
we also assume the coupling is a constant $c_{g}$, rather than a
dynamical axion scalar. Then it is a boundary term as well: 
\begin{equation}
\mathcal{\bar{I}}_{\textrm{PY}}=\frac{c_{g}}{96\pi}\int_{r=0}d^{3}x\sqrt{\bar{G}}\bar{\varepsilon}^{ij}\bar{\mathcal{P}}_{ij}\:,\label{eq:ChernSimons:NBAction_grav}
\end{equation}
where $\bar{\varepsilon}^{ij}\bar{\mathcal{P}}_{ij}$ is computed
explicitly in Appendix (C). Unlike the boundary form of the gauge
Chern-Simons term, which involves only $\partial_{t}$ and $\partial_{i}$
derivatives, $\bar{\mathcal{P}}_{ij}$ contains $\partial_{r}$ derivative
terms as well. However, all and only these $\partial_{r}$ terms will
be removed by adding a Gibbons-Hawking type boundary term \cite{Grumiller:2008ie}:
\begin{equation}
\mathcal{\bar{I}}_{\partial\textrm{PY}}=\frac{c_{g}}{48\pi}\int_{r=0}d^{3}x\sqrt{-\gamma}\left(n_{M}\varepsilon^{MNPQ}\mathcal{K}_{N}^{\phantom{N}R}\nabla_{P}\mathcal{K}_{QR}\right)\:,\label{eq:ChernSimons:GibbonsHawking}
\end{equation}
where $\gamma_{MN}$ is the induced full spacetime metric at the boundary,
$n_{M}$ the out-going unit normal vector of the boundary and $\mathcal{K}_{MN}=\gamma_{M}^{\phantom{M}P}\gamma_{N}^{\phantom{N}Q}\nabla_{P}n_{Q}$
the extrinsic curvature of the boundary. This term is added such that
the Dirichlet boundary value problem is well posed, in a similar fashion
as Gibbons-Hawking term is added to the Einstein-Hilbert action. Under
the gauge conditions (\ref{eq:BulkGaugeCondition_1})-(\ref{eq:BulkGaugeCondition_3}),
$\mathcal{K}_{MN}=-\frac{L}{r}\Gamma_{MN}^{r}$ contains only $\partial_{r}$
derivative terms, thus it will only remove $\partial_{r}$ terms from
the Chern-Simons action, and it is straightforward to check that it
does remove all such terms.

From now on for the discussion of gravitational Chern-Simons term
we assume 
\[
z>1\:.
\]
In Appendix (C) we have computed 
\begin{eqnarray*}
\bar{\varepsilon}^{ij}\bar{\mathcal{P}}_{ij} & = & -2\bar{\varepsilon}^{ij}\left[\omega_{t}\partial_{i}\omega_{j}+\omega_{i}\left(\partial_{j}\omega_{t}-\partial_{t}\omega_{j}\right)\right]+\bar{\varepsilon}^{ij}\left[\left(\bar{\varepsilon}^{kl}\partial_{k}\bar{N}_{l}\right)\partial_{i}\omega_{j}+\omega_{i}\partial_{j}\left(\bar{\varepsilon}^{kl}\partial_{k}\bar{N}_{l}\right)\right]\\
 &  & +2\bar{\varepsilon}^{ij}\left[\left(\nabla_{j}\nabla_{k}\bar{N}\right)\bar{K}_{i}^{k}-\left(\nabla_{k}\bar{N}\right)\nabla_{j}\bar{K}_{i}^{k}\right]+O\left(\partial_{r}\right)\:,
\end{eqnarray*}
where 
\[
\bar{K}_{ij}=\frac{1}{2\bar{N}}\left(\partial_{t}\bar{G}_{ij}-\nabla_{i}\bar{N}_{j}-\nabla_{j}\bar{N}_{i}\right)
\]
is the extrinsic curvature of the global time foliation at the boundary.
Then the gravitational Chern-Simons term, including the contribution
from the boundary term (\ref{eq:ChernSimons:GibbonsHawking}), is
\begin{eqnarray*}
\mathcal{\bar{I}}_{\textrm{PY}} & = & -\frac{c_{g}}{48\pi}\int d^{3}x\sqrt{\bar{G}}\bar{\varepsilon}^{ij}\left[\omega_{t}\partial_{i}\omega_{j}+\omega_{i}\left(\partial_{j}\omega_{t}-\partial_{t}\omega_{j}\right)\right]\\
 &  & +\frac{c_{g}}{24\pi}\int d^{3}x\sqrt{\bar{G}}\left\{ \bar{\varepsilon}^{ij}\bar{K}_{ik}\left(\nabla_{j}\nabla^{k}\bar{N}\right)+\frac{1}{2}\mathcal{R}^{(2)}\left(\bar{\varepsilon}^{ij}\partial_{i}\bar{N}_{j}\right)\right\} \:.
\end{eqnarray*}
Applying the holographic map, we have 
\begin{equation}
\mathcal{W}_{\textrm{PY}}=-\frac{c_{g}}{48\pi}\int d^{3}x\epsilon^{\rho\mu\nu}\omega_{\rho}\partial_{\mu}\omega_{\nu}+\frac{c_{g}}{24\pi}\int d^{3}x\sqrt{g}\left\{ \varepsilon^{ij}K_{ik}\left(\nabla_{j}\nabla^{k}e^{-\Phi}\right)-\frac{1}{2}\mathcal{R}^{(2)}\left(\varepsilon^{ij}\partial_{i}v_{j}\right)\right\} \:,\label{eq:W_PY}
\end{equation}
where 
\begin{equation}
K_{ij}=\frac{1}{2}e^{\Phi}\left(\partial_{t}g_{ij}+\nabla_{i}v_{j}+\nabla_{j}v_{i}\right)\:.
\end{equation}

This term (\ref{eq:W_PY}) shall now be added to the total effective
action (\ref{eq:BoundaryAction_total}). By our power counting scheme,
every term in (\ref{eq:W_PY}) is of order $O\left(\epsilon^{4}\right)$.
The gravitational Chern-Simons term only gives subleading order contributions
to $\langle J^{\mu}\rangle$ when the spacetime is not flat, and does
not alter the flat-spacetime current 2-point functions we compute
earlier, but they will give non-trivial contributions to energy-stress
correlation functions.

The gravitational Chern-Simons term describes the thermal Hall (Leduc-Righi)
effect \cite{Kane1997} and other parity-violating thermal transport
phenomena \cite{Gromov:2014vla,Geracie:2014oca,Geracie:2014TBA}.
The thermal Hall conductivity is proportional to the coefficient $c_{g}$,
which in turn is related to the gravitational anomaly and corresponds
to the central charge of the chiral CFT on the boundary \cite{Read:1999fn,Cappelli:2001mp,Ryu:2010ah,Wang:2010xh,Stone:2012ud}.
The first term of (\ref{eq:W_PY}) has appeared in the effective action
of \cite{Abanov:2014ula,Gromov:2014gta} and the whole action appeared
in a different form written in Lorentzian notations with Christoffel
symbol in \cite{Bradlyn:2014wla}, in addition to some of the aforementioned
references. In the rest of this paper, we will not go further in this
direction to pursue the thermal Hall effect.

\bigskip{}

\section{Gravitational Response and Geometric Properties of FQHE}

In this section, we will use the effective actions and correlation
functions derived in the last section to analyze geometric properties
and gravitational response of the fractional quantum Hall fluids they
describe, focusing on Hall viscosity, angular momentum density, momentum
density and their relations. The following discussions on viscosities
and angular momentum density are carried out only for constant $B$.
The results do not depend on $f_{2}$, thus the choice of the two
constraints (\ref{eq:PathIntegral_Condition}) or (\ref{eq:Field_Constraint})
does not matter in this section. We will work in flat space with $g_{ij}=\delta_{ij}$,
$\Phi=0$ and $m$, $\mathrm{g}$, $s$ and $s^{\prime}$ are all
constants, unless otherwise stated.

\subsection{Wen-Zee Shift and Hall Viscosity}

The first line of the effective action (\ref{eq:EffectiveAction_template})
includes the three well known topological terms which do not depend
on metric, and is in agreement with \cite{Wen:1992ej} and many subsequent
works. Of particular interest to us is the second term $\omega\wedge\mathrm{d}A$,
the so-called Wen-Zee term, whose coefficient we can identify as the
shift $\mathcal{S}$: 
\begin{equation}
\mathcal{S}=-2\left(s+s^{\prime}\right)\:.
\end{equation}
Thus the physical meaning of the constant $s^{\prime}$ arising from
our holographic dictionary (\ref{eq:HolographicMap_Vt}) and (\ref{eq:HolographicMap_Vi})
is clear: it is related to the shift $\mathcal{S}$ by the above equation,
while $s$ is the intrinsic spin of the microscopic field $\psi$
in the action (\ref{eq:FTAction_S0}). A similar relation and the
relation between the shift and the conformal field theory construction
\cite{MooreRead1991} of fractional quantum Hall effect can be found
in \cite{Read:2008rn,Read:2010epa}. The shift is the offset between
the total charge $Q=\int d^{2}\vec{x}\sqrt{g}\langle J^{t}\rangle$
and the total magnetic flux $N_{\phi}=\frac{1}{2\pi}\int d^{2}\vec{x}\sqrt{g}B$
if the manifold where the quantum Hall fluid lives on has a non-trivial
topology \cite{Haldane1983,Wen:1992ej,Son:2013rqa}:
\begin{equation}
Q=\nu\left(N_{\phi}+\frac{\mathcal{S}}{2}\chi\right)\:,
\end{equation}
where $\chi=\frac{1}{2\pi}\int d^{2}\vec{x}\sqrt{g}\mathcal{R}^{(2)}$
is the Euler characteristic of the spatial manifold.

Another important aspect of the shift is that even the manifold's
topology is trivial and the system lives in flat space, it is still
related to Hall viscosity \cite{Read:2008rn,Read:2010epa,Hoyos:2011ez,Son:2013rqa}.
Hall viscosity is one of the parity-odd dissipationless first order
transport coefficients appearing in (2+1)-dimensional parity violating
hydrodynamics \cite{Jensen:2011xb,Cai:2012mg,Haehl:2013kra,Kaminski:2013gca,Lucas:2014sia}.
Using $\omega_{t}=\frac{1}{2}\delta^{ij}\epsilon^{kl}\left(\delta g_{ik}\right)\partial_{t}\left(\delta g_{jl}\right)$
and $\mathcal{W}=2\int d^{3}x\eta_{H}\omega_{t}+\ldots$, the Hall
viscosity $\eta_{H}$ can be read off from the $\omega\wedge\mathrm{d}A$
term in the effective action: 
\begin{equation}
\eta_{H}=-\frac{\nu}{4\pi}\left(s+s^{\prime}\right)B=\frac{1}{4}\mathcal{S}\langle J^{t}\rangle\:.
\end{equation}
Here we have assumed flat space with constant $B$ field and vanishing
$E_{i}$ field.

\subsection{Magnetization and Momentum Density}

From (\ref{eq:1PointFunction_Ji}), we can identify the long expression
inside $\varepsilon^{ij}\nabla_{j}\left[\ldots\right]$ as the magnetization
defined in \cite{CHM1997}: 
\begin{equation}
\langle M\rangle=f_{0}^{\prime}-\frac{\nu mE^{2}}{4\pi B^{2}}-\frac{f_{1}}{B}\nabla_{k}E^{k}-f_{2}^{\prime}\left(\nabla_{k}\log B\right)^{2}-2\frac{f_{2}}{B}\nabla^{2}\log B\:.
\end{equation}
For constant $B$ and $E_{i}=0$, this reduces to 
\begin{equation}
\langle M\rangle=f_{0}^{\prime}=f^{\prime}\left[B\right]+\frac{\nu(\mathrm{g}-2)}{4\pi m}B\:.
\end{equation}
The first half of the above result $\langle M\rangle=f_{0}^{\prime}$
is in agreement with the recent result of \cite{Bradlyn:2014wla}.%
\footnote{Our $f_{0}$ plays the role of $f$ in \cite{Bradlyn:2014wla}, which
is minus of the unperturbed energy density.%
}

Using (\ref{eq:1PointFunction_Jt}), (\ref{eq:1PointFunction_Ji})
and (\ref{eq:Momentum-Current}) we can compute the momentum density
\begin{eqnarray}
\langle p^{i}\rangle & = & \frac{\nu m}{2\pi}\varepsilon^{ij}E_{j}+\partial_{t}\left(\frac{\nu m^{2}E^{i}}{2\pi B}+mf_{1}\partial^{i}\log B\right)+\varepsilon^{ij}\partial_{j}\Big[mf_{0}^{\prime}-\frac{\mathrm{g}-2s}{4}\frac{\nu}{2\pi}B\nonumber \\
 &  & -\frac{\nu m^{2}E^{2}}{4\pi B^{2}}-\frac{\mathrm{g}-2s}{4}\frac{\nu m}{2\pi B}E^{k}\partial_{k}\log B+\left(\frac{\mathrm{g}-2s}{4}\frac{\nu}{2\pi}-f_{1}\right)\frac{m}{B}\partial_{k}E^{k}\label{eq:TotalMomentumDensity}\\
 &  & +\left(\frac{\mathrm{g}-2s}{4}Bf_{1}^{\prime}-mf_{2}^{\prime}\right)\left(\partial_{k}\log B\right)^{2}+\left(\frac{\mathrm{g}-2s}{4}f_{1}-2\frac{mf_{2}}{B}\right)\partial_{k}^{2}\log B\Big]+O\left(\epsilon^{4}\right)\:.\nonumber 
\end{eqnarray}
The total angular momentum $\langle L_{\mathrm{tot}}\rangle$ and
its density $\ell_{\mathrm{tot}}(x)$ are defined as 
\begin{equation}
\langle L_{\mathrm{tot}}\rangle=\int d^{2}\vec{x}\epsilon_{ij}x^{i}\langle p^{j}(x)\rangle=\int d^{2}\vec{x}\ell_{\mathrm{tot}}(x)+\ldots\:.
\end{equation}
where the $\ldots$ denotes the contribution to total angular momentum
that is not translational invariant, thus can not be attribute to
angular momentum density. In the above expression of $\langle p^{i}\rangle$,
the quantity inside $\varepsilon^{ij}\nabla_{j}\left[\ldots\right]$
is half of the total angular momentum density $\ell_{\mathrm{tot}}$,
as can be seen by plugging the expression into the definition of angular
momentum and integrating by parts. For constant $B$ field and $E_{i}=0$,
the total angular momentum density is 
\begin{equation}
\ell_{\mathrm{tot}}=2\left(mf_{0}^{\prime}-\frac{\mathrm{g}-2s}{4}\frac{\nu}{2\pi}B\right)=-\frac{\nu}{\pi}\left(1-\frac{\mathrm{g}}{4}-\frac{s}{2}\right)B+2mf^{\prime}\left[B\right]\:.
\end{equation}
Its relation to magnetization is 
\begin{equation}
\ell_{\mathrm{tot}}=2m\langle M\rangle+\left(s-\frac{\mathrm{g}}{2}\right)\langle J^{t}\rangle\:.
\end{equation}

\subsection{Vorticity and Guiding Center Angular Momentum Density}

Next, let us look at the vorticity. From (\ref{eq:DriftVelocity_PIC})
and (\ref{eq:DriftVelocity_FC}) we can see that $v^{i}$ has a form
of the drift velocity ($\varepsilon^{ij}E_{j}/B$, plus corrections
for inhomogeneous $B$ field). Having the velocity, the vorticity
\emph{associated with this motion} is usually defined as curl of the
velocity: 
\begin{equation}
\Omega=\varepsilon^{ij}\partial_{i}v_{j}\:.
\end{equation}
Using (\ref{eq:DriftVelocity_PIC}) or (\ref{eq:DriftVelocity_FC})
this can be rewritten as $\Omega=-\nabla_{i}\left(E^{i}/B\right)+\ldots$.
Thus the vorticity is proportional to divergence of electric field
in a constant magnetic field. One possible source of this divergence
could be the impurity, i.e. $\nabla_{i}E^{i}$ equals to the density
of impurity $\rho_{\mathrm{imp}}$. Such identification has been made,
for example, in \cite{Stone1990}. Integrating by parts the $f_{1}\left[B\right]$
term in (\ref{eq:EffectiveAction_template}), this term can be written
as
\begin{equation}
\mathcal{W}=\int d^{3}x\sqrt{g}\left\{ \left(\int f_{1}\left[B\right]dB\right)\Omega+\ldots\right\} \:.
\end{equation}
This is the effective action $\mathcal{W}$'s dependence on $\Omega$.
We can see that in the effective action, terms involving divergence
of the the electric field $\nabla_{i}E^{i}$ are related to vorticity.
In holography, the Euclidean on-shell action $\bar{\mathcal{I}}_{\mathrm{Eucl}}$
with $t$ integrated from $0$ to $1/T$ is $-T$ multiplying the
grand potential $\Omega_{\mathrm{grand}}$ (not to confused with the
vorticity $\Omega$). Here $T$ is temperature and $\Omega_{\mathrm{grand}}=-T\log\mathcal{Z}_{\mathrm{grand}}$
where $\mathcal{Z}_{\mathrm{grand}}$ is the partition function of
the grand canonical ensemble. After going through this procedure,
we see that the integrand (Lagrangian density) of $\mathcal{W}$ is
the thermodynamical pressure $P=-\delta\Omega_{\mathrm{grand}}/\delta V$
where $V=\int d^{2}\vec{x}\sqrt{g}$ is the volume. The same identification
is used in \cite{Hartnoll:2007ai} to study the classical Hall effect.
\cite{Jensen:2011xb} has given the relation between the partition
function, angular momentum $L_{\mathrm{orb}}$ and vorticity: $\mathcal{Z}_{\mathrm{grand}}=\exp\left(L_{\mathrm{gc}}\Omega/2T+\ldots\right)$,
then we have 
\begin{equation}
P=\frac{1}{2}\ell_{\mathrm{gc}}\Omega+\ldots\:,\qquad\textrm{where}\qquad\mathcal{W}=\int d^{3}x\sqrt{g}P\:.
\end{equation}
A similar relation has been derived in \cite{Hoyos:2014lla}. Now
we have the angular momentum density conjugate to vorticity: 
\begin{equation}
\ell_{\mathrm{gc}}=2\int f_{1}\left[B\right]dB=-\frac{\nu}{2\pi}\left(2-\frac{\mathrm{g}}{2}+s^{\prime}\right)B+2mf^{\prime}\left[B\right]\:.\label{eq:GCAngMomDensity}
\end{equation}
Notice that we have added a subscript ``$_{\mathrm{gc}}$'' (stands
for ``\emph{guiding center}'', whose meaning will be explained later)
for $L_{\mathrm{gc}}$ and $\ell_{\mathrm{gc}}$ to distinguish them
from the total angular momentum $L_{\mathrm{tot}}$ and its density
$\ell_{\mathrm{tot}}$ discussed earlier. The explicit expressions
of $\ell_{\mathrm{tot}}$ and $\ell_{\mathrm{gc}}$ also show that
they are different. In \cite{Hoyos:2014lla} it has been shown that
$L_{\mathrm{tot}}$ defined from $\langle p^{i}\rangle$ is the corrected
total kinetic angular momentum, and the authors also notice that the
quantity $2\delta P/\delta\Omega$ as what we call $\ell_{\mathrm{gc}}$
here does not have a direct interpretation as the total angular momentum
density $\ell_{\mathrm{tot}}$. To keep our focus on the derivations,
we will postpone the explanation of the physical meaning of this so-called
guiding center angular momentum density in the context of the quantum
Hall effect to the end of the section. At this moment, from theoretical
mechanics\textquoteright{}s point of view, vorticity $\Omega$ can
be viewed as a temporal derivative of some generalized coordinate
(some angle) and $\ell_{\mathrm{gc}}$ is the generalized momentum
conjugate to it, which does not equal to the total angular momentum
density. In terms of magnetization and charge density, $\ell_{\mathrm{gc}}$
can be expressed as 
\begin{equation}
\ell_{\mathrm{gc}}=2m\langle M\rangle+\left(-s^{\prime}-\frac{\mathrm{g}}{2}\right)\langle J^{t}\rangle\:.
\end{equation}

It is worthy to note that the identification of $\Omega=\varepsilon^{ij}\partial_{i}v_{j}$
and $\ell_{\mathrm{gc}}=2\int f_{1}\left[B\right]dB$ does not depend
on the explicit expression (\ref{eq:DriftVelocity_PIC}) or (\ref{eq:DriftVelocity_FC}).
It can be directly made from (\ref{eq:W_CS_with_vi}) and (\ref{eq:W_loc_with_vi}),
thus depends only on the holographic dictionary (\ref{eq:HolographicMap_N})-(\ref{eq:HolographicMap_phi}).
Hence our results for guiding center angular momentum densities equally
apply to the case when $\bar{N}_{i}$ is constrained by the bulk dynamics
to be a functional of other boundary fields and $v^{i}$ takes some
other form other than the drift velocity given in (\ref{eq:DriftVelocity_PIC})
or (\ref{eq:DriftVelocity_FC}).

\subsection{Internal Angular Momentum Density}

Since the guiding center angular momentum density is different from
the total angular momentum density, we can define their difference
as the \emph{internal} angular momentum density: $\ell_{\mathrm{int}}=\ell_{\mathrm{tot}}-\ell_{\mathrm{gc}}$.
Thus we have 
\begin{equation}
\ell_{\mathrm{int}}=\frac{\nu}{2\pi}\left(s+s^{\prime}\right)B=-\frac{\mathcal{S}}{2}\langle J^{t}\rangle\:,\label{eq:IntAngMomDensity}
\end{equation}
where the subscript ``$_{\mathrm{int}}$'' stands for ``internal''.
This agrees with the results found in \cite{Read:2008rn,Read:2010epa},
where $-\mathcal{S}/2$ is interpreted as conformal spin or mean orbital
spin per particle. Here to clarify some potential confusion about
terminology, we note that the quantity $\ell_{\mathrm{int}}$ has
different names in the literature. In \cite{Read:2008rn} it is called
``conformal spin density'', in \cite{Wen:1992ej,Read:2010epa} ``mean
orbital spin density'', in \cite{Nicolis:2011ey} ``intrinsic angular
momentum density'' and recently in \cite{Haldane1403} related to
``Landau orbit spin''. We will call it ``internal angular momentum
density'' and avoid using the terms ``intrinsic'' and ``spin''
so as not to confuse with the intrinsic spin of particles in the usual
sense in high energy physics. In this paper, the quantity $s$ will
be called the intrinsic spin of the microscopic field $\psi$, which
can be either a fundamental particle like the electron or a composite
particle like composite boson or fermion. For example, if $\psi$
denotes an Dirac electron in vacuum, $s=1/2$. For Laughlin states
with filling factor $\nu=1/(2n+1)$ where $n\in\mathbb{N}$, $\mathcal{S}=2n+1$
\cite{Wen:1992ej}. Then the mean internal angular momentum per particle
$-\mathcal{S}/2=-n-1/2$. \cite{Cho:2014vfl} derives the same result
using both composite fermion and boson theories. Comparing our action
(\ref{eq:FTAction_S0}) with those in \cite{Cho:2014vfl}, $\psi$
can denote composite fermions with intrinsic spin%
\footnote{The quantity we call $s$ here is called topological spin in \cite{Cho:2014vfl}.
We also notice there are different sign conventions in the literature,
thus some of our expressions like those for spins have different signs
compared with other references. This is due to various normalization
choices for the charge, spin, Hall viscosity, alignment of the angular
momentum, or combinations of them. For example, in \cite{Cho:2014vfl}
the charge is set to $1$ while we set to $-1$, hence our conventions
for the intrinsic (topological) spin $s$ will also differ by a sign.
They have $s=n$ and $n+1/2$ for composite fermions and bosons, respectively.
Equivalently, the sign of Wen-Zee shift can be flipped.%
} $s=-n$ for the former theory and composite bosons with $s=-n-1/2$
for the latter. Hence the parameter $s^{\prime}$ arising from the
map (\ref{eq:HolographicMap_Vt}) and (\ref{eq:HolographicMap_Vi})
takes value $s^{\prime}=-1/2$ for composite fermions and $s^{\prime}=0$
for composite bosons. The former corresponds to orbital spin of composite
fermions at filling factor $\nu=1$. The intrinsic spin $s$ and the
orbital spin $s^{\prime}$ of the composite particles together give
the total internal angular momentum given by (\ref{eq:IntAngMomDensity}).

Now we can immediately recognize the simple relation between Hall
viscosity and the internal angular momentum density: 
\begin{equation}
\eta_{H}=-\frac{1}{2}\ell_{\mathrm{int}}\:.\label{eq:HallSpin_int}
\end{equation}
This is the well-known relation of \cite{Read:2010epa,Nicolis:2011ey,Bradlyn:2012ea}.
The key point of reproducing this relation is to correctly identify
the internal angular momentum density. Only the internal part satisfies
this relation, as pointed out in \cite{Read:2010epa,Nicolis:2011ey,Bradlyn:2012ea},
but not the total angular momentum density, unless the guiding center
part vanishes.%
\footnote{\cite{Nicolis:2011ey} ensures this by assuming total angular momentum
density is proportional to particle density ($\langle J^{t}\rangle$
here), which could be in general violated by $f^{\prime}\left[B\right]$
term in $\ell_{\mathrm{gc}}$.%
}\textsuperscript{,}%
\footnote{For a more detailed discussion on the shift and angular momentum of
paired states, see \cite{Read:2010epa} and \cite{StoneRoy:2003,Tsutsumi:2012us,Sauls:2011}.%
} The Chern-Simons model generally has non-vanishing guiding center
angular momentum density, as is shown here. So it is crucial to identify
and distinguish the internal part from the guiding center part. Total
angular momentum density generally does not satisfy this simple $1/2$
relation with Hall viscosity.

\subsection{Bulk and Total Hall Viscosities}

There is a further generalization of the above relation between Hall
viscosity and internal angular momentum density. In \cite{Jensen:2011xb,Cai:2012mg,Kaminski:2013gca,Hoyos:2014lla}
a parity-odd thermodynamic transport coefficient $\tilde{\chi}_{\Omega}$
is defined as $T^{ij}=-\tilde{\chi}_{\Omega}\Omega g^{ij}+\ldots$
and it has been shown that 
\begin{equation}
\tilde{\chi}_{\Omega}=\frac{\delta P}{\delta\Omega}\:.
\end{equation}
$\tilde{\chi}_{\Omega}$ is called ``Hall bulk viscosity'' in \cite{Hoyos:2014lla}
and ``curl viscosity'' in \cite{Cai:2012mg} (denoted by $\zeta_{A}$
there). Then we have 
\begin{equation}
\tilde{\chi}_{\Omega}=\frac{1}{2}\ell_{\mathrm{gc}}\:.\label{eq:HallSpin_gc}
\end{equation}
As the Hall viscosity is (minus) half of the internal angular momentum
density, this relation tells us that the Hall bulk viscosity is half
the guiding center angular momentum density for fractional quantum
Hall fluids. We can combine these two relations to find a relation
with the total angular momentum density: 
\begin{equation}
\tilde{\chi}_{\Omega}-\eta_{H}=\frac{1}{2}\ell_{\mathrm{tot}}\:.\label{eq:HallSpin_tot}
\end{equation}
We can view this as a statement that the ``total'' parity-odd viscosity
is half of the total angular momentum density in fractional quantum
Hall fluids. The minus sign is just an accident of the definitions.

\subsection{Relations to Hall Conductivity}

From (\ref{eq:HallConductivity_template}) where $\Pi_{1}$ is just
the Hall conductivity $\sigma_{H}$ defined in the usual way, $\ell_{\mathrm{gc}}$
($\tilde{\chi}_{\Omega}$) enters as $O\left(\vec{\partial}^{2}\right)$
order coefficient: 
\begin{equation}
\sigma_{H}=\frac{\nu}{2\pi}-\frac{1}{2B}\frac{\delta\ell_{\mathrm{gc}}}{\delta B}\vec{\partial}^{2}+O\left(\epsilon^{3}\right)=\frac{\nu}{2\pi}-\frac{1}{B}\frac{\delta\tilde{\chi}_{\Omega}}{\delta B}\vec{\partial}^{2}+O\left(\epsilon^{3}\right)\:.
\end{equation}
The same numeric result is derived in \cite{Son:2013rqa}, without
explicit identification of $\ell_{\mathrm{gc}}$. We can further compute
the energy density $\langle\mathcal{E}^{0}\rangle$, internal pressure
$P_{\mathrm{int}}=\frac{1}{2}g_{ij}\langle T^{ij}\rangle$ and inverse
internal compressibility $\kappa_{\mathrm{int}}^{-1}=-\delta P_{\mathrm{int}}/\delta\log V$,
as defined in \cite{Bradlyn:2012ea} (where $\langle\mathcal{E}^{0}\rangle$
is denoted by $\varepsilon$). They are computed directly from (\ref{eq:ConservedQuantities}).
For flat space with constant $B$ field and vanishing $E_{i}$ field,
the results are
\begin{eqnarray}
\langle\mathcal{E}^{0}\rangle & = & -f_{0}\left[B\right]=-f\left[B\right]-\frac{\nu(\mathrm{g}-2)}{8\pi m}B^{2}\:,\\
P_{\mathrm{int}} & = & f_{0}\left[B\right]-Bf_{0}^{\prime}\left[B\right]=f\left[B\right]-Bf^{\prime}\left[B\right]-\frac{\nu(\mathrm{g}-2)}{8\pi m}B^{2}\:,\\
\kappa_{\mathrm{int}}^{-1} & = & -B^{2}f_{0}^{\prime\prime}\left[B\right]=-B^{2}f^{\prime\prime}\left[B\right]-\frac{\nu(\mathrm{g}-2)}{4\pi m}B^{2}\:.
\end{eqnarray}
The central column agrees with the relations given in \cite{Bradlyn:2012ea}.
For the scale-invariant cases such as that in (\ref{eq:f[B]_4Fermi}),
$Bf^{\prime}\left[B\right]=2f\left[B\right]$, the conformal Ward
identity (\ref{eq:ConformalWardID}) is indeed satisfied. Using the
right column, we get the following relation:
\begin{equation}
\frac{1}{2}\frac{\delta\ell_{\mathrm{gc}}}{\delta\log B}=\eta_{H}-\frac{\mathrm{g}-2s}{4}\langle J^{t}\rangle-\frac{m}{B}\kappa_{\mathrm{int}}^{-1}\:.
\end{equation}
Then
\begin{equation}
\sigma_{H}=\frac{\nu}{2\pi}-\frac{1}{B^{2}}\left(\eta_{H}-\frac{\mathrm{g}-2s}{4}\langle J^{t}\rangle-\frac{m}{B}\kappa_{\mathrm{int}}^{-1}\right)\vec{\partial}^{2}+O\left(\epsilon^{3}\right)\:.
\end{equation}
This is exactly the relation derived in \cite{Bradlyn:2012ea} when
$\mathrm{g}=s=0$, because there these parameters are not considered.

\subsection{Interpretation of Two Types of Angular Momentum Density}

Now we present the overdue explanation of why there are two types
of angular momentum density that we have identified from the total
angular momentum density. For simplicity, in this subsection we will
ignore the gyromagnetic factor $\mathrm{g}$ and intrinsic spin $s$,
unless otherwise stated. Their existence do not change the physical
interpretation the angular momentum densities, but add additional
parts to their expressions, as can be see from the above results.
Since the guiding center angular momentum density $\ell_{\mathrm{gc}}$
and vorticity $\Omega$ are closely related to the drift velocity
$v^{i}$, which is a classical concept for particle moving in electromagnetic
field, it is intuitive to start with classical mechanics, and then
proceed to its quantum version.

\subsubsection{Classical Charged Particles in Electromagnetic Field}

Let us consider the classical motion of a charged particle moving
in mutually parallel magnetic and electric fields. The Newtonian equation
of motion is 
\[
m\frac{d\vec{v}}{dt}=q\left(\vec{E}+\vec{v}\times\vec{B}\right)\:.
\]
By the following substitution: 
\[
\vec{v}=\vec{v}_{\mathrm{drift}}+\vec{v}_{\mathrm{cycl}}\:,\qquad\textrm{where}\qquad\vec{v}_{\mathrm{drift}}=\frac{\vec{E}\times\vec{B}}{B^{2}}\:,
\]
the total kinetic motion described by velocity $\vec{v}$ is decomposed
into two parts: a cyclotron motion $\vec{v}_{\mathrm{cycl}}$ (with
frequency $\omega=qB/m$) around a point center and a linear motion
of the cyclotron center described by the drift velocity $\vec{v}_{\mathrm{drift}}$.
Similarly, the position vector of the particle can also be decomposed
into these two parts
\[
\vec{x}=\vec{x}_{\mathrm{drift}}+\vec{x}_{\mathrm{cycl}}\:.
\]
To be in agreement with the quantum Hall literature, we will call
the cyclotron center the ``guiding center''. Thus $\vec{x}_{\mathrm{drift}}$
and $\vec{v}_{\mathrm{drift}}$ are the position and velocity of the
guiding center and $\vec{x}_{\mathrm{cycl}}$ and $\vec{v}_{\mathrm{cycl}}$
describe the relative motion of the charged particle as seen in the
guiding center frame. Furthermore, if we view the particle in cyclotron
motion plus the magnetic flux enclosed by the circular orbit as a
``composite particle'' (in analog to the composite fermion theory
\cite{Jain1989} of fractional quantum Hall effect, but of course
here is no quantization), then $\vec{x}_{\mathrm{drift}}$ and $\vec{v}_{\mathrm{drift}}$
describe the trajectorial motion of the center of mass of the composite
particle. Meanwhile, $\vec{x}_{\mathrm{cycl}}$ and $\vec{v}_{\mathrm{cycl}}$
are no longer directly visible to an external observer, but they will
manifest themselves as some internal properties of the composite particle.
For example, the orbital angular momentum of the cyclotron motion
in the guiding center frame, i.e. $m\vec{v}_{\mathrm{cycl}}\times\vec{x}_{\mathrm{cycl}}$,
will now looks like the ``intrinsic'' spin of the composite particle
(in addition to the actual intrinsic spin of the charged particle).
This is the origin of our terminology ``internal'', which in classical
sense means the ``intrinsic to the composite particle'', or ``viewed
in the guiding center frame''. Thus the total angular momentum of
the charged particle is the sum of the angular momentum of the guiding
center and the internal angular momentum of the composite particle.
Of course, here we have only been talking about a single particle.
The generalization to fluid case is straightforward, and this explains
the split between $\ell_{\mathrm{gc}}$ and $\ell_{\mathrm{int}}$
in $\ell_{\mathrm{tot}}$ in classical sense. Particularly, if the
velocity field of the composite particle fluid $\vec{v}_{\mathrm{drift}}(\vec{x})$
has a non-vanishing curl $\vec{\Omega}=\nabla\times\vec{v}_{\mathrm{drift}}\neq0$
at some point, then this means the guiding centers are circulating
around this point and the fluid will have some non-zero angular momentum
density at this point. This is the guiding center angular momentum
density $\ell_{\mathrm{gc}}$.

\subsubsection{Quantum Mechanics: Two Hilbert Spaces}

The fractional quantum Hall effect is a complicated quantum many body
problem with interactions between charged particles playing a huge
role. The quantum picture is quite different from the classical one
we have just discussed. However, there are structural similarities,
and one of them is the split of the total Hilbert space into the guiding
center part and cyclotron part, which is in direct analog to the classical
case. A discussion on angular momentum and Hall viscosity from this
perspective can be found first in \cite{Read:2010epa} and later more
systematically in \cite{Haldane:2009ke,Haldane:2011ia,Haldane:1112,Haldane1403}.
Other related works and references can be found in the thesis review
\cite{Yang:2013xha}.

To discuss the structure of Hilbert space, let us first focus on 2-dimensional
single-particle quantum mechanics in the presence of perpendicular
magnetic field. The coordinates $x^{i}$ and momentum $p_{i}=-i\partial_{i}$
form the 4-dimensional phase space. The covariant momentum is $\pi_{i}=p_{i}-qA_{i}$
with $\left[\pi_{i},\pi_{j}\right]=iqB\epsilon_{ij}$. Similar to
the example in classical mechanics, $x^{i}$ can be split into the
two parts
\begin{equation}
x^{i}=R^{i}+\tilde{R}^{i}\:,\qquad\textrm{where}\qquad\tilde{R}^{i}=-\frac{\epsilon^{ij}\pi_{j}}{qB}\:,
\end{equation}
where $R^{i}$ are the guiding center coordinates and $\tilde{R}^{i}$
the Landau orbit coordinates, the quantum version of our $\vec{x}_{\mathrm{drift}}$
and $\vec{x}_{\mathrm{cycl}}$. They satisfy commutation relations
\begin{eqnarray}
\left[R^{i},R^{j}\right] & = & -\frac{i}{qB}\epsilon^{ij}\:,\\
\left[\tilde{R}^{i},\tilde{R}^{j}\right] & = & \frac{i}{qB}\epsilon^{ij}\:,\\
\left[R^{i},\tilde{R}^{j}\right] & = & 0\:.
\end{eqnarray}
The 4-dimensional Hilbert space is now mapped into a tensor product
of two copies of 2-dimensional Hilbert spaces whose phase space coordinates
are $R^{i}$ and $\tilde{R}^{i}$ respectively. A notable feature
of the two sub-Hilbert spaces is that within each one the two spatial
coordinates are non-commutative \cite{Girvin1984,Dunne:1991cs,Susskind:2001fb,Fradkin:2002qw}.
As $\{R^{i},R^{j}\}$ and $\{\tilde{R}^{i},\tilde{R}^{j}\}$ form
two sets of $SL(2,\mathbb{R})$ algebra, the guiding center angular
momentum $L_{\mathrm{gc}}$ and Landau orbit angular momentum $\tilde{L}$
can be defined as%
\footnote{Our definition of $L_{\mathrm{gc}}$ differs by a sign from that in
\cite{Haldane1403} such that its positive direction aligns in the
same direction as $\tilde{L}$'s. In \cite{Haldane1403} they align
in opposite directions because their two sets of $SL(2,\mathbb{R})$
algebra, eq. (6a) and eq. (6b), differ by a sign.%
} 
\begin{equation}
L_{\mathrm{gc}}=\frac{1}{2}qBg_{ij}^{\mathrm{gc}}R^{i}R^{j}\:,\qquad\tilde{L}=-\frac{1}{2}qB\tilde{g}_{ij}\tilde{R}^{i}\tilde{R}^{j}\:,
\end{equation}
where according to \cite{Haldane1403} $g_{ij}^{\mathrm{gc}}$ and
$\tilde{g}_{ij}$ are the guiding center and Landau orbit metrics.
For us we always assume Galilean and rotational invariance, thus $g_{ij}^{\mathrm{gc}}=\tilde{g}_{ij}=g_{ij}$.
It is straightforward to check that the total angular momentum defined
in the usual way is the sum of the above two angular momenta: 
\begin{equation}
L_{\mathrm{tot}}=\epsilon^{ij}x_{i}p_{j}=L_{\mathrm{gc}}+\tilde{L}\:.
\end{equation}
$\tilde{L}$ is the angular momentum associated with the cyclotron
motion and $L_{\mathrm{gc}}$ associated with the drift motion of
the guiding center. In term of the more familiar language of creation
and annihilation operators in symmetric gauge \cite{Girvin1984,Dunne:1991cs},
we have 
\begin{eqnarray*}
L_{\mathrm{tot}} & = & b^{\dagger}b-a^{\dagger}a\:,\\
L_{\mathrm{gc}} & = & b^{\dagger}b+\frac{1}{2}\:,\\
\tilde{L} & = & -\left(a^{\dagger}a+\frac{1}{2}\right),
\end{eqnarray*}
Here our definitions of $a$, $a^{\dagger}$, $b$ and $b^{\dagger}$
follow those in \cite{Dunne:1991cs}: $a$ and $a^{\dagger}$ shift
the Landau level quantum number by one and $b$ and $b^{\dagger}$
shift the angular quantum number by one within each Landau level.
It is very clear from these expressions that $\tilde{L}$ is associated
with the cyclotron motion and $L_{\mathrm{gc}}$ is the rest part
of the motion. In fact, $\tilde{L}$ is proportional to the free Hamiltonian
$H_{0}=\frac{qB}{m}\left(a^{\dagger}a+\frac{1}{2}\right)$.

So far we have only talked about single-particle quantum mechanics,
for simplicity of the definitions. In the discussion of \cite{Haldane:2009ke,Haldane:2011ia,Haldane:1112,Haldane1403}
they are all generalized to many-particle case with interactions.
The eigenvalue of (many-particle) Landau orbit angular momentum $\tilde{L}$
is extensive, while that of guiding center $L_{\mathrm{gc}}$ (thus
$L_{\mathrm{tot}}$ as well) has both super-extensive and extensive
parts.%
\footnote{``Extensive'' means proportional to the total electron number $N$,
and ``super-extensive'' quadratic in $N$.%
} The notion of angular momentum density can be defined from the extensive
part of the angular momentum. Then the coefficients of extensive part
of $L_{\mathrm{gc}}$, $\tilde{L}$ and $L_{\mathrm{tot}}$ define
the guiding center, Landau orbit and total angular momentum density
$\ell_{\mathrm{gc}}$, $\tilde{\ell}$ and $\ell_{\mathrm{tot}}$,
respectively. Divided by $\langle J^{t}\rangle$, $\ell_{\mathrm{gc}}$
and $\tilde{\ell}$ are minus the ``guiding center spin'' (denoted
by $s$ in \cite{Haldane1403}, not to be confused with our $s$)
and ``Landau orbit spin'' $\tilde{s}_{n}$ defined in \cite{Haldane1403}.
$\ell_{\mathrm{gc}}$ is exactly the quantity we defined and computed
earlier, and $\tilde{\ell}$ only differs from our $\ell_{\mathrm{int}}$
by the intrinsic spin $s$, which we have ignored here. This is the
quantum interpretation of the meaning of what we called guiding center
angular momentum density $\ell_{\mathrm{gc}}$ and internal angular
momentum density $\ell_{\mathrm{int}}$.

In additional to deriving the analog of the relation (\ref{eq:HallSpin_int})
for $\tilde{\ell}$ , \cite{Haldane1403} defines a guiding center
Hall viscosity and shows that it is proportional to the guiding center
spin. This is the analog of our (\ref{eq:HallSpin_gc}). By comparison,
the guiding center Hall viscosity is essentially the bulk Hall viscosity
$\tilde{\chi}_{\Omega}$. \cite{Haldane1403} also has an equation
that the sum of the two Hall viscosities equals to the sum%
\footnote{``Minus'' in \cite{Haldane1403} again because their definitions
of the two angular momenta differ by a sign.%
} of the guiding center and Landau orbit spins. This is the analog
of our relation (\ref{eq:HallSpin_tot}). Our results can be viewed
as a derivation of these results from Chern-Simons effective theory.

\subsection{Guiding Center Momentum Density}

Since $v^{i}$ has been identified with drift velocity, i.e. the guiding
center velocity field, its natural to identify the ``charge density''
conjugate to it as the guiding center momentum density $p_{\mathrm{gc}}^{i}$,
similar to the relation between $\ell_{\mathrm{gc}}$ and $\Omega$:
\begin{equation}
\langle p_{\mathrm{gc}}^{i}\rangle=\frac{e^{\Phi}}{\sqrt{g}}\frac{\delta\mathcal{I}}{\delta v_{i}}\Bigg|_{A_{\mu},g_{ij},\Phi}\:.\label{eq:GCMomentum_def}
\end{equation}
Notice that this equation shall be computed before the path-integral
constraint or field constraint is implemented, otherwise there would
be no $v_{i}$ at all and the quasi-effective action $\mathcal{I}\left[v^{i},A_{\mu},g_{ij},\Phi,\ldots\right]$
would become the effective action $\mathcal{W}\left[A_{\mu},g_{ij},\Phi,\ldots\right]$.
In other words, $\langle p_{\mathrm{gc}}^{i}\rangle$ is computed
from the quasi-effective potential $\mathcal{I}$, not the actual
effective potential $\mathcal{W}$. In this sense, we can not say
that $v_{i}$ is a source field that sources certain momentum density,
because source fields are the independent functional variables appearing
in $\mathcal{W}$. This is a consequence that there is no $v_{i}$
in the microscopic action (\ref{eq:FTAction_S0}). The above equation
can be viewed as a definition for $\langle p_{\mathrm{gc}}^{i}\rangle,$
and there is no conservation law associated with it. This is different
from the total momentum density (\ref{eq:Momentum-Current}), which
enters the conservation of stress-tensor \cite{Geracie:2014nka}.
On contrary, quantity like $v_{i}$ does enter effective action (and
even the microscopic action) as a source field to momentum density
$p^{i}$ in some formalisms in the literature, for example in \cite{Geracie:2014nka,Gromov:2014vla,Brauner:2014jaa}.
This is case, the definition of $A_{i}$ is also shifted (usually
by $mv_{i}$), then the current $J^{i}$ obtained through functional
derivative with respect to $A_{i}$ has not exactly the same meaning
as ours defined in (\ref{eq:ConservedQuantities}). It is related
to our current by some redefinition, as illustrated in \cite{Geracie:2014nka}.
Thus a relation like $p^{i}=mJ^{i}$ usually given in this case (where
$p^{i}$ is obtained from functional derivative with respect to $v_{i}$)
is in no contradiction with any relation presented in this paper. 

Using (\ref{eq:W_CS_with_vi}) and (\ref{eq:W_loc_with_vi}), we have
\begin{eqnarray}
\langle p_{\mathrm{gc}}^{i}\rangle & = & \frac{\nu}{2\pi}\Big\{ me^{2\Phi}\left(\varepsilon^{ij}E_{j}-Bv^{i}\right)-\varepsilon^{ij}\partial_{j}\left[\left(1-\frac{\mathrm{g}}{4}+\frac{s^{\prime}}{2}\right)B\right]\nonumber \\
 &  & +\left(1-\frac{\mathrm{g}}{4}+\frac{s^{\prime}}{2}\right)e^{\Phi}B\varepsilon^{ij}\partial_{j}\log\left(me^{\Phi}\right)\Big\}+me^{2\Phi}\varepsilon^{ij}\partial_{j}\left(e^{-\Phi}f^{\prime}\left[B\right]\right)\\
 &  & +O\left(\epsilon^{2}\right)\:.\nonumber 
\end{eqnarray}
For $\Phi=0$ and $m$, $\mathrm{g}$, $s$ and $s^{\prime}$ constant,
it becomes 
\[
\langle p_{\mathrm{gc}}^{i}\rangle=\varepsilon^{ij}\partial_{j}\left\{ -\frac{\nu}{2\pi}\left(1-\frac{\mathrm{g}}{4}+\frac{s^{\prime}}{2}\right)B+mf^{\prime}\left[B\right]\right\} +O\left(\epsilon^{2}\right)\:.
\]
It has no ``transport'' part%
\footnote{According to the terminology of \cite{CHM1997}, this is the part
that can not be written as a curl.%
} and the only non-vanishing part is a curl which accounts exactly
for the guiding center angular momentum density (\ref{eq:GCAngMomDensity}).
Subtracting it from the total momentum density (\ref{eq:TotalMomentumDensity}),
we obtain the rest, the internal momentum density: 
\begin{equation}
\langle p_{\mathrm{int}}^{i}\rangle=\frac{\nu m}{2\pi}\varepsilon^{ij}E_{j}+\varepsilon^{ij}\partial_{j}\Big[\frac{\nu}{4\pi}\left(s+s^{\prime}\right)B\Big]+O\left(\epsilon^{2}\right)\:.
\end{equation}
The second term is the curl part which accounts for all the internal
angular momentum density (\ref{eq:IntAngMomDensity}), as expected.
The first term is the transport part, which is a product of number
density $\langle J^{t}\rangle$, mass $m$ and the drift velocity
$\varepsilon^{ij}E_{j}/B$. This form is exactly what one would expect
from classical mechanics. Notably, this accounts for \emph{all} the
transport part of the total momentum density in the homogeneous case. 

\bigskip{}

\section{Summary and Comments}

In this paper we have shown that Ho\v{r}ava-Lifshitz gravity theory
can be used as a covariant formalism for the effective field theory
of the fractional quantum Hall effect in 2+1 dimensions. It guarantees
the non-relativistic spacetime symmetries possessed by the quantum
Hall system. Lying in the heart of this formalism is the map between
the field degrees of freedom of the two theories, (\ref{eq:HolographicMap_N})-(\ref{eq:HolographicMap_phi}).
We originally derive this map as a holographic dictionary, but its
existence and form are independent of the holographic duality. It
can be directly applied on (2+1)-dimensional Ho\v{r}ava-Lifshitz gravity
theory to constructed the low energy effective action for the fractional
quantum Hall effect in a phenomenological way. It also serves as the
holographic dictionary for (3+1)-dimensional Ho\v{r}ava-Lifshitz gravity
theory with asymptotic Lifshitz background (\ref{eq:AdS-Lifshitz_metric}).
The latter paves the way for a systematic study on the fractional
quantum Hall effect using the approach of gauge/gravity duality (AdS/CMT),
that has been well-developed over the last decade. Of particular interest
is that this formalism naturally produces the Wen-Zee term at the
boundary, which is otherwise hard to obtain using the conventional
relativistic (3+1)-dimensional holography.

We have shown that the universal electromagnetic and geometric properties
of the fractional quantum Hall effect that are previously studied
over decades in various contexts using different methods can be encoded
in a simple Chern-Simons model using the Ho\v{r}ava-Lifshitz gravity
formalism. The effective action we obtain in this paper is by no means
a complete one. Additional terms can be added phenomenologically based
on symmetry considerations, but the connection of Ho\v{r}ava-Lifshitz
formalism to holography allows them to be computed systematically
using gauge/gravity duality. The shift function in Ho\v{r}ava-Lifshitz
gravity is identifies with (minus) the guiding center velocity field
(drift velocity) of the quantum Hall fluid, whose conjugate momentum
density is also that of the guiding center. Through this identification,
we can further identify the guiding center angular momentum density
conjugate to the guiding center vorticity is half of the Hall bulk
viscosity defined in the context of parity-violating first order hydrodynamics
\cite{Jensen:2011xb,Kaminski:2013gca,Hoyos:2014lla}. The rest of
the total angular momentum density is the internal angular momentum
density of the composite particles, which includes the Landau orbit
spin and the intrinsic spin of the composite particles. It is proportional
to the Wen-Zee shift and is minus twice of the Hall viscosity. These
relations are well-known in the literature and are derived using various
methods \cite{Read:2010epa,Nicolis:2011ey,Bradlyn:2012ea}. Here we
show that they can be produced via non-relativistic diffeomorphism
invariant Chern-Simons effective field theory.

We close our paper with some concluding remarks and comments:
\begin{enumerate}
\item \emph{Various versions of Ho\v{r}ava-Lifshitz gravity}: There exist
different versions of Ho\v{r}ava-Lifshitz gravity theory in the literature.
What we employ here is a minimal version of the non-projectable Ho\v{r}ava-Lifshitz
gravity, where the graviton sector includes only the lapse function
$N$, shift function $N_{i}$ (or $N_{i}$) and spatial metric $g_{ij}$
(or $G_{IJ}$), all being functions of both spatial coordinates and
time, and the spacetime symmetries contain primarily only the foliation
preserving diffeomorphism (FPD) and Weyl scaling. There are extended
versions of the theory, notably that in \cite{Horava:2010zj,daSilva:2010bm}.
There an additional $U(1)_{\Sigma}$ gauge symmetry associated with
the co-dimension one foliation of constant time $\Sigma(t)$ is introduced.
This is not to be confused with the electromagnetic $U(1)$ symmetry
represented by the vector field $V_{\mu}$ or $V_{M}$ in this paper.
But sometimes they are identified in the literature, for example in
\cite{Son:2005rv,Andreev:2013qsa}. Consequently additional fields
are needed to construct a manifest $U(1)_{\Sigma}$ gauge-invariant
theory. This kind of extended version of Ho\v{r}ava-Lifshitz gravity
can also be used as a bulk theory for holography \cite{Janiszewski:2012nb}
as well as covariant formalism for quantum Hall effects. The covariant
map can be built in a similar way as we have done in this paper to
incorporate the additional fields. But what are the field theory dual
of this additional symmetry and fields is a question that usually
has no obvious answer, at least in the context of quantum Hall effects.
\cite{Janiszewski:2012nb} gives two possible holographic scenarios
for the extended version with $U(1)_{\Sigma}$ symmetry. We choose
to use the minimal version without $U(1)_{\Sigma}$ symmetry symmetry
and additional fields. A fundamental reason behind these different
choices is how the shift function is interpreted in the field theory,
because the $U(1)_{\Sigma}$ is an Abelian gauge symmetry of the shift
function. If the shift function is treated like a gauge field in the
dual field theory, such as in \cite{Son:2005rv,Andreev:2013qsa},
it is natural to include this symmetry and employ the $U(1)_{\Sigma}$
extended version of Ho\v{r}ava-Lifshitz gravity as bulk theory, following
\cite{Janiszewski:2012nb}'s prescription regarding the role of this
symmetry and additional fields in the holographic correspondence.
However, if the shift function is interpreted as a velocity field,
as in \cite{Son:2013rqa} and this paper, it is more of the nature
of a current%
\footnote{We recall in non-relativistic theories, current $\vec{j}$ is proportional
to velocity $\vec{v}$, with coefficient being the density $\rho$:
$\vec{j}=\rho\vec{v}$. %
} which does not transform under a $U(1)$ symmetry. In this case,
the $U(1)_{\Sigma}$ extended version will not help much in building
the effective theory based on symmetries. Thus we choose the minimal
version.
\item \emph{Behavior of the shift function}: Different versions of Ho\v{r}ava-Lifshitz
gravity may have different dynamics and numbers of degrees of freedom,
depending on the numbers of fields and numbers of first-class and
second-class constraints (for example, see \cite{Bellorin:2010je,Kluson:2010nf,Bellorin:2011ff,Bellorin:2012di}).
Even in the low-energy minimal version we use here, there are still
free tunable parameters $\alpha$, $\beta$ and $\hat{\lambda}$ in
the action (\ref{eq:HoraveAction_Graviton}), which can give rise
to different dynamics and solutions. Thus the shift function (or the
boundary field of the shift function in the case of holography) can
have different behaviors depending on what the model action one start
with. This feature is particularly relevant for holography. This does
not affect the the holographic dictionary since the symmetries are
the same, but gives different physical results. In some versions of
Ho\v{r}ava-Lifshitz gravity, the shift function (to be precise, its
boundary field in holography) may totally drop off after the holographic
dictionary is applied. This is suggested as one scenario in \cite{Janiszewski:2012nb}.
In some other versions, it may be constrained to be just a functional
of other fields. In either case, the question of how to deal with
it is trivial. We do not focus on these cases in this paper, but are
aware of these possibilities. What we are mostly interested in is
the case that the boundary shift function is independent of the other
boundary fields. This is similar to that of the relativistic holography.
From the computation we have done, it also seems to be the most relevant
case for application to quantum Hall effects. It gives rise to the
non-trivial questions of how to interpret and determine the shift
function. This leads to the next comment. We hope to give some case-by-case
study in the future to show some examples of what model actions can
give what possible cases among these three possibilities for the shift
function.
\item \emph{The two constraints}: We identify the shift function arising
in Ho\v{r}ava-Lifshitz formalism as minus the guiding center velocity
field, but how to determine it is a different question. Similar problem
exists in the Newton-Cartan formalism as well. There is no clear answer
to this question in recent studies of the quantum Hall effective action,
among which many do not provide an answer at all. In this paper, we
provide two slightly different answers, the path-integral constraint
(\ref{eq:PathIntegral_Condition}) and the field constraint (\ref{eq:Field_Constraint}),
both satisfying the requirements from symmetries and LLL projection.
They are both meaningful and applicable in (2+1)-dimensional effective
field theory and (3+1)-dimensional holography. The former is more
in the spirit of effective theory, similar to what is done in \cite{Son:2013rqa},
while the latter is more commonly seen in holography as a Dirichlet
boundary condition imposed at the boundary. For the fractional quantum
Hall effect where at low energy scale the Chern-Simons term dominates,
the two constraints are equivalent in producing the universal features,
but can give different results for non-universal features related
to local dynamics. The question that which one is more favorable has
to be answered by computing more non-universal features and comparing
with other approaches and the experimental results. It is beyond the
reach of this paper. It is worthy to note that in the absence of gauge
Chern-Simons term and strong background magnetic field, i.e. out of
the context of quantum Hall effects (e.g. in the context of \cite{Janiszewski:2012nb}),
these two constraints will yield quite different results.
\item \emph{(2+1)-dimensional Chern-Simons effective theory}: The fractional
quantum Hall effective action we obtain in this paper is through holographic
non-dynamical Chern-Simons model in 3+1 dimensions. This is the simplest
holographic model one can write down for the fractional quantum Hall
effect. But the same results can be equally obtained without holography,
by just starting with a (2+1)-dimensional Chern-Simons model, as is
typically done in the traditional effective field theory approach.
The reason is that the non-dynamical Chern-Simons terms are boundary
terms in holography. From this point of view, it seems holography
does not make a difference. This is only true so far as only the non-dynamical
Chern-Simons terms and the universal features of certain fixed fractional
quantum Hall states are concerned, and the universality of quantum
Hall physics at low energy guarantees that all methods converge at
this point. The true power of holography really lies in the analysis
of local dynamics which governs the non-universal properties, and
of quantum phase transitions such as the transitions between different
Hall plateaus. This leads to the next two remarks.
\item \emph{Local dynamics}: The majority of the quantum Hall research focuses
on the universal properties that is related to topology and Chern-Simons
field theory, so is the Chern-Simons model studied in this paper.
Properties related to local dynamics such as the Coulomb interaction
are much harder to study for such a strongly correlated quantum system
and the methods available are quite limited. Holography offers a strong
and efficient toolkit to deal with such a difficulty, and the holographic
dictionary we build in this paper opens the doorway to it. It will
be interesting to solve a specific holographic Ho\v{r}ava-Lifshitz
model with local dynamics, for example, specified by (\ref{eq:HoraveAction_Graviton})
and (\ref{eq:HoravaAction_Gauge}), maybe with additional scalar fields,
to see what local dynamics and non-universal properties it can generate
for the fractional quantum Hall effect. We will leave this in a forthcoming
study.
\item \emph{Holographic dynamical Chern-Simons and axion-dilaton models}:
In the non-dynamical Chern-Simons model we have studied in this paper,
the coefficient of the gauge Chern-Simons term is a constant. It can
be turned into a dynamical axion field and thus enters the bulk dynamics.
So is the coefficient of the Maxwell action (\ref{eq:HoravaAction_Gauge})
that can be turned into a dynamical dilaton. The relativistic version
of this model and its variations have been studied in \cite{Bayntun:2010nx,Fujita:2012fp}.
The additional $SL(2,\mathbb{Z})$ symmetry of the axion-dilation
system captures the transport features of the Hall plateau transitions.
This is another example of Holography as a powerful tool for studying
quantum phase transitions. A similar study can be carried out using
Ho\v{r}ava-Lifshitz holography using our holographic dictionary. We
expect it will give more information and features, both universal
and non-universal, than the simple Chern-Simons model we have studied
here.
\item \emph{Global time and energy flux}: In this paper, we work exclusively
in the global time coordinates and prohibit the foliation mixing diffeomorphism
$\partial_{i}\xi^{t}\neq0$. This keeps our formalism simple and convenient.
But a major drawback is the lack of the source field for energy flux,
thus our current formalism can not compute energy flux. The source
can be restored in non-relativistic field theory \cite{Geracie:2014nka}
and dealt with in non-relativistic holography by introducing additional
vector field in the bulk as shown in \cite{Janiszewski:2012nb}. A
closely related question is about the existence of global time. Clearly
it is needed for any genuine non-relativistic theory as part of the
causal structure. But for holography, this does not necessarily imply
the existence of global time in the whole bulk, as long as there is
another mechanism that provides the global time structure on the boundary.
In this paper, we assume global time for the whole bulk and thus employ
the Ho\v{r}ava-Lifshitz gravity theory as the holographic dual. This
is a simple and convenient choice conceptually, and practically is
enough for many purposes as long as the energy flux is not concerned.
But there can be other choices. For example, \cite{Christensen:2013lma,Christensen:2013rfa}
has shown that certain $z=2$ Lifshitz holography with relativistic
bulk theory can have a boundary whose geometry can be identified with
various versions of non-relativistic Newton-Cartan geometry, depending
on the boundary condition for the time-like vielbein. In a forthcoming
paper we will discuss the possibility of giving up bulk global time
and restoring the energy flux source and try to generalize our formalism
to a relativistic bulk.
\item \emph{Generalization to multiple species / hierarchical states}: Our
formalism, starting from the microscopic field theory action (\ref{eq:FTAction_S0}),
describes a single species of spin-polarized particles carrying mass
$m$, Abelian charge $e$ (which has been set to unity in this paper),
gyromagnetic factor $\mathrm{g}$ and intrinsic spin $s$. It thus
mostly naturally describe the Laughlin states of fractional quantum
Hall fluids with inverse filling factor equal to an odd integer. A
large number of other quantum Hall states can be described by Abelian
hierarchical states \cite{Halperin1983,Haldane1983,Jain1989} where
different types of quantum Hall fluids of quasiholes/quasiparticles
coexist. A field theoretical framework, the K-matrix formulation,
is introduced in \cite{Wen:1992uk}. In this formulation, the Abelian
gauge fields carry additional flavor index; the charges becomes a
flavor space vector and Chern-Simons couplings a matrix -- the K-matrix.
Such a formulation is employed recently in the construction of low
energy effective action in \cite{Abanov:2014ula,Gromov:2014gta}.
To accommodate this picture, our formalism need to be extended to
include multiple species of particles carrying different characteristics
such as mass and charge. Since the non-relativistic spacetime symmetries
(\ref{eq:FTDiffeo_psi})-(\ref{eq:FTDiffeo_hij}) and (\ref{eq:FTWeyl_psi})-(\ref{eq:FTWeyl_m})
depend on parameters such as mass and charge, they may need modifications
first. Thus this generalization is quite non-trivial. Generalization
to multi-species in non-relativistic holography is a non-trivial question
as well. A first trial is given in \cite{Balasubramanian:2010uw},
but this construction involves adding more extra dimensions to the
bulk, which is not a method suitable to the approach we employ in
this paper. We will leave this task to a forthcoming study.
\end{enumerate}
\bigskip{}

\section*{Acknowledgments\textmd{\normalsize{ \addcontentsline{toc}{section}{Acknowledgments}}}}

We are very grateful to Dam Thanh Son for constant support and inspiring
discussions throughout the progress of this work. We thank Stefan
Janiszewski and Andreas Karch for reading the draft of this paper
and making valuable and encouraging comments. We also thank Stefan
Janiszewski and Kristan Jensen for useful discussions during their
visits to EFI. This work is supported, in part, by the US DOE grant
No. DE-FG02-13ER41958, NSF DMS-1206648 and a Simons Investigator grant
from the Simons Foundation. S.-F. Wu is supported by NNSFC No. 11275120
and the China Scholarship Council.

\bigskip{}

\bigskip{}

\begin{appendices}

\section{General Formulae for Correlation Functions  }

If the effective action takes the following form: 
\begin{align}
\mathcal{W} & =\int dtd^{2}\vec{x}\Big\{\frac{\nu}{4\pi}\epsilon^{\rho\mu\nu}A_{\rho}\partial_{\mu}A_{\nu}-\frac{\nu}{2\pi}(s+s^{\prime})\epsilon^{\rho\mu\nu}A_{\rho}\partial_{\mu}\omega_{\nu}+\frac{\nu}{4\pi}(s+s^{\prime})^{2}\epsilon^{\rho\mu\nu}\omega_{\rho}\partial_{\mu}\omega_{\nu}\Big\}\nonumber \\
 & +\int dtd^{2}\vec{x}\sqrt{g}\left\{ \alpha\left[B,E^{2}\right]+f_{0}\left[B,\mathcal{R}^{(2)}\right]+f_{1}\left[B\right]E^{i}\nabla_{i}\log B+f_{2}\left[B\right]\left(\nabla_{i}\log B\right)^{2}+O\left(\epsilon^{3}\right)\right\} \:.
\end{align}
with $\Phi=0$ and $m$, $\mathrm{g}$, $s$ and $s^{\prime}$ are
all constants, then the correlation functions defined by (\ref{eq:ConservedQuantities})
can be computed as following. The non-equilibrium 1-point function
of the conserved current is 
\begin{eqnarray}
\langle J^{t}\rangle & = & \frac{\nu}{2\pi}\left[B-\left(s+s^{\prime}\right)\mathcal{R}^{(2)}\right]-\nabla_{i}\left(2E^{i}\frac{\delta\alpha}{\delta E^{2}}+f_{1}\nabla^{i}\log B\right)+O\left(\epsilon^{3}\right)\:,\\
\langle J^{i}\rangle & = & \frac{\nu}{2\pi}\varepsilon^{ij}\left[E_{j}-\left(s+s^{\prime}\right)\left(\partial_{j}\omega_{t}-\partial_{t}\omega_{j}\right)\right]+\frac{1}{\sqrt{g}}\partial_{t}\left[\sqrt{g}\left(2E^{i}\frac{\delta\alpha}{\delta E^{2}}+f_{1}\nabla^{i}\log B\right)\right]\nonumber \\
 &  & +\varepsilon^{ij}\nabla_{j}\left[\alpha^{\prime}-\frac{f_{1}}{B}\nabla_{k}E^{k}+f_{0}^{\prime}-f_{2}^{\prime}\left(\nabla_{i}\log B\right)^{2}-2\frac{f_{2}}{B}\nabla^{2}\log B\right]+O\left(\epsilon^{4}\right)\:,\\
 & = & \frac{\nu}{2\pi}\varepsilon^{ij}\left[E_{j}-\left(s+s^{\prime}\right)\left(\partial_{j}\omega_{t}-\partial_{t}\omega_{j}\right)\right]+\frac{1}{\sqrt{g}}\partial_{t}\left[\sqrt{g}\left(2E^{i}\frac{\delta\alpha}{\delta E^{2}}+f_{1}\nabla^{i}\log B\right)\right]\nonumber \\
 &  & +\Big\{\alpha^{\prime\prime}+f_{0}^{\prime\prime}-\left(\frac{f_{1}}{B}\right)^{\prime}\nabla_{k}E^{k}-2\left(\frac{f_{2}}{B^{2}}\right)^{\prime}\nabla^{2}B-\left[f_{2}^{\prime\prime}-2\left(\frac{f_{2}}{B}\right)^{\prime}\right]\left(\nabla_{k}\log B\right)^{2}\Big\}\varepsilon^{ij}\nabla_{j}B\nonumber \\
 &  & -2\frac{f_{2}}{B}\varepsilon^{ij}\nabla_{j}\nabla^{2}B-\left(f_{2}^{\prime}-2\frac{f_{2}}{B}\right)\varepsilon^{ij}\nabla_{j}\left[\left(\nabla_{k}\log B\right)^{2}\right]\nonumber \\
 &  & +\frac{\delta\alpha^{\prime}}{\delta E^{2}}\varepsilon^{ij}\nabla_{j}E^{2}-\frac{f_{1}}{B}\varepsilon^{ij}\nabla_{j}\nabla_{k}E^{k}+O\left(\epsilon^{4}\right)\:,
\end{eqnarray}
where ``$^{\prime}$'' is derivative with respect to $B$. The 2-point
functions are
\begin{eqnarray}
\langle J^{t}(x)J^{t}(0)\rangle & = & i\nabla^{i}\left[2\frac{\delta\alpha}{\delta E^{2}}\partial_{i}\delta(x)\right]+O\left(\epsilon^{3}\right)\:,\\
\langle J^{t}(x)J^{i}(0)\rangle & = & i\frac{\nu}{2\pi}\varepsilon^{ij}\partial_{j}\delta(x)-2i\nabla^{i}\left[\frac{\delta\alpha}{\delta E^{2}}\partial_{t}\delta(x)\right]\\
 &  & -i\varepsilon^{ij}\nabla^{k}\left[\left(2E_{k}\frac{\delta\alpha^{\prime}}{\delta E^{2}}+f_{1}^{\prime}\nabla_{k}\log B\right)\partial_{j}\delta(x)+f_{1}\nabla_{k}\left(\frac{\partial_{j}\delta(x)}{B}\right)\right]+O\left(\epsilon^{4}\right)\:,\nonumber \\
\langle J^{i}(x)J^{t}(0)\rangle & = & -i\frac{\nu}{2\pi}\varepsilon^{ij}\partial_{j}\delta(x)-\frac{2i}{\sqrt{g}}\partial_{t}\left[\sqrt{g}\frac{\delta\alpha}{\delta E^{2}}\partial^{i}\delta(x)\right]\nonumber \\
 &  & -i\varepsilon^{ij}\nabla_{j}\left[2E^{k}\frac{\delta\alpha^{\prime}}{\delta E^{2}}\partial_{k}\delta(x)-\frac{f_{1}}{B}\nabla^{2}\delta(x)\right]+O\left(\epsilon^{4}\right)\:,\\
\langle J^{i}(x)J^{j}(0)\rangle & = & i\frac{\nu}{2\pi}\varepsilon^{ij}\partial_{t}\delta(x)+\frac{i}{\sqrt{g}}\partial_{t}\left[2\sqrt{g}\frac{\delta\alpha}{\delta E^{2}}g^{ij}\partial_{t}\delta(x)\right]\nonumber \\
 &  & +\frac{i}{\sqrt{g}}\partial_{t}\left\{ \sqrt{g}\varepsilon^{jk}\left[\left(2E^{i}\frac{\delta\alpha^{\prime}}{\delta E^{2}}+f_{1}^{\prime}\nabla^{i}\log B\right)\partial_{k}\delta(x)+f_{1}\nabla^{i}\left(\frac{\partial_{k}\delta(x)}{B}\right)\right]\right\} \nonumber \\
 &  & +i\varepsilon^{ik}\nabla_{k}\left[2E^{j}\frac{\delta\alpha^{\prime}}{\delta E^{2}}\partial_{t}\delta(x)-\frac{f_{1}}{B}\partial^{j}\partial_{t}\delta(x)\right]\nonumber \\
 &  & +i\varepsilon^{ik}\varepsilon^{jl}\nabla_{k}\Big\{\Big[\alpha^{\prime\prime}-\left(\frac{f_{1}}{B}\right)^{\prime}\nabla\cdot\vec{E}+f_{0}^{\prime\prime}-f_{2}^{\prime\prime}\left(\nabla\log B\right)^{2}\nonumber \\
 &  & \qquad-2\left(\frac{f_{2}}{B}\right)^{\prime}\nabla^{2}\log B\Big]\partial_{l}\delta(x)\Big\}\nonumber \\
 &  & -i\varepsilon^{ik}\varepsilon^{jl}\nabla_{k}\left[2f_{2}^{\prime}\left(\nabla^{n}\log B\right)\nabla_{n}\left(\frac{\partial_{l}\delta(x)}{B}\right)+2\frac{f_{2}}{B}\nabla^{2}\left(\frac{\partial_{l}\delta(x)}{B}\right)\right]+O\left(\epsilon^{5}\right)\:,\nonumber \\
\end{eqnarray}
where we have assumed $\delta^{2}\alpha/\left(\delta E^{2}\right)^{2}=0$.
For flat background with constant $B$ and $E_{i}$ fields, the 2-point
functions simplify to 
\begin{eqnarray}
\langle J^{t}(x)J^{t}(0)\rangle & = & 2i\frac{\delta\alpha}{\delta E^{2}}\vec{\partial}^{2}\delta(x)+O\left(\epsilon^{3}\right)\:,\\
\langle J^{t}(x)J^{i}(0)\rangle & = & i\frac{\nu}{2\pi}\epsilon^{ij}\partial_{j}\delta(x)-2i\frac{\delta\alpha}{\delta E^{2}}\partial^{i}\partial_{t}\delta(x)\nonumber \\
 &  & -i\epsilon^{ij}\left[2\frac{\delta\alpha^{\prime}}{\delta E^{2}}E^{k}\partial_{k}\partial_{j}\delta(x)+\frac{f_{1}}{B}\vec{\partial}^{2}\partial_{j}\delta(x)\right]+O\left(\epsilon^{4}\right)\:,\\
\langle J^{i}(x)J^{t}(0)\rangle & = & -i\frac{\nu}{2\pi}\epsilon^{ij}\partial_{j}\delta(x)-2i\frac{\delta\alpha}{\delta E^{2}}\partial^{i}\partial_{t}\delta(x)\nonumber \\
 &  & -i\epsilon^{ij}\left[2\frac{\delta\alpha^{\prime}}{\delta E^{2}}E^{k}\partial_{k}\partial_{j}\delta(x)-\frac{f_{1}}{B}\vec{\partial}^{2}\partial_{j}\delta(x)\right]+O\left(\epsilon^{4}\right)\:,\\
\langle J^{i}(x)J^{j}(0)\rangle & = & i\frac{\nu}{2\pi}\epsilon^{ij}\partial_{t}\delta(x)+2i\delta^{ij}\frac{\delta\alpha}{\delta E^{2}}\partial_{t}^{2}\delta(x)-i\epsilon^{ij}\frac{f_{1}}{B}\vec{\partial}^{2}\partial_{t}\delta(x)\nonumber \\
 &  & +2i\left(\epsilon^{jk}E^{i}+\epsilon^{ik}E^{j}\right)\frac{\delta\alpha^{\prime}}{\delta E^{2}}\partial_{k}\partial_{t}\delta(x)\nonumber \\
 &  & +i\left(\alpha^{\prime\prime}+f_{0}^{\prime\prime}-2\frac{f_{2}}{B^{2}}\vec{\partial}^{2}\right)\left(\delta^{ij}\vec{\partial}^{2}-\partial^{i}\partial^{j}\right)\delta(x)+O\left(\epsilon^{5}\right)\:.
\end{eqnarray}
They have the following structures: 
\begin{eqnarray}
\langle J^{t}(x)J^{t}(0)\rangle & = & i\Pi_{0}\vec{\partial}^{2}\delta(x)\:,\\
\langle J^{t}(x)J^{i}(0)\rangle & = & -i\Pi_{0}\partial^{i}\partial_{t}\delta(x)+i\Pi_{1}\epsilon^{ij}\partial_{j}\delta(x)-i\Pi_{3}\epsilon^{ij}E^{k}\partial_{k}\partial_{j}\delta(x)\:,\\
\langle J^{i}(x)J^{t}(0)\rangle & = & -i\Pi_{0}\partial^{i}\partial_{t}\delta(x)-i\Pi_{1}\epsilon^{ij}\partial_{j}\delta(x)-i\Pi_{3}\epsilon^{ij}E^{k}\partial_{k}\partial_{j}\delta(x)\:,\\
\langle J^{i}(x)J^{j}(0)\rangle & = & i\Pi_{0}\delta^{ij}\partial_{t}^{2}\delta(x)+i\Pi_{1}\epsilon^{ij}\partial_{t}\delta(x)+i\Pi_{2}\left(\delta^{ij}\vec{\partial}^{2}-\partial^{i}\partial^{j}\right)\delta(x)\nonumber \\
 &  & +i\Pi_{3}\left(\epsilon^{jk}E^{i}+\epsilon^{ik}E^{j}\right)\partial_{k}\partial_{t}\delta(x)\:,
\end{eqnarray}
with 
\begin{eqnarray}
\Pi_{0} & = & 2\frac{\delta\alpha}{\delta E^{2}}+O\left(\epsilon\right)\:,\\
\Pi_{1} & = & \frac{\nu}{2\pi}-\frac{f_{1}}{B}\vec{\partial}^{2}+O\left(\epsilon^{3}\right)\:,\label{eq:HallConductivity_template}\\
\Pi_{2} & = & \alpha^{\prime\prime}+f_{0}^{\prime\prime}-2\frac{f_{2}}{B^{2}}\vec{\partial}^{2}+O\left(\epsilon^{3}\right)\:,\\
\Pi_{3} & = & 2\frac{\delta\alpha^{\prime}}{\delta E^{2}}+O\left(\epsilon\right)\:.
\end{eqnarray}
Notice that $\Pi_{1}$ is actually the Hall conductivity $\sigma_{H}$
defined in the usual way. The above expressions of 2-point functions
are usually written in momentum space, where $\partial_{t}\rightarrow-i\omega$,
$\partial_{i}\rightarrow ik_{i}$ and $\delta(x)\rightarrow1$.

\bigskip{}

\section{Near-Boundary Vectorial Structures and Diffeomorphism }

There are additional vectorial structures we can make up using the
boundary fields $\bar{N}$, $\bar{N}_{i}$, $\bar{G}_{ij}$ and $\bar{\phi}$.
Their diffeomorphism transformations at the boundary are 
\begin{eqnarray}
\delta\left(\bar{\varepsilon}^{jk}\partial_{j}\bar{N}_{k}\right) & = & \bar{\xi}^{\mu}\partial_{\mu}\left(\bar{\varepsilon}^{jk}\partial_{j}\bar{N}_{k}\right)+\left(\bar{\varepsilon}^{jk}\partial_{j}\bar{N}_{k}\right)\partial_{t}\bar{\xi}^{t}\nonumber \\
 &  & +\bar{\varepsilon}^{jk}\partial_{j}\left(\bar{G}_{kl}\partial_{t}\bar{\xi}^{l}\right)+2\bar{\varepsilon}^{jk}\bar{N}_{k}\partial_{j}\bar{\sigma}\:,\\
\delta\left(\bar{\varepsilon}^{jk}\bar{N}_{k}\partial_{j}\log\frac{\bar{N}}{\bar{\phi}}\right) & = & \bar{\xi}^{\mu}\partial_{\mu}\left(\bar{\varepsilon}^{jk}\bar{N}_{k}\partial_{j}\log\frac{\bar{N}}{\bar{\phi}}\right)+\left(\bar{\varepsilon}^{jk}\bar{N}_{k}\partial_{j}\log\frac{\bar{N}}{\bar{\phi}}\right)\partial_{t}\bar{\xi}^{t}\nonumber \\
 &  & +\left(\bar{\varepsilon}^{jk}\bar{G}_{kl}\partial_{j}\log\frac{\bar{N}}{\bar{\phi}}\right)\partial_{t}\bar{\xi}^{l}+(z-\Delta_{\phi})\bar{\varepsilon}^{jk}\bar{N}_{k}\partial_{j}\bar{\sigma}\:,\\
\delta\left(\bar{\varepsilon}^{jk}\bar{G}_{ki}\partial_{j}\log\frac{\bar{N}}{\bar{\phi}}\right) & = & \bar{\xi}^{\mu}\partial_{\mu}\left(\bar{\varepsilon}^{jk}\bar{G}_{ki}\partial_{j}\log\frac{\bar{N}}{\bar{\phi}}\right)+\left(\bar{\varepsilon}^{jk}\bar{G}_{kl}\partial_{j}\log\frac{\bar{N}}{\bar{\phi}}\right)\partial_{i}\bar{\xi}^{l}\nonumber \\
 &  & +(z-\Delta_{\phi})\bar{\varepsilon}^{jk}\bar{G}_{ki}\partial_{j}\bar{\sigma}\:.\\
\delta\left(\bar{\phi}\frac{\bar{G}^{jk}\bar{N}_{j}\bar{N}_{k}}{2\bar{N}}\right) & = & \bar{\xi}^{\mu}\partial_{\mu}\left(\bar{\phi}\frac{\bar{G}^{jk}\bar{N}_{j}\bar{N}_{k}}{2\bar{N}}\right)+\left(\bar{\phi}\frac{\bar{G}^{jk}\bar{N}_{j}\bar{N}_{k}}{2\bar{N}}\right)\partial_{t}\bar{\xi}^{t}\nonumber \\
 &  & +\left(\bar{\phi}\frac{\bar{N}_{k}}{\bar{N}}\right)\partial_{t}\bar{\xi}^{k}+(2-z+\Delta_{\phi})\bar{\sigma}\left(\bar{\phi}\frac{\bar{G}^{jk}\bar{N}_{j}\bar{N}_{k}}{2\bar{N}}\right)\:,\label{eq:NBVectorStruct_1}\\
\delta\left(\bar{\phi}\frac{\bar{N}_{i}}{\bar{N}}\right) & = & \bar{\xi}^{\mu}\partial_{\mu}\left(\bar{\phi}\frac{\bar{N}_{i}}{\bar{N}}\right)+\left(\bar{\phi}\frac{\bar{N}_{k}}{\bar{N}}\right)\partial_{i}\bar{\xi}^{k}\nonumber \\
 &  & +\frac{\bar{\phi}}{\bar{N}}\bar{G}_{ik}\partial_{t}\bar{\xi}^{k}+(2-z+\Delta_{\phi})\bar{\sigma}\left(\bar{\phi}\frac{\bar{N}_{i}}{\bar{N}}\right)\:.\label{eq:NBVectorStruct_2}
\end{eqnarray}
Here $\bar{\varepsilon}^{ij}$ is the Levi-Civita tensor associated
with the boundary metric $\bar{G}_{ij}$.

\bigskip{}

\section{Calculating the Gravitational Chern-Simons Term }

In this appendix, we add a ``$\bar{\phantom{a}}$'' to an index
to denote the corresponding tangent space (vielbein frame) index.
We extend the tangent space to 4 dimensions to include the temporal
direction as well.

\subsection{Bulk Spin Connection}

We define the vielbein 1-form in the bulk as: 
\begin{eqnarray}
E^{\bar{t}} & = & N\textrm{d}t\:,\\
E^{\bar{i}} & = & \frac{L}{r}e_{i}^{\bar{i}}\left(N^{i}\textrm{d}t+\textrm{d}x^{i}\right)\:,\\
E^{\bar{r}} & = & \frac{L}{r}\left(N^{r}\textrm{d}t+\textrm{d}r\right)\:,
\end{eqnarray}
where $\bar{H}_{ij}\equiv\delta_{\bar{i}\bar{j}}e_{i}^{\bar{i}}e_{j}^{\bar{j}}$
and $\bar{H}^{ij}$ is the matrix inverse of $\bar{H}_{ij}$. This
is compatible with the gauge condition (\ref{eq:BulkGaugeCondition_1})-(\ref{eq:BulkGaugeCondition_3}),
with $\Upsilon(r)=L/r$. So far we have no identified $e_{i}^{\bar{i}}$
with the boundary vielbein $e_{i}^{a}$. It is a bulk quantity and
a function of $r$ in general. $\bar{\varepsilon}_{ij}$ is the Levi-Civita
tensor defined via $\bar{H}_{ij}$. In this section we defined the
\emph{spatial} covariant derivative $\nabla_{i}$ is the one compatible
with the spatial metric $\bar{H}_{ij}$: $\nabla_{k}\bar{H}_{ij}=0$
and the indices involved in $\nabla_{i}$ only run through $x$ and
$y$. The spin connection 1-form $\Omega^{\bar{M}\bar{N}}=\Omega_{M}^{\bar{M}\bar{N}}\textrm{d}x^{M}$
is defined as 
\begin{equation}
\textrm{d}E^{\bar{M}}+\Omega_{\phantom{\bar{M}}\bar{N}}^{\bar{M}}\wedge E^{\bar{N}}=0\:,\qquad\Omega^{\bar{M}\bar{N}}=-\Omega^{\bar{N}\bar{M}}\:.
\end{equation}
Components of the spin connection can be computed using these two
equations. The results are 
\begin{eqnarray}
\Omega_{t}^{\bar{t}\bar{i}} & = & -\frac{1}{2N}\left(\frac{L}{r}\right)\Bigg[e_{i}^{\bar{i}}N^{j}\nabla_{j}N^{i}+\bar{H}_{jk}N^{k}e^{\bar{i}i}\nabla_{i}N^{j}+N^{r}e^{\bar{i}i}\partial_{r}\left(\bar{H}_{ij}N^{j}\right)+N^{r}e^{\bar{i}i}\partial_{i}N^{r}\nonumber \\
 &  & -e^{\bar{i}i}N^{j}\partial_{t}\bar{H}_{ij}-\frac{2}{r}N^{r}e_{i}^{\bar{i}}N^{i}\Bigg]+\frac{r}{L}e^{\bar{i}i}\partial_{i}N\:,\\
\Omega_{j}^{\bar{t}\bar{i}} & = & -\frac{1}{2N}\left(\frac{L}{r}\right)\left[e_{i}^{\bar{i}}\nabla_{j}N^{i}+e^{\bar{i}i}\nabla_{i}\left(\bar{H}_{jk}N^{k}\right)+N^{r}e^{\bar{i}i}\partial_{r}\bar{H}_{ij}-e^{\bar{i}i}\partial_{t}\bar{H}_{ij}-\frac{2}{r}N^{r}e_{j}^{\bar{i}}\right]\:,\\
\Omega_{r}^{\bar{t}\bar{i}} & = & -\frac{1}{2N}\left(\frac{L}{r}\right)\left[e_{i}^{\bar{i}}\partial_{r}N^{i}+e^{\bar{i}i}\partial_{i}N^{r}\right]\:,\\
\Omega_{t}^{\bar{t}\bar{r}} & = & -\frac{1}{2N}\left(\frac{L}{r}\right)\left[\bar{H}_{ij}N^{i}\partial_{r}N^{j}+N^{i}\partial_{i}N^{r}+2N^{r}\partial_{r}N^{r}-\frac{2}{r}\left(N^{r}\right)^{2}\right]+\frac{r}{L}\partial_{r}N\:,\\
\Omega_{i}^{\bar{t}\bar{r}} & = & -\frac{1}{2N}\left(\frac{L}{r}\right)\left[\bar{H}_{ij}\partial_{r}N^{j}+\partial_{i}N^{r}\right]\:,\\
\Omega_{r}^{\bar{t}\bar{r}} & = & -\frac{1}{N}\partial_{r}\left(\frac{L}{r}N^{r}\right)\:,\\
\Omega_{t}^{\bar{i}\bar{j}} & = & e^{\bar{i}k}\partial_{t}e_{k}^{\bar{j}}-\frac{1}{2}e^{\bar{i}i}e^{\bar{j}j}\partial_{t}\bar{H}_{ij}+\frac{1}{2}e^{\bar{i}i}e^{\bar{j}j}\left(\bar{H}_{ik}\nabla_{j}N^{k}-\bar{H}_{jk}\nabla_{i}N^{k}\right)\:,\\
\Omega_{k}^{\bar{i}\bar{j}} & = & e^{\bar{i}l}\nabla_{k}e_{l}^{\bar{j}}\:,\\
\Omega_{r}^{\bar{i}\bar{j}} & = & e^{\bar{i}k}\partial_{r}e_{k}^{\bar{j}}-\frac{1}{2}e^{\bar{i}i}e^{\bar{j}j}\partial_{r}\bar{H}_{ij}=-e^{\bar{j}k}\partial_{r}e_{k}^{\bar{i}}+\frac{1}{2}e^{\bar{i}i}e^{\bar{j}j}\partial_{r}\bar{H}_{ij}\:,\\
\Omega_{t}^{\bar{i}\bar{r}} & = & \frac{1}{2}e_{i}^{\bar{i}}\partial_{r}N^{i}-\frac{1}{2}e^{\bar{i}i}\partial_{i}N^{r}+\frac{1}{2}e^{\bar{i}i}N^{j}\partial_{r}\bar{H}_{ij}-\frac{1}{r}e_{i}^{\bar{i}}N^{i}\:,\\
\Omega_{j}^{\bar{i}\bar{r}} & = & \frac{1}{2}e^{\bar{i}i}\partial_{r}\bar{H}_{ij}-\frac{1}{r}e_{j}^{\bar{i}}\:,\\
\Omega_{r}^{\bar{i}\bar{r}} & = & 0\:.
\end{eqnarray}
Using $\Omega_{M}^{\bar{i}\bar{j}}=\frac{1}{2}\epsilon^{\bar{i}\bar{j}}\left(\epsilon_{\bar{k}\bar{l}}\Omega_{M}^{\bar{k}\bar{l}}\right)$,
$\Omega_{M}^{\bar{i}\bar{j}}$ can also be written as 
\begin{eqnarray}
\Omega_{t}^{\bar{i}\bar{j}} & = & \epsilon^{\bar{i}\bar{j}}\left(\frac{1}{2}\epsilon_{\bar{k}\bar{l}}e^{\bar{k}j}\partial_{t}e_{j}^{\bar{l}}-\frac{1}{2}\bar{\varepsilon}^{kl}\bar{H}_{jl}\nabla_{k}N^{j}\right)\:,\\
\Omega_{k}^{\bar{i}\bar{j}} & = & \epsilon^{\bar{i}\bar{j}}\left(\frac{1}{2}\epsilon_{\bar{k}\bar{l}}e^{\bar{k}j}\nabla_{k}e_{j}^{\bar{l}}\right)=\epsilon^{\bar{i}\bar{j}}\left(\frac{1}{2}\epsilon_{\bar{k}\bar{l}}e^{\bar{k}j}\partial_{k}e_{j}^{\bar{l}}-\frac{1}{2}\bar{\varepsilon}^{jl}\partial_{j}\bar{H}_{lk}\right)\:,\\
\Omega_{r}^{\bar{i}\bar{j}} & = & \epsilon^{\bar{i}\bar{j}}\left(\frac{1}{2}\epsilon_{\bar{k}\bar{l}}e^{\bar{k}j}\partial_{r}e_{j}^{\bar{l}}\right)\:.
\end{eqnarray}

\subsection{Pontryagin Density in the Bulk}

The gravitational Chern-Simons term: 
\begin{equation}
S_{\textrm{PY}}=\frac{c_{g}}{192\pi}\int d^{4}x\sqrt{-G^{(4)}}\left(^{*}\! RR\right)\:,
\end{equation}
where the Pontryagin density is 
\begin{equation}
^{*}\! RR={}^{*}\! R^{MNPQ}R_{NMPQ}\:,\qquad{}^{*}\! R^{MNPQ}=\frac{1}{2}\varepsilon^{PQRS}R_{\phantom{MN}RS}^{MN}\:.
\end{equation}
Here $R_{MNPQ}$ and $G^{(4)}$ are the Riemann tensor and determinant
constructed from the full 4-dimensional metric. The curvature 2-form
is defined as 
\begin{equation}
R_{\phantom{\bar{M}}\bar{N}}^{\bar{M}}=\textrm{d}\Omega_{\phantom{\bar{M}}\bar{N}}^{\bar{M}}+\Omega_{\phantom{\bar{M}}\bar{P}}^{\bar{M}}\wedge\Omega_{\phantom{\bar{P}}\bar{N}}^{\bar{P}}\:.
\end{equation}
The Pontryagin current 3-form is defined as 
\begin{equation}
\mathcal{P}=\Omega_{\phantom{\bar{M}}\bar{N}}^{\bar{M}}\wedge\textrm{d}\Omega_{\phantom{\bar{N}}\bar{M}}^{\bar{N}}+\frac{2}{3}\Omega_{\phantom{\bar{M}}\bar{N}}^{\bar{M}}\wedge\Omega_{\phantom{\bar{N}}\bar{P}}^{\bar{N}}\wedge\Omega_{\phantom{\bar{P}}\bar{M}}^{\bar{P}}=\mathcal{P}_{ij}\textrm{d}t\wedge\textrm{d}x^{i}\wedge\textrm{d}x^{j}+\ldots\wedge\textrm{d}r\:,
\end{equation}
where the part denoted by $\ldots$ is not relevant to our calculation,
so we will keep denoting it as $\ldots$ in the following. Taking
exterior derivative on this equation, we have 
\[
\textrm{d}\mathcal{P}=\left\{ -\partial_{r}\left(\epsilon^{ij}\mathcal{P}_{ij}\right)+\partial_{\mu}\left(\ldots\right)\right\} \textrm{d}t\wedge\textrm{d}x\wedge\textrm{d}y\wedge\textrm{d}r\:.
\]
On the other hand the Pontryagin density can be expressed as 
\[
\textrm{d}\mathcal{P}=\frac{1}{2}\sqrt{-G^{(4)}}\left(^{*}\! RR\right)\textrm{d}t\wedge\textrm{d}x\wedge\textrm{d}y\wedge\textrm{d}r\:.
\]
Thus we arrive at 
\begin{equation}
\sqrt{-G^{(4)}}{}^{*}\! RR=-2\partial_{r}\left(\epsilon^{ij}\mathcal{P}_{ij}\right)+\partial_{\mu}\left(\ldots\right)\:.
\end{equation}
This shows that the gravitational Chern-Simons term is a boundary
term as well: 
\begin{equation}
S_{\textrm{CS}}^{g}=\frac{c_{g}}{96\pi}\int_{r=0}d^{3}x\sqrt{\bar{H}}\bar{\varepsilon}^{ij}\bar{\mathcal{P}}_{ij}\:,
\end{equation}
where $\bar{\mathcal{P}}_{ij}$ is the boundary limit of $\mathcal{P}_{ij}$
and it will be computed explicitly in the next subsection.

\subsection{Pontryagin Current at the Boundary}

We now compute $\bar{\mathcal{P}}_{ij}$ at the boundary. Using the
near-boundary behaviors of the bulk fields: 
\[
N\Rightarrow\left(\frac{L}{r}\right)^{z}\bar{N}\:,\qquad N^{I}\Rightarrow\left(\frac{L}{r}\right)^{0}\bar{N}^{I}\:,\qquad\bar{H}_{ij}\Rightarrow\bar{G}_{ij}\:,
\]
the condition $\bar{N}_{r}=0$, and identify $e_{i}^{\bar{i}}$ at
the boundary with the boundary vielbein, the near-boundary behavior
of the spin connection is 
\begin{eqnarray}
\Omega^{\bar{t}\bar{i}} & \Rightarrow & \left(\frac{L}{r}\right)^{1-z}\alpha_{i}^{\bar{i}}\left(\textrm{d}x^{i}+\bar{N}^{i}\textrm{d}t\right)+\left(\frac{L}{r}\right)^{z-1}\left(e^{\bar{i}i}\partial_{i}\bar{N}\right)\textrm{d}t+\ldots\textrm{d}r\:,\\
\Omega^{\bar{t}\bar{r}} & \Rightarrow & \left(\frac{L}{r}\right)^{1-z}\beta_{i}\left(\textrm{d}x^{i}+\bar{N}^{i}\textrm{d}t\right)+\left(\frac{L}{r}\right)^{z-1}\left(\partial_{r}\bar{N}\right)\textrm{d}t\:,\\
\Omega^{\bar{i}\bar{r}} & \Rightarrow & \gamma_{i}^{\bar{i}}\left(\textrm{d}x^{i}+\bar{N}^{i}\textrm{d}t\right)+\left(\frac{1}{2}e_{i}^{\bar{i}}\partial_{r}\bar{N}^{i}\right)\textrm{d}t\:,\\
\Omega^{\bar{i}\bar{j}} & \Rightarrow & \frac{1}{2}\epsilon^{\bar{i}\bar{j}}\left(\Omega_{t}\textrm{d}t+\Omega_{k}\textrm{d}x^{k}\right)+\ldots\textrm{d}r\:,
\end{eqnarray}
where 
\begin{eqnarray*}
\alpha_{i}^{\bar{i}} & = & -\frac{1}{2\bar{N}}e^{\bar{i}j}\left(\nabla_{i}\bar{N}_{j}+\nabla_{j}\bar{N}_{i}-\partial_{t}\bar{G}_{ij}\right)\:,\\
\beta_{i} & = & -\frac{1}{2\bar{N}}\bar{G}_{ij}\partial_{r}\bar{N}^{j}\:,\\
\gamma_{i}^{\bar{i}} & = & \frac{1}{2}e^{\bar{i}j}\partial_{r}\bar{G}_{ij}-\frac{1}{r}e_{i}^{\bar{i}}\:,\\
\Omega_{t} & = & \epsilon_{\bar{i}\bar{j}}e^{\bar{i}l}\partial_{t}e_{l}^{\bar{j}}-\bar{\varepsilon}^{ij}\partial_{i}\bar{N}_{j}\:,\\
\Omega_{k} & = & \epsilon_{\bar{i}\bar{j}}e^{\bar{i}l}\nabla_{k}e_{l}^{\bar{j}}=\epsilon_{\bar{i}\bar{j}}e^{\bar{i}l}\partial_{k}e_{l}^{\bar{j}}-\bar{\varepsilon}^{ij}\partial_{i}\bar{G}_{jk}\:,
\end{eqnarray*}
and now the \emph{spatial} covariant derivative $\nabla_{i}$ is the
one compatible with the spatial metric $\bar{G}_{ij}$: $\nabla_{k}\bar{G}_{ij}=0$
and the indices involved in $\nabla_{i}$ only run through $x$ and
$y$. Then the relevant term in the Pontryagin current is given by
\begin{equation}
\mathcal{P}_{ij}\Rightarrow\left(\frac{r}{L}\right)^{2(z-1)}\bar{\mathcal{P}}_{ij}^{\textrm{sig}}+\bar{\mathcal{P}}_{ij}^{\textrm{reg}}\:,
\end{equation}
where 
\begin{eqnarray}
\bar{\mathcal{P}}_{ij}^{\textrm{sig}} & = & -\alpha_{i}^{\bar{i}}\left(\partial_{t}\alpha_{j}^{\bar{i}}\right)+\alpha_{i}^{\bar{i}}\partial_{j}\left(\alpha_{k}^{\bar{i}}\bar{N}^{k}\right)-\alpha_{k}^{\bar{i}}\bar{N}^{k}\left(\partial_{j}\alpha_{i}^{\bar{i}}\right)-\frac{1}{2}\epsilon_{\bar{i}\bar{j}}\alpha_{i}^{\bar{i}}\alpha_{j}^{\bar{j}}\Omega_{t}+\epsilon_{\bar{i}\bar{j}}\alpha_{i}^{\bar{i}}\alpha_{k}^{\bar{j}}\bar{N}^{k}\Omega_{j}\nonumber \\
 &  & -\beta_{i}\left(\partial_{t}\beta_{j}\right)+\beta_{i}\partial_{j}\left(\beta_{k}\bar{N}^{k}\right)-\beta_{k}\bar{N}^{k}\left(\partial_{j}\beta_{i}\right)-2\alpha_{i}^{\bar{i}}\beta_{j}\left(\gamma_{k}^{\bar{i}}\bar{N}^{k}+\frac{1}{2}e_{k}^{\bar{i}}\partial_{r}\bar{N}^{k}\right)\nonumber \\
 &  & +2\alpha_{k}^{\bar{i}}\bar{N}^{k}\beta_{j}\gamma_{i}^{\bar{i}}+2\beta_{k}\bar{N}^{k}\alpha_{i}^{\bar{i}}\gamma_{j}^{\bar{i}}-\left(i\leftrightarrow j\right)\:,\\
\bar{\mathcal{P}}_{ij}^{\textrm{reg}} & = & \left[\alpha_{i}^{\bar{i}}\left(\partial_{j}e^{\bar{i}k}\right)-e^{\bar{i}k}\left(\partial_{j}\alpha_{i}^{\bar{i}}\right)+\epsilon_{\bar{i}\bar{j}}\alpha_{i}^{\bar{i}}e^{\bar{j}k}\Omega_{j}+2e^{\bar{i}k}\beta_{j}\gamma_{i}^{\bar{i}}\right]\left(\partial_{k}\bar{N}\right)+\alpha_{i}^{\bar{i}}e^{\bar{i}k}\left(\partial_{j}\partial_{k}\bar{N}\right)\nonumber \\
 &  & +\left[-\gamma_{i}^{\bar{i}}\left(\partial_{j}\gamma_{k}^{\bar{i}}\right)+\gamma_{k}^{\bar{i}}\left(\partial_{j}\gamma_{i}^{\bar{i}}\right)-\epsilon_{\bar{i}\bar{j}}\gamma_{i}^{\bar{i}}\gamma_{k}^{\bar{j}}\Omega_{j}\right]\bar{N}^{k}-\gamma_{i}^{\bar{i}}\gamma_{k}^{\bar{i}}\left(\partial_{j}\bar{N}^{k}\right)+\gamma_{i}^{\bar{i}}\left(\partial_{t}\gamma_{j}^{\bar{i}}\right)\nonumber \\
 &  & +\frac{1}{2}\epsilon_{\bar{i}\bar{j}}\gamma_{i}^{\bar{i}}\gamma_{j}^{\bar{j}}\Omega_{t}-\frac{1}{4}\left[\Omega_{t}\left(\partial_{i}\Omega_{j}\right)+\Omega_{i}\left(\partial_{j}\Omega_{t}-\partial_{t}\Omega_{j}\right)\right]+\beta_{i}\left(\partial_{j}\partial_{r}\bar{N}\right)\nonumber \\
 &  & -\frac{1}{2}\gamma_{i}^{\bar{i}}e_{k}^{\bar{i}}\left(\partial_{j}\partial_{r}\bar{N}^{k}\right)+\frac{1}{2}\left[-\gamma_{i}^{\bar{i}}\left(\partial_{j}e_{k}^{\bar{i}}\right)+e_{k}^{\bar{i}}\left(\partial_{j}\gamma_{i}^{\bar{i}}\right)-\epsilon_{\bar{i}\bar{j}}\gamma_{i}^{\bar{i}}e_{k}^{\bar{j}}\Omega_{j}\right]\left(\partial_{r}\bar{N}^{k}\right)\nonumber \\
 &  & +\left(2\alpha_{i}^{\bar{i}}\gamma_{j}^{\bar{i}}-\partial_{j}\beta_{i}\right)\left(\partial_{r}\bar{N}\right)-\left(i\leftrightarrow j\right)\:,
\end{eqnarray}
and $\bar{i}$ index in terms such as $\alpha_{i}^{\bar{i}}\left(\partial_{t}\alpha_{j}^{\bar{i}}\right)$
are understood to be summed over, which is the same as $\delta_{\bar{i}\bar{j}}\alpha_{i}^{\bar{i}}\left(\partial_{t}\alpha_{j}^{\bar{j}}\right)$.

There are simplifications that can dramatically reduce the number
of terms in the above expressions. First, we assume $z>1$, then $\bar{\mathcal{P}}_{ij}^{\textrm{sig}}$
completely drops off near the boundary $r\rightarrow0$. This can
also be understood by thinking of the non-relativistic limit $c\rightarrow\infty$.
We can restore the dependence of speed of light $c$ by just replacing
$N$ by $cN$, hence $\bar{N}$ by $c\bar{N}$ in the above expressions.
Both $\alpha_{i}^{\bar{i}}$ and $\beta_{i}$ are proportional to
$\bar{N}^{-1},$ thus are of order $O(c^{-1})$. Every term in $\bar{\mathcal{P}}_{ij}^{\textrm{sig}}$
is either quadratic in $\alpha_{i}^{\bar{i}}$ or $\beta_{i}$, or
proportional to a product of them. Thus we have 
\[
\bar{\mathcal{P}}_{ij}^{\textrm{sig}}\sim O\left(\frac{1}{c^{2}}\right)\:,\qquad\bar{\mathcal{P}}_{ij}=\bar{\mathcal{P}}_{ij}^{\textrm{reg}}+O\left(\frac{1}{c^{2}}\right)\:.
\]
Of course when $z=1$, corresponding to the Lorentzian case, it gives
a finite contribution, but for simplicity, we will not consider this
case here. The second simplification comes from adding the Gibbons-Hawking
term (\ref{eq:ChernSimons:GibbonsHawking}). As we have argued before,
it completely removes all $\partial_{r}$ terms without changing any
other term. Thus all $\partial_{r}$ terms in $\bar{\mathcal{P}}_{ij}^{\textrm{reg}}$,
including all terms containing $\beta_{i}$ and $\gamma_{i}^{\bar{i}}$,
drop off.%
\footnote{The $\frac{1}{r}e_{i}^{\bar{i}}$ term in $\gamma_{i}^{\bar{i}}$
is a consequence of $\partial_{r}$ hitting $\frac{L}{r}$, thus is
part of the $\partial_{r}$ terms that are removed by the Gibbons-Hawking
term.%
} By using the identity 
\[
\epsilon_{\bar{i}\bar{j}}e^{\bar{i}l}\nabla_{k}e_{l}^{\bar{j}}=-\bar{\varepsilon}_{jl}e^{\bar{i}j}\nabla_{k}e^{\bar{i}l}
\]
the first line in the expression of $\bar{\mathcal{P}}_{ij}^{\textrm{reg}}$
can also be simplified. At the end, we have 
\begin{align}
 & \bar{\varepsilon}^{ij}\bar{\mathcal{P}}_{ij}\nonumber \\
= & \bar{\varepsilon}^{ij}\bar{G}^{kl}\left\{ \left(\nabla_{k}\bar{N}\right)\nabla_{j}\left[\frac{1}{\bar{N}}\left(\nabla_{i}\bar{N}_{l}+\nabla_{l}\bar{N}_{i}-\partial_{t}\bar{G}_{il}\right)\right]-\left(\nabla_{j}\nabla_{k}\bar{N}\right)\frac{1}{\bar{N}}\left(\nabla_{i}\bar{N}_{l}+\nabla_{l}\bar{N}_{i}-\partial_{t}\bar{G}_{il}\right)\right\} \nonumber \\
 & -\frac{1}{2}\bar{\varepsilon}^{ij}\left[\Omega_{t}\partial_{i}\Omega_{j}+\Omega_{i}\left(\partial_{j}\Omega_{t}-\partial_{t}\Omega_{j}\right)\right]+O\left(\partial_{r}\right)\:,
\end{align}
for $z>1$.

\bigskip{}

\end{appendices}

\bigskip{}

\end{document}